\documentclass[twocolumn]{aastex63}
\usepackage{rotating}
\usepackage{multirow}

\usepackage{textcomp}
\usepackage{soul}

\received{}
\revised{}
\accepted{}
\submitjournal{ApJ}
\shorttitle{First high-contrast images of X-ray binaries}
\shortauthors{Prasow-\'{E}mond et al.}

\graphicspath{{./}{figures/}}

\begin{document}

\title{The First High-Contrast Images of Near High-Mass X-Ray Binaries with Keck/NIRC2}

\correspondingauthor{Myriam Prasow-\'{E}mond}
\email{m.prasow-emond22@imperial.ac.uk}

\author[0000-0002-2457-3431]{M. Prasow-Émond}
\affiliation{Department of Earth Science and Engineering, Imperial College London, Prince Consort Rd, London SW7 2BP, United Kingdom}
\affiliation{D\'{e}partement de Physique, Universit\'{e} de Montr\'{e}al, C.P. 6128, Succ. Centre-Ville, Montr\'{e}al, QC H3C 3J7, Canada}
\affiliation{Trottier Institute for Research on Exoplanets, Université de Montréal, Département de Physique, C.P. 6128 Succ. Centre-ville, Montréal, QC H3C 3J7, Canada}

\author[0000-0001-7271-7340]{J. Hlavacek-Larrondo}
\affiliation{D\'{e}partement de Physique, Universit\'{e} de Montr\'{e}al, C.P. 6128, Succ. Centre-Ville, Montr\'{e}al, QC H3C 3J7, Canada}

\author[0000-0002-2691-2476]{K. Fogarty}
\affiliation{Division of Physics, Math, and Astronomy, California Institute of Technology, Pasadena, CA, USA}
\affiliation{NASA Ames Research Center, Moffett Field, CA 94035, USA}

\author[00000-0003-3506-5667]{É. Artigau}
\affiliation{Trottier Institute for Research on Exoplanets, Université de Montréal, Département de Physique, C.P. 6128 Succ. Centre-ville, Montréal, QC H3C 3J7, Canada}

\author[0000-0002-8895-4735]{D. Mawet}
\affiliation{Division of Physics, Math, and Astronomy, California Institute of Technology, Pasadena, CA, USA}
\affiliation{Jet Propulsion Laboratory, California Institute of Technology, Pasadena, CA 91109, USA}

\author[0000-0003-3105-2615]{P. Gandhi}
\affiliation{Department of Physics and Astronomy, University of Southampton, Highfield, Southampton, SO17 1BJ, United Kingdom}

\author[0000-0002-5872-6061]{J. F. Steiner}
\affiliation{Harvard-Smithsonian Center for Astrophysics, Cambridge, MA 02138, United States}

\author[0000-0003-0029-0258]{J. Rameau}
\affiliation{Trottier Institute for Research on Exoplanets, Université de Montréal, Département de Physique, C.P. 6128 Succ. Centre-ville, Montréal, QC H3C 3J7, Canada}
\affiliation{Univ. Grenoble Alpes, CNRS, IPAG, F-38000 Grenoble, France}

\author[0000-0002-6780-4252]{D. Lafreni\`{e}re}
\affiliation{D\'{e}partement de Physique, Universit\'{e} de Montr\'{e}al, C.P. 6128, Succ. Centre-Ville, Montr\'{e}al, QC H3C 3J7, Canada}
\affiliation{Trottier Institute for Research on Exoplanets, Université de Montréal, Département de Physique, C.P. 6128 Succ. Centre-ville, Montréal, QC H3C 3J7, Canada}

\author[0000-0002-9378-4072]{A. Fabian}
\affiliation{Institute of Astronomy, Cambridge University, Madingley Road, Cambridge CB3 0HA, United Kingdom}

\author[0000-0001-5819-3552]{D. J. Walton}
\affiliation{Centre for Astrophysics Research, University of Hertfordshire, College Lane, Hatfield AL10 9AB, United Kingdom}

\author[0000-0001-5485-4675]{R. Doyon}
\affiliation{D\'{e}partement de Physique, Universit\'{e} de Montr\'{e}al, C.P. 6128, Succ. Centre-Ville, Montr\'{e}al, QC H3C 3J7, Canada}
\affiliation{Trottier Institute for Research on Exoplanets, Université de Montréal, Département de Physique, C.P. 6128 Succ. Centre-ville, Montréal, QC H3C 3J7, Canada}

\author[0000-0003-1698-9696]{B. B. Ren}
\affiliation{Division of Physics, Math, and Astronomy, California Institute of Technology, Pasadena, CA, USA}

\begin{abstract}
Although the study of X-ray binaries has led to major breakthroughs in high-energy astrophysics, their circumbinary environment at scales of $\sim$100--10,000 astronomical units has not been thoroughly investigated. In this paper, we undertake a novel and exploratory study by employing direct and high-contrast imaging techniques on a sample of X-ray binaries, using adaptive optics and the vortex coronagraph on Keck/NIRC2. High-contrast imaging opens up the possibility to search for exoplanets, brown dwarfs, circumbinary companion stars, and protoplanetary disks in these extreme systems. Here, we present the first near-infrared high-contrast images of 13 high-mass X-ray binaries located within $\sim$2--3 kpc. The key results of this campaign involve the discovery of several candidate circumbinary companions ranging from sub-stellar (brown dwarf) to stellar masses. By conducting an analysis based on galactic population models, we discriminate sources that are likely background/foreground stars and isolate those that have a high probability ($\gtrsim 60 - 99\%$) of being gravitationally bound to the X-ray binary. This publication seeks to establish a preliminary catalog for future analyses of proper motion and subsequent observations. With our preliminary results, we calculate the first estimate of the companion frequency and the multiplicity frequency for X-ray binaries: $\approx$0.6 and 1.8 $\pm$ 0.9 respectively, considering only the sources that are most likely bound to the X-ray binary. In addition to extending our comprehension of how brown dwarfs and stars can form and survive in such extreme systems, our study opens a new window to our understanding of the formation of X-ray binaries.
\end{abstract}


\section{Introduction} \label{sec:intro}

X-ray binaries are semidetached binary systems in which a compact object (white dwarf; WD, neutron star; NS, or stellar-mass black hole; BH) accretes material from a donor star. These systems undergo several extreme physical phenomena, such as processes acting predominantly in soft X-rays (e.g., \citealt{2010ApJ...716.1105K,2021MNRAS.501.3406T}), and detectable X-ray pulsations (e.g., \citealt{2005A&A...444..821L}). 

The variations in physical processes among different X-ray binaries are directly linked to the mass of the donor star. Over 90\% of these systems can be classified into two distinct categories: high-mass X-ray binaries (HMXB; $M_\mathrm{donor} \gtrsim $ 8 $M_\odot$) and low-mass X-ray binaries (LMXB; $M_\mathrm{donor} \lesssim $ 1,5 $M_\odot$; e.g., \citealt{2006csxs.book..623T}). LMXBs are relatively old systems ($> 10^9$ yr) harboring a K-M spectral type donor star, where the process of mass transfer is believed to be triggered by Roche-lobe overflow (RLO; e.g., \citealt{1978A&A....62..317S}). RLO is triggered either by stellar evolution or by angular momentum loss (e.g., \citealt{1967AcA....17..287P,1981A&A...100L...7V,2002ApJ...581..577S,2006MNRAS.366.1415J,2016ApJ...830..131C,2018MNRAS.475.1392S,2019MNRAS.483.5595V}). The transferred mass then agglomerates to form an accretion disk around the compact object, giving rise to transient accretion and X-ray emission (e.g., \citealt{2006csxs.book..215C}).

As for HMXBs, they are generally thought to be younger systems ($\lesssim 10^7$ yr) harboring a massive O-B spectral type donor star. The transferred and accreted matter is thought to predominantly come from the capture of a fraction of the stellar winds ejected from the donor star (e.g., \citealt{2007ASPC..372..397M,2013A&A...552A..26A,2019A&A...622L...3E}). There are two sub-categories of HMXBs relevant to this work. Firstly, we emphasize Be/X-ray binaries (BeXRB), wherein the donor star is a fast-rotating Be star. In these systems, the X-ray emission is mainly triggered by the compact object passing through a diffuse and gaseous circumstellar disk surrounding the Be star (known as a decretion disk; e.g., \citealt{2002MNRAS.337..967O, 2011MNRAS.416.2827M, 2018MNRAS.476.3555R, 2020A&A...643A.170K}). Secondly, we highlight supergiant fast X-ray transients (SFXT; \citealt{2006ESASP.604..165N}), characterized by the presence of a supergiant donor star, and by fast transient X-ray flaring activity within the system (likely induced by a NS; e.g., \citealt{2013arXiv1301.7574S, 2019A&A...631A.135D}).

X-ray binaries are important touchstone objects for high-energy phenomena in astrophysics. They have been widely used to study several high-energy astronomical phenomena, including accretion physics (e.g., \citealt{2007A&ARv..15....1D,kara_corona_2019}) and outflow/jet processes (e.g., \citealt{2001A&A...372L..25M,2004MNRAS.355.1105F, 2018Natur.554..207M}). However, the immediate surroundings of X-ray binaries, at the scale of $\sim$100-10,000 astronomical units (hereafter au), have been poorly studied. This paper undertakes a pioneering exploration of the circumstellar environments of X-ray binaries through the application of adaptive optics (AO) and direct/high-contrast imaging techniques. The goal is to probe a variety of phenomena ranging from protoplanetary disks to debris disks and fallback disks, and particularly to search for wide-orbiting circumbinary companions (CBCs) -- be they exoplanets, brown dwarfs, or stars.

Considering the discovery of planetary-mass CBCs orbiting both binary systems (e.g., \citealt{2007ApJ...656..552B,2007A&A...462..345D,2020A&A...638L...6E}) and compact objects (WD or pulsars; e.g., \citealt{1992Natur.355..145W, 2003Sci...301..193S, 2018MNRAS.475..469S,2020Natur.585..363V,2021Natur.598..272B}),  it is not unfounded to expect CBCs orbiting X-ray binaries. A recent study argued that X-ray binaries could host planetary systems in close orbits detectable via X-ray eclipses \citep{imara_searching_2018}. In this paper, we explore wider orbits ($\sim$100--10,000 au), as the increased number of interactions within the system could lead to the ejection of potential CBCs from the direct environment of the X-ray binary (e.g., \citealt{bonavita_spots_2016}).

In \cite{2022AJ....164....7P}, we presented the first set of observations from a pilot study aiming to survey all X-ray binaries amenable for direct imaging within $\sim$3 kpc. We first targeted a $\gamma$ Cassiopeiae-like X-ray binary harboring a Be donor star, RX J1744.7$-$2713, for which we had observations from two different bands and two epochs. We unveiled the presence of three potential CBCs within this system, exhibiting a strong likelihood of being stellar-mass CBCs. Here, we present the first $L'$-band high-contrast images of 13 other systems and conduct a preliminary statistical analysis derived from the results of the first epochs of observations.

The paper is organized as follows: Section \ref{sec:sample} presents the sample and how it was constructed. Section \ref{sec:data} presents the near-infrared observations, and the data reduction and processing. Section \ref{sec:images} presents the first high-contrast images of the observed X-ray binaries. In Section \ref{sec:analysis}, we analyze the images and explore the nature of the detection. Finally, in Section \ref{sec:discussion}, we discuss our results and their implications.

\section{The Sample} \label{sec:sample}

Despite the ongoing active search for new X-ray binaries both within and beyond our Galaxy (e.g., \citealt{2022MNRAS.510.3885G}), their presence remains relatively scarce. Our Galaxy hosts $\sim$300 identified X-ray binaries known to date \citep{2006A&A...455.1165L, 2007A&A...469..807L}. We drew upon this list of X-ray binaries as our initial dataset; however, not all of these systems are suitable for direct imaging with Keck/NIRC2. In order to build a sample of X-ray binaries that would yield optimal statistical constraints and mitigate potential biases, we used four selection criteria:

\begin{enumerate}
    \item \textbf{Distance.} The system must be close enough to resolve the direct environment at $\sim 100-10,000$ au scales. We chose a distance limit of $\sim$3 kpc within our Galaxy, which enables the detection of structures and objects located within a couple of thousands of au from the X-ray binary. The outer limit (10,000 au) corresponds to the approximate limit of the Keck/NIRC2 field of view (foV).
    
    \item \textbf{Brightness and Adaptive optics.} The donor star must be bright enough ($I < 9-10$ mag) for the AO loop to be closed.
    
    \item \textbf{Age.} We targeted young ($\lesssim$ 100 Myr) X-ray binaries to favor the detection of sub-stellar CBCs, considering the steep decline in planet brightness with time (e.g., \citealt{2001RvMP...73..719B}). This limited us to bright LMXBs, and HMXBs with a massive O/B donor star. 
    
    \item \textbf{Visibility.} We selected X-ray binaries visible from the W. M. Keck Observatory at the time of our observations (Keck Observatory Semesters 2017B and 2020A).
\end{enumerate}

Applying these criteria narrowed down the initial list from $\sim$300 X-ray binaries to 19, out of which 14 were observed between 2017 and 2020 using Keck/NIRC2 (see Section \ref{sec:observations} for more details). Our sample includes both HMXBs and LMXBs (e.g., MAXI J1820+070, V404 Cyg, 1A0620-00); however, we have only observed HMXBs to date due to observational constraints. Note that X-ray binary surveys are far from complete, as X-ray binaries can be undetectable in quiescence (e.g., \citealt{2007ApJS..170..175B,2009ApJ...707..870B}). New X-ray binaries have also been discovered since we did our sampling \citep{2023A&A...675A.199A, 2023A&A...677A.134N}. Nonetheless, we assume that our sample is as complete as possible and that our results accurately represent the known population of X-ray binaries.

\begin{figure*}
    \centering
    \includegraphics[width=\textwidth]{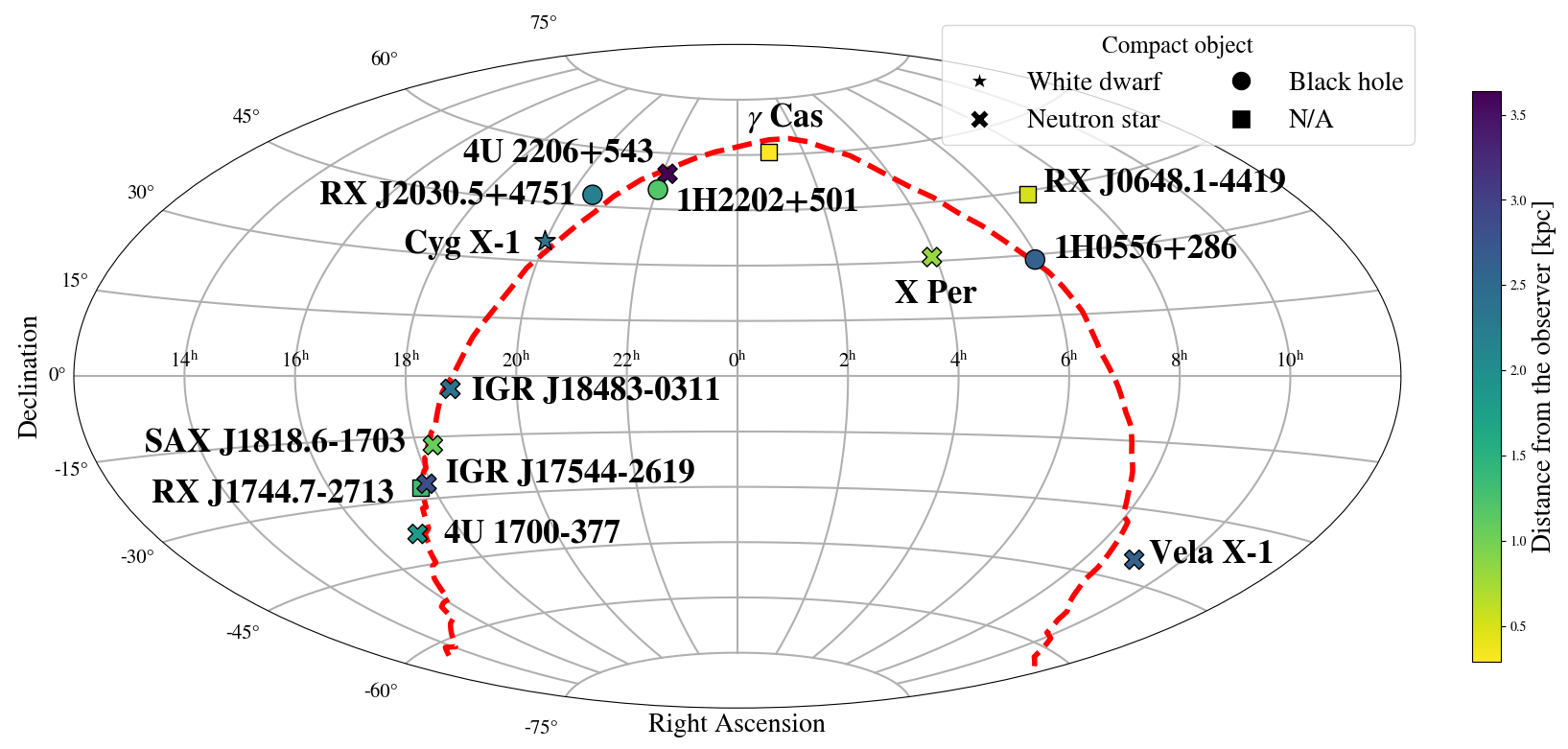}
    \caption{Position of the 14 observed X-ray binaries on the sky with equatorial coordinates and Aitoff projection, color coded with distance from the observer in kpc. The nature of the compact object is illustrated with different markers: squares for white dwarfs, crosses for neutron stars, asterisks for black holes and circles for cases where the nature is either unavailable or uncertain. The red dotted line shows the approximate coordinates of the galactic plane.}
    \label{fig:skymap}
\end{figure*}

Figure \ref{fig:skymap} displays the position of the 14 observed targets onto the sky (Aitoff projection). It also indicates the distance from the observer (in kpc) and, if known, the nature of the system's compact object. In Section \ref{sec:litrev}, we present a brief literature review for each of the observed X-ray binaries. Table \ref{tab:XRB_properties} summarizes the known relevant physical properties of the systems, namely the X-ray binary type, X-ray emission class, donor star spectral type and compact object type. Additional relevant physical properties can be found in the Appendix (see Table \ref{tab:XRB_add_properties}).

\begin{table*}
\caption{Information on the binary nature for the 14 observed X-ray binaries. The columns are: 1. Name of the target; 2. Sub-class/type of the X-ray binary, as found in \cite{2016ApJS..223...15B}; 3. Class of X-ray emission, as found in \cite{2013ApJS..209...14K}; 4. The spectral type of the donor star and 5. the reference; 6. The nature of the compact object and 7. the reference. $^\dagger$Indicates an uncertain nature. \textbf{Reference code}: (1) \cite{sarty_photometric_2011}, (2) \cite{2006A&A...454..265L}, (3) \cite{2008A&A...484..783C}, (4) \cite{2010A&A...510A..61T}, (5) \cite{2007A&A...467..249S}, (6) \cite{1973A&A....23..433M}, (7) \cite{2005A&AT...24..151R}, (8) \cite{2017MNRAS.465L.119P}, (9) \cite{2007A&A...476..335W}, (10) \cite{2019KPCB...35...38S}, (11) \cite{2001A&A...371.1056N}, (12) \cite{2004A&A...423..301T}, (13) \cite{2014ApJS..211...10S}, (14) \cite{1999A&A...349..873R}, (15) \cite{2006A&A...455..653P}, (16) \cite{2016A&A...591A..26G}, (17) \cite{2017AstL...43..664B}, (18) \cite{2005A&A...441L...1I}, (19) \cite{1997A&A...323..853M}, (20) \cite{2011ApJS..193...24S}, (21) \cite{1972NPhS..240..124B}, (22) \cite{1997MNRAS.286..549L}, (23) \cite{1977Natur.267..229W}, (24) \cite{1950ApJ...111..495P}, (25) \cite{2006A&A...455.1165L}, (26) \cite{1963PASP...75..365J}, (27) \cite{2018MNRAS.474.2750P}, (28) \cite{1972ApJ...175L..19H}.}
\begin{tabular}{ccccccc}
\hline\hline
Target & Type & Class & Donor star & Ref. & Compact Object & Ref. \\ \hline
\multirow{3}{*}{RX J1744.7$-$2713} & HMXB & \multirow{3}{*}{N/A} & \multirow{3}{*}{B0.5 V-IIIe} & \multirow{3}{*}{(1)} & \multirow{3}{*}{WD$^\dagger$} & \multirow{3}{*}{(2)} \\
 & BeXRB &  &  &  &  &  \\
 & $\gamma$ Cas analog &  &  &  &  &  \\ \hline
\multirow{3}{*}{IGR J18483$-$0311} & HMXB & \multirow{3}{*}{Outburst} & \multirow{3}{*}{B0.5-BI/B0-1 Iab} & \multirow{3}{*}{(3, 4)} & \multirow{3}{*}{NS} & \multirow{3}{*}{(5)} \\
 & SFXT &  &  &  &  &  \\
 & XP &  &  &  &  &  \\ \hline
\multirow{2}{*}{$\gamma$ Cas} & HMXB & \multirow{2}{*}{Steady} & \multirow{2}{*}{B0.5 IVe} & \multirow{2}{*}{(6, 7)} & \multirow{2}{*}{WD$^\dagger$} & \multirow{2}{*}{(8)} \\
 & BeXRB &  &  &  &  &  \\ \hline
\multirow{2}{*}{SAX J1818.6$-$1703} & HMXB & \multirow{2}{*}{Variable} & \multirow{2}{*}{B0.5 Iab} & \multirow{2}{*}{(4)} & \multirow{2}{*}{NS$^\dagger$} & \multirow{2}{*}{(9)} \\
 & SFXT &  &  &  &  & \\ \hline
 \multirow{2}{*}{1H2202+501} & HMXB & \multirow{2}{*}{N/A} & \multirow{2}{*}{B3Ve} & \multirow{2}{*}{(10)} & \multirow{2}{*}{N/A} & \multirow{2}{*}{} \\
 & BeXRB$^\dagger$ &  &  &  &  & \\ \hline
 4U 2206+543 & HMXB & Variable & O9.5 V & (11) & NS & (12) \\ \hline
 4U 1700$-$377 & sgHMXB & Variable & O6Iafpe & (13) & NS & (14) \\ \hline
\multirow{2}{*}{IGR J17544$-$2619} & HMXB & \multirow{2}{*}{Flaring} & \multirow{2}{*}{O9Ib/O9IV-V} & \multirow{2}{*}{(15, 16, 17)} & \multirow{2}{*}{NS} & \multirow{2}{*}{(18)} \\ 
 & SFXT &  &  &  &  &  \\ \hline
\multirow{3}{*}{RX J2030.5+4751} & HMXB & \multirow{3}{*}{N/A} & \multirow{3}{*}{B0.5 III-Ve} & \multirow{3}{*}{(19)} & \multirow{3}{*}{N/A} & \multirow{3}{*}{} \\
 & BeXRB &  &  &  &  &  \\ 
 & $\gamma$ Cas analog &  &  &  &  &  \\ \hline
\multirow{2}{*}{Cyg X-1} & HMXB & \multirow{2}{*}{Variable} & \multirow{2}{*}{O9.7Iabpvar} & \multirow{2}{*}{(20)} & \multirow{2}{*}{BH} & \multirow{2}{*}{(21)}  \\
 & Microquasar &  &  &  &  &  \\ \hline
 \multirow{3}{*}{X Per} & HMXB & \multirow{3}{*}{Variable} & \multirow{3}{*}{B0Ve} & \multirow{3}{*}{(22)} & \multirow{3}{*}{NS} & \multirow{3}{*}{(23)} \\
 & BeXRB &  &  &  &  &  \\ 
 & XP &  &  &  &  &  \\ \hline
1H0556+286 & HMXB$^\dagger$ & N/A & B5ne & (24) & WD or NS & (25) \\ \hline
\multirow{2}{*}{RX J0648.1$-$4419} & HMXB & \multirow{2}{*}{N/A} & \multirow{2}{*}{sdO5.5} & \multirow{2}{*}{(26)} & \multirow{2}{*}{WD} & \multirow{2}{*}{(27)} \\
 & XP &  &  &  &  &  \\ \hline
\multirow{2}{*}{Vela X-1} & HMXB & \multirow{2}{*}{Variable} & \multirow{2}{*}{B0.5 Ib} & \multirow{2}{*}{(28)} & \multirow{2}{*}{NS} & \multirow{2}{*}{(28)} \\
 & XP &  &  &  &  & \\ \hline
\end{tabular}
\label{tab:XRB_properties}
\end{table*}

\section{Observations and Data Reduction} \label{sec:data}
\subsection{Keck/NIRC2 Observations}\label{sec:observations}
On September 8, 2017, we observed four HMXBs from our survey sample using the Keck/NIRC2 vortex coronagraph \citep{2005ApJ...633.1191M, 2016OptCo.379...64S} in pupil-tracking mode in $L'$-band ($\lambda$ = 3.776 $\mu$m, $\Delta \lambda$ = 0.700 $\mu$m; PI: Mawet), and with the narrow camera (plate scale of 9.971 $\pm$ 0.004 mas pixel$^{-1}$; \citealt{2016PASP..128i5004S}). On January 3, 2018, we observed three additional HMXBs using a similar setting. Due to the successful and promising preliminary results of this initial campaign, we were awarded three supplementary nights of observation on July 11, 12, and 13, 2020 (PI: Fogarty). On the first night, we observed one additional target and re-observed two targets (RX J1744.7$-$2713 and $\gamma$ Cas) using a similar setting. However, due to saturation, we had to downscale the frame size of $\gamma$ Cas from 1024$\times$1024 pixel$^2$ to 512$\times$512 pixel$^2$. On July 12, 2020, we obtained data for three other HMXBs in $L'$-band, in addition to re-observing RX J1744.7$-$2713 in $K_s$-band ($\lambda$ = 2.146 $\mu$m, $\Delta \lambda$ = 0.311 $\mu$m). Finally, on July 13, 2020, we obtained data for three other HMXBs, totaling observations for 14 out of the 19 X-ray binaries in the sample.

During the observations, we used the Quadrant Analysis of Coronagraphic Images for Tip-tilt sensing (QACITS; \citealt{2017A&A...600A..46H}) to make tip-tilt adjustments to maintain precise centering of the target on the vortex focal plane mask. The observations were AO-assisted using the Shack-Hartmann wavefront sensor (which performs wavefront sensing in $R$-band) in 2017, 2018 and the last night of 2020. For the first two nights of our 2020 observations, we opted for the Pyramid wavefront sensor (PyWFS) instead. It performs wavefront sensing in $H$ \citep{2000PASP..112..315W, 2018SPIE10703E..1ZB}, which is more advantageous for the redder targets in our sample.

A summary of the observing log is presented in Table \ref{tab:observations}.

\begin{table*}
\caption{Keck/NIRC2 Observing Log. Acronyms are: wavefront sensor (WFS), Shack-Hartmann (SH), pyramid WFS (py), integration time ($t_\mathrm{int}$), number of frames ($N_\mathrm{frames}$), parallactic angle coverage (P. A. cov.).}
\begin{tabular}{cccccccc}
\hline\hline
UT Date & Target & Filter & WFS & $t_\mathrm{int}$ (s) & Coadds & $N_\mathrm{frames}$ & P. A. cov. ($^\mathrm{o}$) \\ \hline
 {2017 September 8} & RX J1744.7$-$2713 & $L'$ & SH & 0.5 & 60 & 40 & 14.3\\ 
 & Cygnus X-1 & $L'$ & SH & 0.5 & 60 & 140 & 60.6\\ 
 & $\gamma$ Cassiopeiae & $L'$ & SH & 0.18 & 150 & 130 & 61.2\\ 
 & X Persei & $L'$ & SH & 0.5 & 60 & 89 & 43.9\\ \hline
 {2018 January 3} & X Persei & $L'$ & SH & 1 & 45 & 54 & 36.7\\ 
 & 1H0556+286 & $L'$ & SH & 1 & 45 & 50 & 48.6\\ 
 & RX J0648.1$-$4419 & $L'$ & SH & 1 & 45 & 17 & 6.5 \\ 
 & Vela X-1 & $L'$ & SH & 1 & 45 & 60 & 26.0 \\ \hline
 {2020 July 11} & RX J1744.7$-$2713 & {$L'$} & py & 0.5 & 60 & 120 & 38.3 \\ 
 & IGR J18483$-$0311 & {$L'$} & py & 0.4 & 60 & 125 & 42.6 \\ 
 & $\gamma$ Cassiopeiae & {$L'$} & py & 0.0528 & 400 & 150 & 48.1 \\ \hline
{2020 July 12} & RX J1744.7$-$2713 & $K_s$ & py & 0.6 & 45 & 92 & 39.6 \\ 
 & SAX J1818.6$-$1703 & {$L'$} & py & 0.4 & 60 & 90 & 26.3 \\ 
 & 1H2202+501 & {$L'$} & py & 0.4 & 60 & 89 & 36.9 \\ 
 & 4U2206+543 & {$L'$} & py & 0.4 & 60 & 27 & 49.5 \\ \hline
{2020 July 13} & 4U1700$-$37 & {$L'$} & SH & 0.4 & 50 & 69 & 15.1 \\ 
 & IGR J17544$-$2619 & {$L'$} & SH & 0.4 & 50 & 94 & 36.5 \\ 
 & RX J2030.5+4751 & {$L'$} & SH & 0.4 & 60 & 132 & 53.9 \\ 
 & 4U2206+543 & {$L'$} & SH & 0.4 & 60 & 149 & 49.5 \\ \hline
\end{tabular}
\label{tab:observations}
\end{table*}

\subsection{Data Reduction}\label{sec:datareduction}
Similarly to \cite{2022AJ....164....7P}, we performed data reduction using the Vortex Image Processing (\texttt{VIP}) and \texttt{NIRC2 Preprocessing} packages \citep{2017AJ....154....7G}. To obtain a preprocessed data cube, we proceeded as follows: (1) flat-fielding of the frames, (2) bad pixel masking using the dark frames, (3) determination of the vortex center for each frame followed by cropping the science cube around the mean center, (4) removal of sky contribution via a Principal Component Analysis (PCA)-based technique, and (5) image registration to align the quasi-static speckle pattern across frames.

After acquiring the preprocessed data cube, we applied a PCA-based Angular Differential Imaging (ADI; \citealt{2006ApJ...641..556M}) algorithm to obtain high-contrast images. Subsequently, we generated several images using two algorithms in \texttt{VIP} (annular PCA and full-frame PCA) for a broad range of principal components (from 1 to 50). This was done to ensure consistency in the detection of sources within the images (i.e., that the source was detected regardless of the number of principal components). We then listed sources with a signal-to-noise ratio (SNR) greater than 5 (or with a signal exceeding 4$\sigma$) and determined the optimal number of principal components ($n_\mathrm{comp}$) that maximized the SNR for each source. 

\subsection{Source Magnitude Calculation}\label{sec:magnitude}
To calculate the apparent and absolute magnitudes of the detected sources, we proceeded as follows: (1) we fit a 2D Gaussian profile to the Point Spread Function (PSF) cube to obtain the Full Width at Half Maximum (FWHM) in milliarcseconds (hereafter mas) and to recenter the PSF frames. (2) The PSF cube was reduced into a single 2D PSF by computing the median of the frames. (3) We normalized the PSF so that the flux within a radius of 1 FWHM equated to 1. (4) Given a thermal artifact affecting the quality of the real PSF in our 2020 observations (as discussed in \citealt{2022AJ....164....7P}), we generated a normalized synthetic 2D PSF using the FWHM. (5) From the list of sources with SNR $>$ 5, we generated a list of approximate coordinates using \texttt{ds9}. (6) Using the preprocessed cube, the optimal number of principal components, and the approximate coordinates, we fit the astrometric parameters ($\theta$, the position angle in degrees and $\rho$, the relative separation from the X-ray binary in pixels) and photometric parameters ($f_1$, the number of counts within an aperture radius of 3 FWHM, i.e., the relative flux) using a Nelder-Mead optimization algorithm from \texttt{VIP}. This fitting process, involving the injection of synthetic sources with a negative flux at source location \citep{2010Sci...329...57L}, aimed to minimize $\chi^2$ residuals. (7) Once the parameters were determined, the position angle in the images was converted into the true position angle using the celestial north of NIRC2, which is 0\textdegree.262 ± 0\textdegree.018 (\citealt{2016PASP..128i5004S} and see \citealt{2022AJ....163...50F} for an example and more details). (8) We converted the units of $\rho$ from pixel to mas using the plate scale (see Section \ref{sec:observations}). (9) To derive the apparent magnitude of the sources ($m_\mathrm{cc}$), we applied the following equation:

\begin{equation}
    m_\mathrm{cc} = -2.5 \log_{10} (f_1/f_2) + m_\mathrm{XRB}
\end{equation}

where $f_2$ is the PSF flux within the same aperture as the cube (3 FWHM) and $m_\mathrm{XRB}$ is the apparent magnitude of the X-ray binary (see Section \ref{sec:appmag}). (10) Using the known distance in parsec (see Section \ref{sec:distance}), the apparent magnitude of the candidate CBCs ($m_\mathrm{cc}$) was converted to absolute magnitude ($M_\mathrm{cc}$) via the distance modulus equation. (11) Finally, to estimate the mass of the sources, we compared the absolute magnitude ($M_\mathrm{cc}$) with evolutionary models from MESA Isochrones \& Stellar Tracks (MIST; \citealt{2011ApJS..192....3P, 2013ApJS..208....4P,2018ApJS..234...34P, 2015ApJS..220...15P, 2016ApJS..222....8D, 2016ApJS..225...10F, 2016ApJ...823..102C}) at the system's age (see Section \ref{sec:age}).

Due to the poor quality of the PSF in our 2020 observations, we used the synthetic 2D PSF for fitting processes related to these observations. However, we kept the original PSF for our 2017 and 2018 observations. We conducted tests using both synthetic and real PSFs on the 2017 and 2018 data, yielding consistent results. Consequently, the use of a synthetic PSF does not affect significantly the derived parameters.

The upcoming sections detail the acquisition of parameters used in the magnitude calculations.

\subsubsection{Determining the Distance from the Observer}\label{sec:distance}
The distance from the observer is presented in the second column of Table \ref{tab:XRB_add_properties} in the Appendix. Distances for 1H2202+501, 4U 2206+543, 4U 1700$-$377, IGR J17544$-$2619, Cyg X-1, X Per and Vela X-1 were obtained from \cite{2023MNRAS.525.1498Z}, which uses parallax measurements to infer distances either using an inversion or Bayesian approach for a catalog of X-ray binaries. For the other X-ray binaries, the distance was estimated through a photogeometric calculation \citep{2021AJ....161..147B}. It was calculated using the parallax measurement and its uncertainty (geometric), as well as the $G$ magnitude and the BP-RP color (photometric) from the Gaia Data Release 3 (DR3; \citealt{2023A&A...674A...2D, 2023A&A...674A...3M}). Table \ref{tab:gaia_ID} presents the Gaia DR3 ID for each target. In cases where an object's distance was sourced from the literature, the respective reference is cited in the literature review of Section \ref{sec:sample}. As previously mentioned, this study targets X-ray binaries within $\sim$2--3 kpc accessible with Keck/NIRC2.

\begin{table}[ht!]
    \caption{Gaia DR3 ID for each target.}
    \centering
    \begin{tabular}{cc}
    \hline\hline
    \textbf{Target} & \textbf{Gaia DR3 ID} \\ \hline
    RX J1744.7$-$2713 & 4060784345959549184 \\
    IGR J18483$-$0311 & 4258428501693172736 \\
    $\gamma$ Cas & 426558460884582016 \\
    SAX J1818.6$-$1703 & 4097365235226829312 \\
    1H2202+501 & 1979911002134040960 \\
    4U 2206+543 & 2005653524280214400 \\
    4U 1700$-$377 & 5976382915813535232 \\
    IGR J17544$-$2619 & 4063908810076415872 \\
    RX J2030.5+4751 & 2083644392294059520 \\
    Cyg X-1 & 2059383668236814720 \\
    X Per & 168450545792009600 \\
    1H0556+286 & 3431561565357225088 \\
    RX J0648.1$-$4419 & 5562023884304070000 \\
    Vela X-1 & 5620657678322625920 \\ \hline
    \end{tabular}
    \label{tab:gaia_ID}
\end{table}

\subsubsection{Determining the Apparent Magnitude of the Central X-Ray Binary}\label{sec:appmag}
Observing X-ray binaries in the $L'$-band of Keck/NIRC2 is not standard practice, making the direct determination of the true apparent magnitudes of the central X-ray binaries ($m_\mathrm{XRB}$) unfeasible. Nonetheless, we leveraged the $W1$ filter from the Wide-field Infrared Survey Explorer (WISE), which has a central wavelength similar to Keck/NIRC2 $L'$-band ($\lambda$ = 3.353 $\mu$m). Using the WISE Source Catalog \citep{2010AJ....140.1868W}, we approximated $m_\mathrm{XRB}$ for all X-ray binaries. These values are presented in the fifth column of Table \ref{tab:XRB_add_properties} in the Appendix. To account for the differences between the two filters, we considered an uncertainty of 0.5 mag on $m_\mathrm{XRB}$.

\subsubsection{Determining the Age of the System}\label{sec:age}
Most X-ray binaries in our sample lack age estimates in the literature, except for RX J1744.7$-$2713 (up to $\sim$60 Myr; \citealt{coleiro_distribution_2013}), 4U 1700$-$377 (up to $\sim$80 Myr; \citealt{coleiro_distribution_2013}), $\gamma$ Cas (8.0 $\pm$ 0.4 Myr; \citealt{2005A&A...441..235Z}), Cyg X-1 ($<4$ Myr; \citealt{2021Sci...371.1046M}) and X Per ($\sim$5 Myr; \citealt{1997MNRAS.286..549L}). For the remaining X-ray binaries, we established an upper limit using basic evolutionary models and the spectral type of the donor star (see Table \ref{tab:XRB_properties}). We found the maximum age that the donor star can reach before exploding in supernovae. However, this approach provides only an approximate estimation of the age. These values are presented in the sixth column of Table \ref{tab:XRB_add_properties} in the Appendix.

\subsubsection{Determining the Errors}\label{sec:errors}
Errors in the fit parameters ($\theta$, $\rho$, and $f_1$) were estimated using an injection/recovery approach (see \citealt{2022AJ....164....7P}). This approach relies on injecting synthetic sources with known parameters into the images. Subsequently, the same optimization method was applied (see Section \ref{sec:magnitude}), and the error was determined as the difference between estimated and known parameters. This method was employed across a range of parameter values to ensure consistency; it was observed that errors were more pronounced for smaller offset values (i.e., those closer to the central X-ray binary). Additional sources of uncertainty were taken into consideration in cases where the value of a parameter is expected to remain consistent between two sets of observations, specifically astrometric parameters in two different bands. The dominant source of uncertainty was defined as the total uncertainty.

\section{Key properties of the sample} \label{sec:litrev}
\subsection{$\gamma$ Cassiopeiae}\label{sec:gammaCas}
$\gamma$ Cassiopeiae -- also known as 2S 0053+604 (hereafter $\gamma$ Cas) -- harbors a well-studied optical component classified as a Be star \citep{1973A&A....23..433M}. Its X-ray luminosity ($\sim$10$^{32}$--10$^{33}$ erg s$^{-1}$; \citealt{2005A&AT...24..151R}) is higher than the typical luminosity for O/B stars ($\sim$10$^{30}$ erg s$^{-1}$), but too low to be a Be/NS system ($\sim$10$^{34}$ erg s$^{-1}$ in quiescence; \citealt{2015ApJ...799...84S}). The nature of the system can be explained by two hypotheses: (1) the system is a HMXB, involving accretion onto a WD or a fast-spinning NS \citep{2017MNRAS.465L.119P}; or (2) the excess of X-ray emission stems from physical processes in the high atmosphere of $\gamma$ Cas \citep{1998PASJ...50..417K, 2000ApJ...540..474R}. Though $\gamma$ Cas is not confirmed as being a HMXB, its resemblance to other sources, referred to as $\gamma$ Cas analogs, warranted its inclusion in our sample. Located at a distance of 0.19 $\pm$ 0.02 kpc, it is the nearest system in our sample. Its proximity and proper motions (25.7 $\pm$ 0.5 mas/yr in RA, $-$3.9 $\pm$ 0.4 mas/yr in Dec; \citealt{1997A&A...323L..49P}) allowed us to conduct a proper motion analysis within the interval between our two observation sets (see Section \ref{sec:astrometry}).

\subsection{RX J1744.7-2713}
RX J1744.7-2713 is classified as a BeXRB \citep{1997ApJ...474L..53I} and is composed of a B0.5 III-Ve star \citep{1997A&A...323..853M, 1999A&AS..137..147S, 2006A&A...454..265L} and a WD. However, the origin of X-ray emission is still uncertain and debated in the literature \citep{2006A&A...454..265L}. It is also known as a $\gamma$ Cas analog, due to the similarities in their X-ray properties (e.g., \citealt{2015ApJ...799...84S}). The first high-contrast images of this HMXB were presented in \cite{2022AJ....164....7P}, in which more comprehensive information on the system can be found. 

\subsection{4U 1700$-$377}
4U 1700$-$377, discovered with the \textit{Uhuru} X-ray satellite, is classified as a HMXB \citep{1973ApJ...181L..43J}. The system contains a supergiant donor star of spectral type O6Iafpe \citep{2014ApJS..211...10S} and a magnetized NS (e.g., \citealt{1999A&A...349..873R, 2020MNRAS.493.3045B,2021A&A...655A..31V}) exhibiting strong flaring activity (e.g., \citealt{2007ATel.1266....1K}). This HMXB was observed with, e.g., \textit{XMM-Newton} \citep{2005A&A...432..999V, 2015A&A...576A.108G}, \textit{Chandra} \citep{2003ApJ...592..516B, 2021MNRAS.501.5646M}, \textit{Hubble Space Telescope} (HST; \citealt{2020A&A...634A..49H}), \textit{Fiber-fed Extended Range Optical Spectrograph} (FEROS; \citealt{2020A&A...634A..49H}), \textit{IUE} \citep{1978Natur.275..400D}, \textit{EXOSAT} \citep{1989ApJ...343..409H}, \textit{BATSE} \citep{1996ApJ...459..259R}, and \textit{BeppoSAX} \citep{1999A&A...349..873R}. NGC 6231, located within the OB association Sco OB1, has been recently confirmed as the parent cluster of 4U 1700$-$377 \citep{2021A&A...655A..31V}. Moreover, \cite{coleiro_distribution_2013} inferred an estimated age of $\sim$80 Myr for the system, a value we have regarded as an upper limit as in \cite{2022AJ....164....7P}. Its X-ray luminosity can reach up to $\sim$7 $\times$ 10$^{36}$ erg s$^{-1}$ \citep{1992A&A...260..237L}.

\subsection{4U 2206+543}
Although the nature of its donor star remains uncertain (first believed to be of spectral type Be; \citealt{1984ApJ...280..688S}, then O9.5V; \citealt{2001A&A...371.1056N}), 4U 2206+543 is a well-studied HMXB harboring a magnetar NS (e.g, \citealt{2004A&A...423..301T, 2005A&A...438..963B,2009A&A...494.1073R,2010Ap.....53..237I,2018MNRAS.479.3366T}). First discovered with \textit{Uhuru} \citep{1972ApJ...178..281G}, the system was subsequently observed with, e.g., \textit{EXOSAT} \citep{1992ApJ...401..678S}, \textit{RXTE} (e.g., \citealt{2001ApJ...562..936C}), IBIS/ISGRI on \textit{INTEGRAL} (e.g., \citealt{2004ApJ...607L..33B}), \textit{BeppoSAX} (e.g., \citealt{2004A&A...423..311M}), $VLA$ \citep{2005A&A...438..963B}, \textit{Swift} \citep{2007ApJ...655..458C} and \textit{Suzaku} \citep{2010ApJ...709.1249F}. Its X-ray luminosity ranges from $\sim$10$^{33}$ erg s$^{-1}$ in quiescence up to $\sim$10$^{35}$--10$^{36}$ erg s$^{-1}$ during more active phases \citep{2006A&A...449..687R}.

\subsection{RX J2030.5+4751}
RX J2030.5+4751, also known as BD+47 3129 and SAO 49725, was first discovered with \textit{ROSAT} \citep{1997A&A...323..853M}. It is identified as a BeXRB and $\gamma$ Cas analog. It has a B0.5 III-Ve spectral type donor star and a maximum X-ray luminosity of $\sim$10$^{33}$ erg s$^{-1}$ \citep{2006A&A...455.1165L,2007BeSN...38...24R}. XMM-Newton/EPIC observations have revealed that RX J2030.5+4751 has a hard X-ray spectrum, suggesting the presence of a dense, large, and stable circumstellar environment surrounding the system \citep{2006A&A...454..265L}. On July 20, 2016, RX J2030.5+4751 underwent type I (i.e., smaller and repetitive) bursts, reaching its maximum luminosity (progressive weakening since; \citealt{2016ATel.8927....1S,2016ATel.9265....1S,2016ATel.9487....1S}). Regarding the long-term variability, \cite{2012int..workE..23S} reported a significant non-periodic variability of approximately one magnitude in the light curve of RX J2030.5+4751 over $\sim$100 years, likely caused by changes in the properties of the decretion disk.

\subsection{1H2202+501}
1H2202+501 is a poorly-studied HMXB, albeit appearing in some surveys (e.g., \citealt{1984ApJS...56..507W, 1996ApJS..107..281H,2000A&AS..147...25L}). It consists of a Be star of spectral type B3Ve \citep{2019KPCB...35...38S}, and the nature of the compact object remains uncertain. Its X-ray luminosity  is estimated to be $\sim$9 $\times$ 10$^{32}$ erg s$^{-1}$ \citep{1998A&A...330..201C}.

\subsection{SAX J1818.6$-$1703}
SAX J1818.6$-$1703 was discovered with \textit{BeppoSAX} while undergoing a strong outburst \citep{1998IAUC.6840....2I}. Subsequent observations were carried out using IBIS/ISGRI onboard \textit{INTEGRAL} (e.g., \citealt{2005AstL...31..672G,2005A&A...444..221S, 2009A&A...493L...1Z,2016MNRAS.457.3693S}), \textit{RossiXTE} \citep{2005A&A...444..221S, 2012MNRAS.422.2661S} and \textit{Swift} (e.g., \citealt{2009MNRAS.393L..11B, 2009MNRAS.400..258S}). The system has similar properties to other SFXT (e.g., \citealt{2005A&A...444..221S,2006ATel..831....1N,2009MNRAS.400..258S,2012A&A...544A.118B}), and its compact object is likely a NS (e.g., \citealt{2007A&A...476..335W,2016MNRAS.456.4111B}). The donor star of the system, confirmed by \textit{Chandra} \citep{2006ATel..915....1I}, is classified as a supergiant star of spectral type B0.5 Iab \citep{2010A&A...510A..61T}. Also observed using \textit{XMM-Newton}, SAX J1818.6$-$1703 has a quiescent X-ray luminosity that can drop to values below $\sim$10$^{32}$ erg s$^{-1}$ (as determined by not being detected in \citealt{2012A&A...544A.118B}) and can reach up to $\sim$8 $\times$ 10$^{35}$ erg s$^{-1}$ \citep{2010A&A...510A..61T}.

\subsection{IGR J18483$-$0311}
Discovered by IBIS/ISGRI onboard \textit{INTEGRAL} by \cite{2003ATel..157....1C}, IGR J18483$-$0311 is a well-studied SFXT \citep{2008A&A...492..163R} composed of an X-ray pulsar \citep{2007A&A...467..249S} and a supergiant donor star of spectral type B0.5 Ia \citep{2008A&A...484..783C} or B0.5-B1 Iab \citep{2010A&A...510A..61T}. The system undergoes multiple short and long outbursts (e.g., \citealt{2007A&A...467..249S, 2010MNRAS.402L..49S, 2013A&A...559A.135D, 2015MNRAS.449.1228S}), resulting in its X-ray luminosity ranging from $\sim$10$^{33}$--10$^{34}$ erg s$^{-1}$ in quiescence \citep{2010MNRAS.401.1564R, 2015MNRAS.449.1228S} up to $\sim$10$^{36}$ erg s$^{-1}$ during its strongest flares \citep{2010A&A...510A..61T}. By considering evolutionary scenarios and exploring the relationship between spin and orbital periods, \cite{2011MNRAS.415.3349L} suggested that the compact object in IGR J18483$-$0311 may originate from an O-type emission line star rather than an average main-sequence star.

\subsection{IGR J17544$-$2619}
IGR J17544$-$2619 was first discovered near the Galactic Center while undergoing short (a few hours) outbursts using IBIS/ISGRI onboard \textit{INTEGRAL} \citep{2003ATel..190....1S, 2003ATel..192....1G,2004ATel..252....1G}. The system was subsequently observed with, e.g., \textit{XMM-Newton} \citep{2004A&A...420..589G,2014MNRAS.439.2175D}, \textit{Chandra} \citep{2005A&A...441L...1I}, EMMI/SOFI/NTT \citep{2006A&A...455..653P}, \textit{Suzaku} \citep{2009ApJ...707..243R}, RXTE \citep{2012A&A...539A..21D}, \textit{Swift} (e.g., \citealt{2015A&A...576L...4R}), and NuSTAR (e.g., \citealt{2015MNRAS.447.2274B}). It is identified as a SFXT \citep{2006ESASP.604..165N}, and the optical/NIR counterpart is a massive star of spectral type O9Ib (25--28 $M_\odot$; \citealt{2006A&A...455..653P,2016A&A...591A..26G}) or O9IV-V (23 $M_\odot$; \citealt{2017AstL...43..664B}). The compact object is a NS \citep{2005A&A...441L...1I}, as inferred by the presence of a cyclotron line at 17 keV and the magnetic field strength ($\sim 1.5 \times 10^{12}$ G, typical for NS in X-ray binaries; \citealt{2015MNRAS.447.2274B}). The change in X-ray luminosity between quiescence and outburst is significant, ranging from $L_X \sim 10^{32} - 10^{34}$ erg s$^{-1}$ (e.g., \citealt{2016A&A...596A..16B}) during quiescence to a maximum of $L_X \sim 3 \times 10^{38}$ erg s$^{-1}$ during outburst (\citealt{2015A&A...576L...4R}).

\subsection{Cyg X-1}
Cygnus X-1 (hereafter Cyg X-1), first discovered in 1964, is one of the most well-studied astronomical objects (e.g., \citealt{1989MNRAS.238..729F, 1998ApJ...505..854E, 2011ApJ...742...84O, 2014ApJ...780...78T, 2015ApJ...808....9P, 2015MNRAS.446.3579S, 2016ApJ...826...87W, 2019MNRAS.488..348M}). The system's compact object, with a mass of 21.2 $\pm$ 2.2 $M_\odot$ \citep{2021Sci...371.1046M}, was the first observed candidate black hole \citep{1971Natur.233..110M, 1972NPhS..240..124B}, leading to significant breakthroughs in the astronomical scientific community. Cyg X-1 is classified as a HMXB \citep{1972Natur.235..271B} and the donor star is characterized as a massive star of spectral type O9.7Iabpvar \citep{2011ApJS..193...24S}. The system is fairly young, with an estimated age of 5 $\pm$ 1.5 Myr in \cite{2003Sci...300.1119M} and $<4$ Myr in \cite{2021Sci...371.1046M}. Cyg X-1 is located at 2.22$_{-0.18}^{+0.17}$ kpc; its distance from the observer was precisely estimated in \cite{2021Sci...371.1046M} based on radio parallax measurements and validated with Gaia DR3. The microquasar undergoes variable X-ray emission \citep{2013ApJS..209...14K}, with a maximum X-ray luminosity of $\sim$ 3 $\times$ 10$^{37}$ erg s$^{-1}$ \citep{2001ApJ...547.1024D}.

\subsection{Vela X-1}
Vela X-1 is a pulsing HMXB discovered with the \textit{Uhuru} X-ray satellite (\citealt{1971ApJ...165L..27G}) and was observed through multiple surveys and with several instruments (e.g., \citealt{1978MNRAS.183..813C, 2003A&A...400..993L, 2014ApJ...780..133F, 2014A&A...563A..70M}). The system is highly variable and undergoes transient outbursts and X-ray eclipses (e.g., \citealt{1984A&A...135..155V}). Its X-ray luminosity can be as high as 4 $\times$ 10$^{36}$ erg s$^{-1}$ (\citealt{2008A&A...492..511K}). It is composed of a B0.5 1b donor star (\citealt{1972ApJ...175L..19H}) and of a magnetized NS (e.g., \citealt{1972ApJ...175L..19H, 1995A&A...303..483V, 2002A&A...395..129K, 2022A&A...660A..19D}). A complete review of this object can be found in \cite{2021A&A...652A..95K}.

\subsection{RX J0648.1$-$4419}
RX J0648.1$-$4419 is a unique X-ray pulsating system, as it is the only HMXB known to date containing a hot sub-dwarf donor star of spectral type O (sdO5.5; \citealt{1963PASP...75..365J, 2009Sci...325.1222M}). The compact object has a mass of 1.28 $\pm$ 0.05 $M_\odot$ \citep{2009Sci...325.1222M} and was initially believed to be a NS (e.g., \citealt{1997ApJ...474L..53I, 2016MNRAS.458.3523M}). \cite{2018MNRAS.474.2750P} suggested that it was likely a young ($\sim$2 Myr) contracting WD. Over the course of almost 30 years of observations (e.g., \citealt{2011ApJ...737...51M, 2015A&A...580A..56L,2021MNRAS.504..920M}), the X-ray luminosity remained stable, maintaining a value of $\sim$10$^{32}$ erg s$^{-1}$.

\subsection{1H0556+286}
1H0556+286 contains a Be star of spectral type B5ne \citep{1950ApJ...111..495P}. It is a poorly-studied system; while it is generally thought to be a HMXB (e.g., \citealt{2001ApJ...554...27H, 2006A&A...455.1165L}), \cite{2001A&A...377..148T} presented \textit{BeppoSAX} observations in which no X-ray emission was detected. As per these results, neither the Be/NS nor Be/WD scenarios appear probable, hence the nature of the system remains unclear.

\subsection{X Per}
X Persei (hereafter X Per), also known as 4U 0352+309, was discovered in 1972 with the \textit{Copernicus} Observatory \citep{1975ApL....16...19H, 1976MNRAS.176..193M}. The system was subsequently observed with, e.g., the High Energy Astrophysical Observatory (HEAO 2/Einstein; \citealt{1984ApJ...278..711W}), \textit{RXTE} (e.g., \citealt{2001ApJ...546..455D, 2002ApJ...580..394C}), \textit{INTEGRAL} (e.g., \citealt{2012MNRAS.423.1978L}), \textit{XMM-Newton} (e.g., \citealt{2007A&A...474..137L}), and \textit{Chandra} (e.g., \citealt{2013ApJ...770...22V}). The system is identified as a HMXB/BeXRB, composed of a magnetized NS as the compact object (e.g., \citealt{1977Natur.267..229W, 2001ApJ...552..738C, 2012A&A...540L...1D,2017MNRAS.470..713M,2018PASJ...70...89Y}) and a Be star of spectral type B0Ve as the donor star \citep{1997MNRAS.286..549L}. In quiescence, its X-ray luminosity is $\sim$ 10$^{34}$ erg s$^{-1}$ (e.g., \citealt{2001ApJ...552..738C}). When undergoing strong outburst activity, its X-ray luminosity can reach up to $\sim$ 2 $\times$ 10$^{35}$ erg s$^{-1}$ \citep{2012MNRAS.423.1978L}. Similarly to Cyg X-1, the system is also relatively young, with an estimated age of $\sim$5 Myr \citep{1997MNRAS.286..549L}. Objects with similar properties are often referred to as X Per analogs.

\section{High-Contrast Images} \label{sec:images}
\subsection{$\gamma$ Cas}
\begin{figure}[ht!]
    \centering
    \includegraphics[width=\columnwidth]{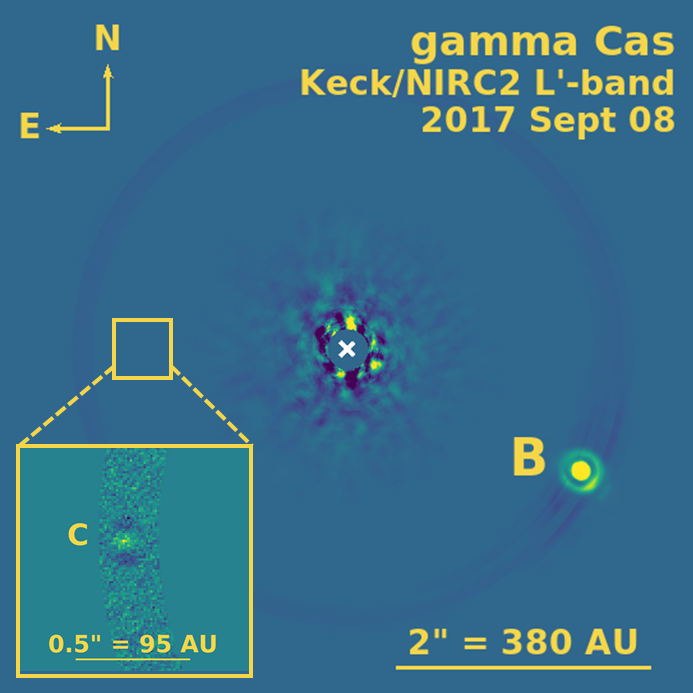}
    \includegraphics[width=\columnwidth]{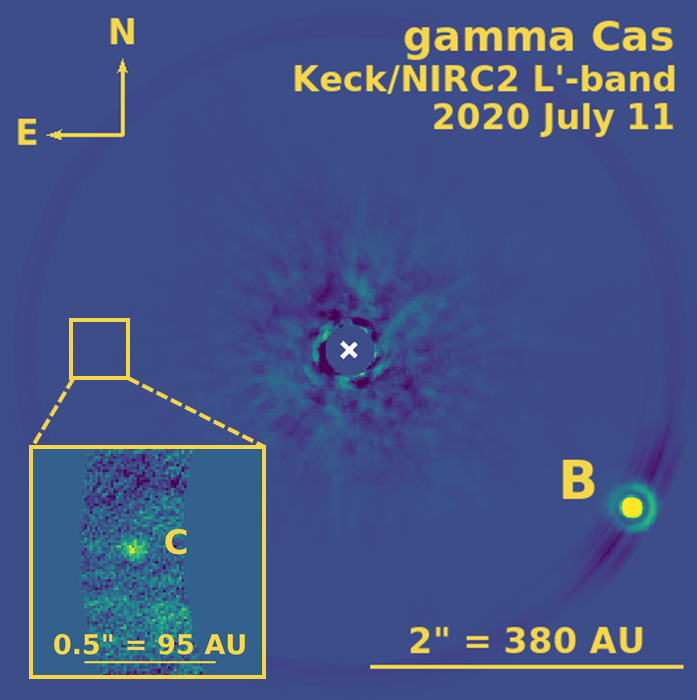}
    \caption{Keck/NIRC2 $L'$-band high-contrast images of $\gamma$ Cas acquired on 2017 Sept 08 (up) and 2020 July 11 (bottom), treated and reduced using a PCA annular ADI algorithm (using \texttt{VIP}; \citealt{2017AJ....154....7G}). The sources detected with SNR $>$ 3 are labeled B and C. The white X symbol denotes the approximate position of $\gamma$ Cas, masked by the coronagraph. The bottom-left insets show zoomed-in high-contrast images with a focus on an annular region (obtained using the PCA annulus algorithm in \texttt{VIP}), highlighting the presence of the fainter source (labeled as C).}
    \label{fig:images_high_contrast_gammaCas}
\end{figure}

As discussed in Section \ref{sec:gammaCas}, observations of $\gamma$ Cas in the $L'$-band were initially made on September 08, 2017, and subsequently revisited almost three years later on July 11, 2020. Figure \ref{fig:images_high_contrast_gammaCas} presents the $L'$-band high-contrast images of $\gamma$ Cas for both epochs. A bright source, labeled B, was detected with a SNR $\gg$ 5. A much fainter source, labeled C, was also detected, with a SNR $\sim$ 3. In Section \ref{sec:astrometry}, we undertake a proper motion analysis to determine whether these sources are more likely to be bound CBCs or background stars.

\subsection{Other X-ray Binaries}
\begin{figure*}
    \centering
    \includegraphics[width=0.30\textwidth]{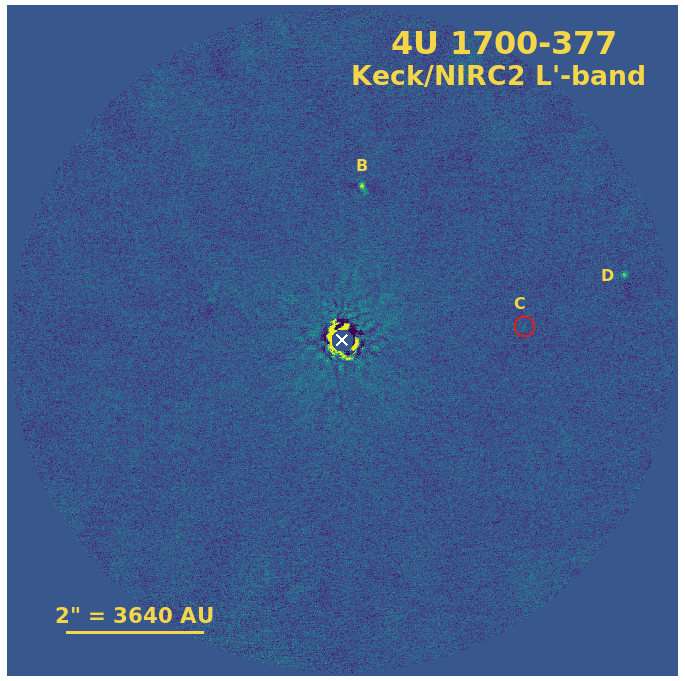}
    \includegraphics[width=0.30\textwidth]{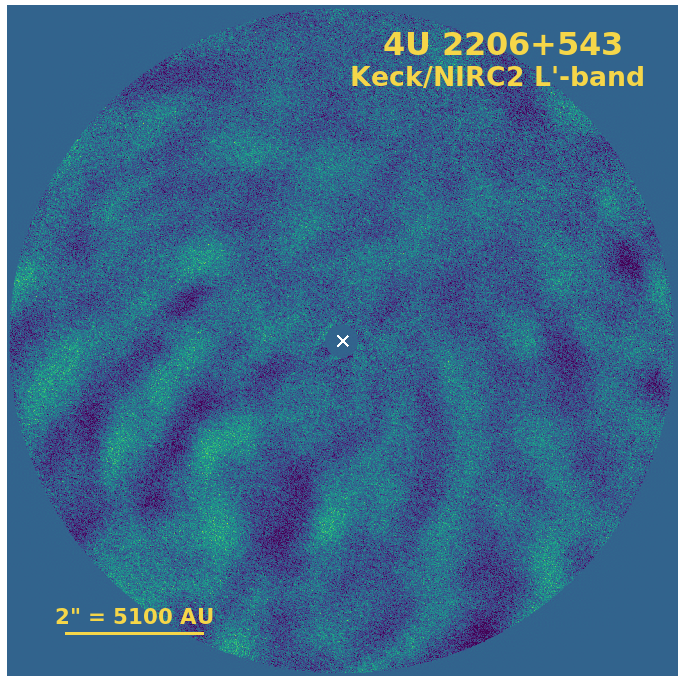}
    \includegraphics[width=0.30\textwidth]{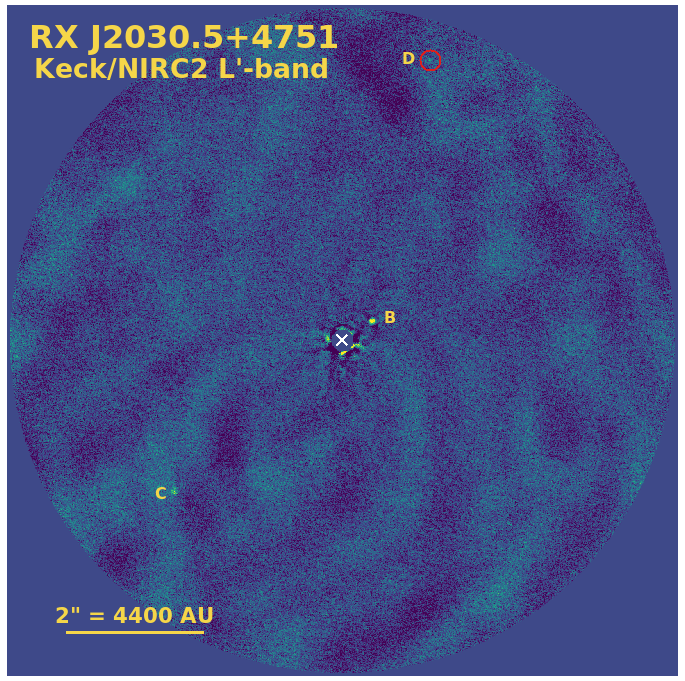}
    \includegraphics[width=0.30\textwidth]{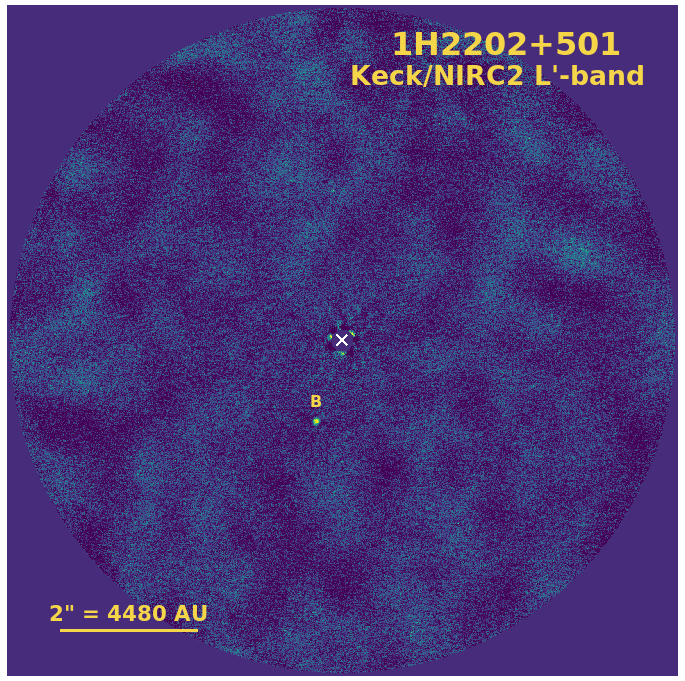}
    \includegraphics[width=0.30\textwidth]{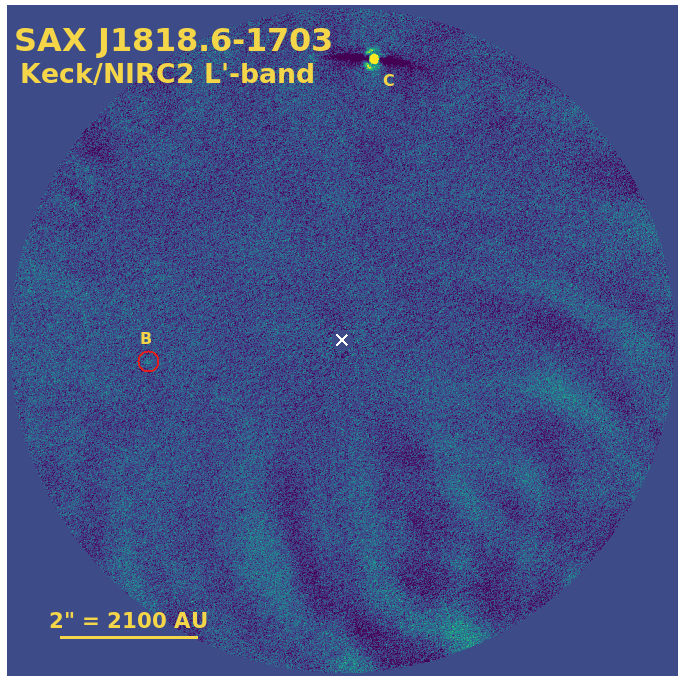}
    \includegraphics[width=0.30\textwidth]{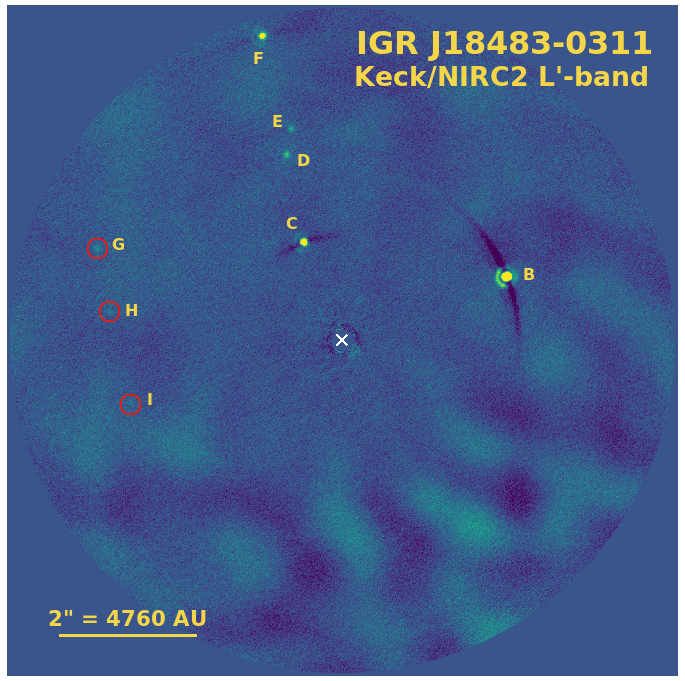}
    \includegraphics[width=0.30\textwidth]{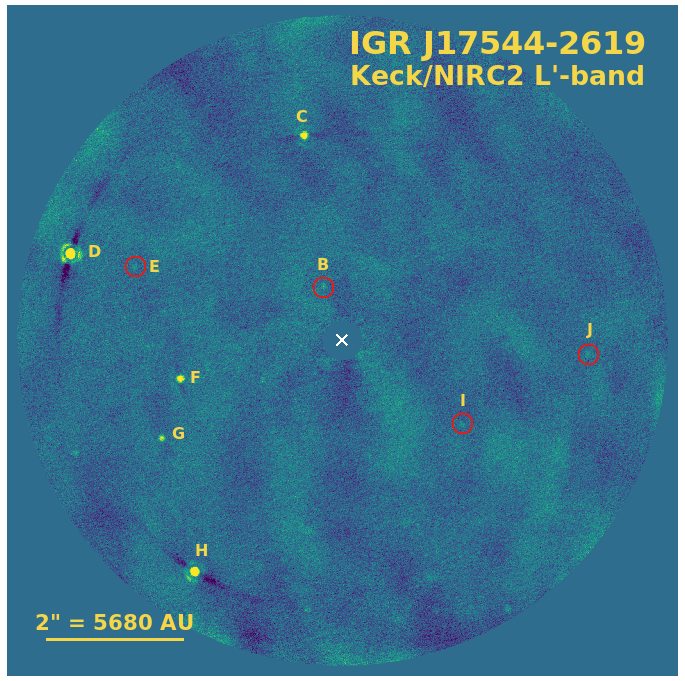}
    \includegraphics[width=0.30\textwidth]{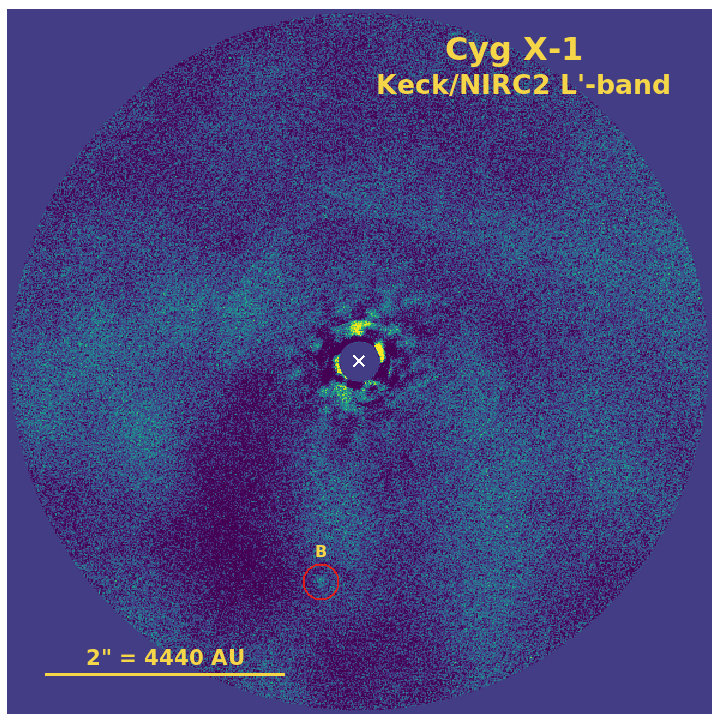}
    \includegraphics[width=0.30\textwidth]{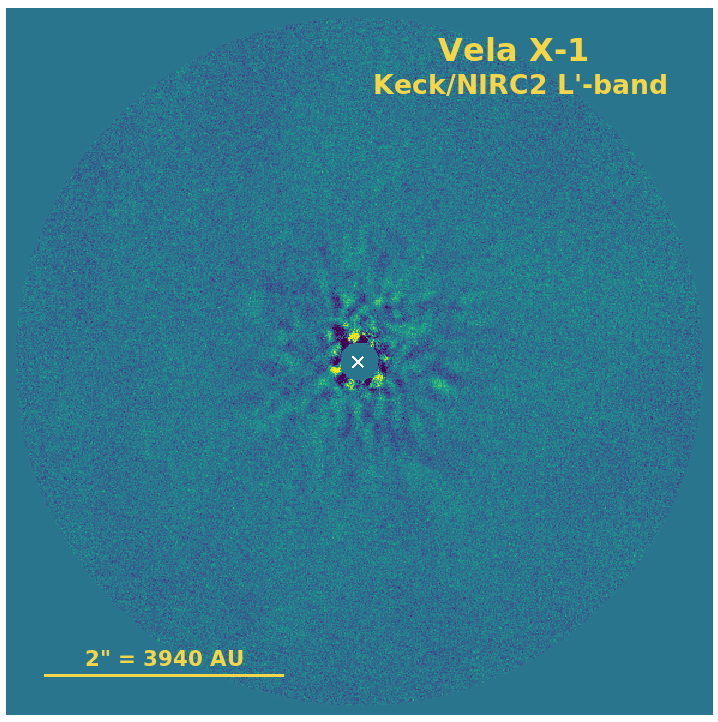}
    \includegraphics[width=0.30\textwidth]{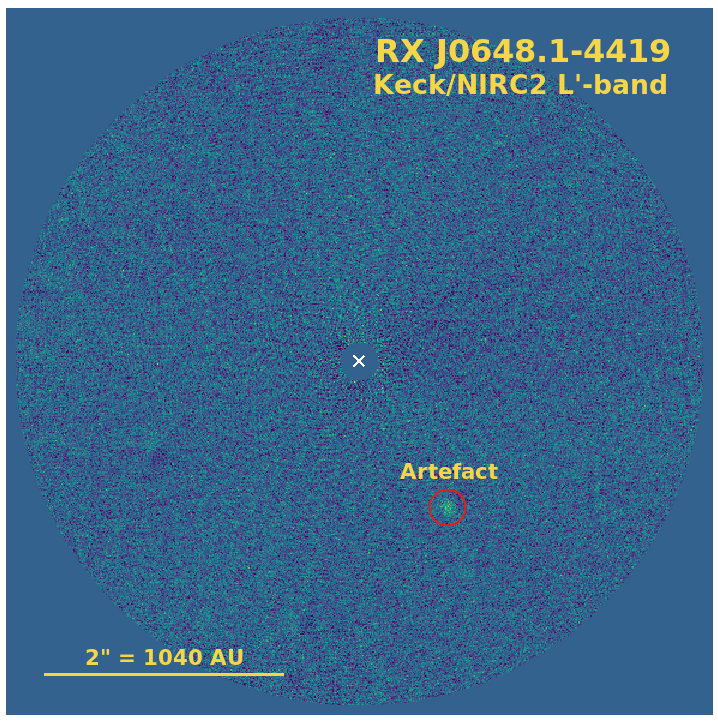}
    \includegraphics[width=0.30\textwidth]{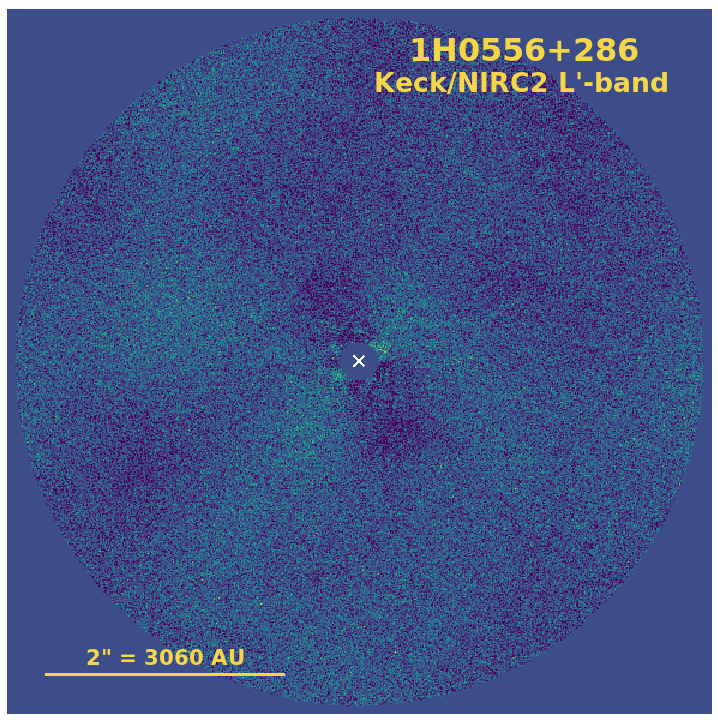}
    \includegraphics[width=0.30\textwidth]{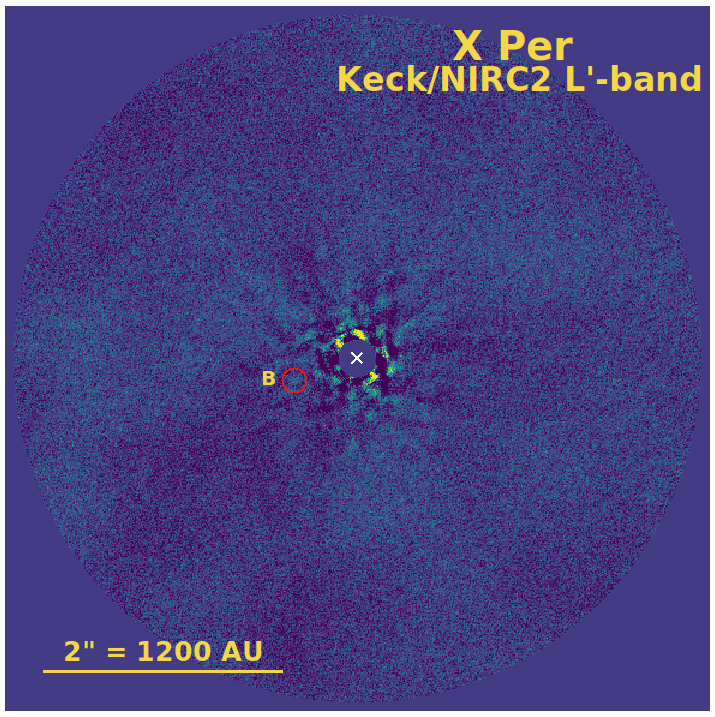}
    \caption{Keck/NIRC2 $L'$-band high-contrast images of all observed X-ray binaries -- except RX J1744.7$-$2713 (see \citealt{2022AJ....164....7P}) and $\gamma$ Cas (see Fig \ref{fig:images_high_contrast_gammaCas}) -- acquired on September 8, 2017, January 3, 2018, and July 11--13, 2020. Images were treated and reduced using a PCA annular ADI algorithm (using \texttt{VIP}; \citealt{2017AJ....154....7G}). The sources detected with SNR $>$ 5 when computing the signal-to-noise map are labeled. The white X symbol denotes the position of the X-ray binary masked by the coronagraph. North points upwards and East points to the left, as in Fig. \ref{fig:images_high_contrast_gammaCas}.}
    \label{fig:images_high_contrast}
\end{figure*}

Figure \ref{fig:images_high_contrast} presents a panel of $L'$-band high-contrast images of all the other X-ray binaries, in order: 4U 1700$-$377, 4U 2206+543, RX J2030.5+4751, 1H2202+501, SAX J1818.6$-$1703, IGR J18483$-$0311, IGR J17544$-$2619, Cyg X-1, Vela X-1, RX J0648.1$-$4419, 1H0556+286, and X Per. The $L'$- and $Ks$-band images of RX J1744.7$-$2713 can be found in \cite{2022AJ....164....7P}. By inspecting the images, we determined that 4U 2206+543, Vela X-1, RX J0648.1$-$4419, and 1H0556+286 do not exhibit any potential candidate CBCs. Consequently, further analysis of these systems will not be pursued, except for the calculation of companion frequency in Section \ref{sec:frequency}. Among the remaining X-ray binaries, we successfully detected at least one source for each system with a significantly large SNR ($>$5). These sources are labeled in the images, starting from the letter B. 

Table \ref{tab:prop_sources_XRB} in the Appendix presents several physical properties of the detected sources, including the angular separation in mas, the position angle in degrees, the apparent magnitude in $L'$-band and the mass estimated from evolutionary models.

In the next sections, we analyze the nature of the detected sources and discuss the implications of the results.

\section{On the Nature of the Detected Sources}\label{sec:analysis}
\subsection{Background Contamination}\label{sec:contamination}
We used TRILEGAL \citep{2005A&A...436..895G} to discriminate background/foreground stars from gravitationally-bound CBCs. TRILEGAL is a 3D model employed to simulate the photometric properties of star fields within the Galaxy (e.g., \citealt{2015A&A...578A..51C, 2018A&A...620A.102D, 2021MNRAS.504..565J, 2021ApJS..253...53W}). We compiled a list of predicted sources within a 1 $\times$ 1 arcmin$^2$ region surrounding the X-ray binary using its RA/Dec coordinates. Subsequently, we applied a geometric rescaling so that the TRILEGAL foV matches the foV of the high-contrast images (10.24 $\times$ 10.24 arcsec$^2$) for statistical consistency. We calculated the cumulative distribution of the expected number of sources in the foV below an apparent magnitude ($m_{L'}$), for the distribution of apparent magnitudes ($\mathcal{L'}$). It is denoted as $n_\mathrm{foV} (\mathcal{L'} \le m_{L'})$.

\begin{figure*}
    \centering
    \includegraphics[width=0.45\textwidth]{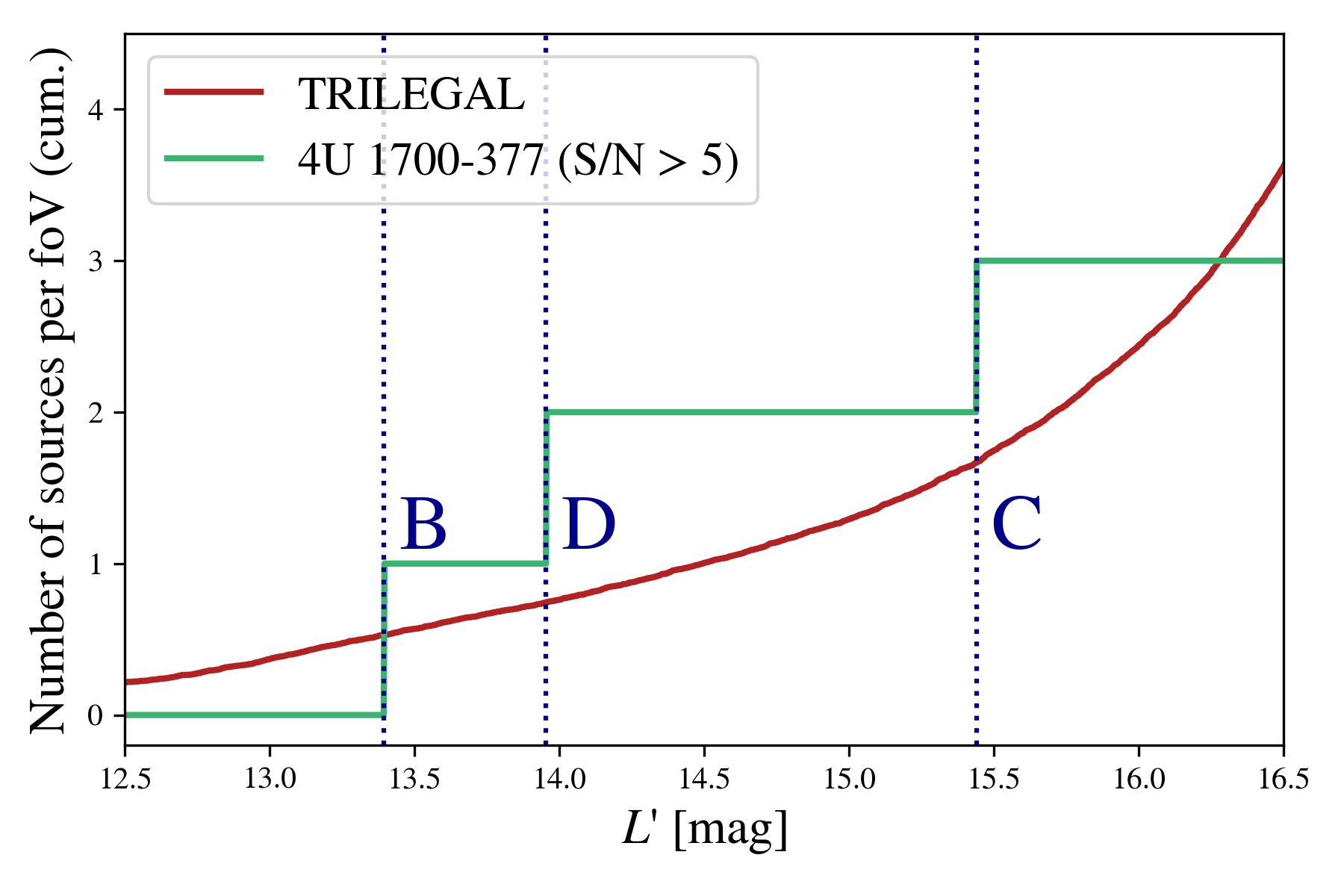}
    \includegraphics[width=0.45\textwidth]{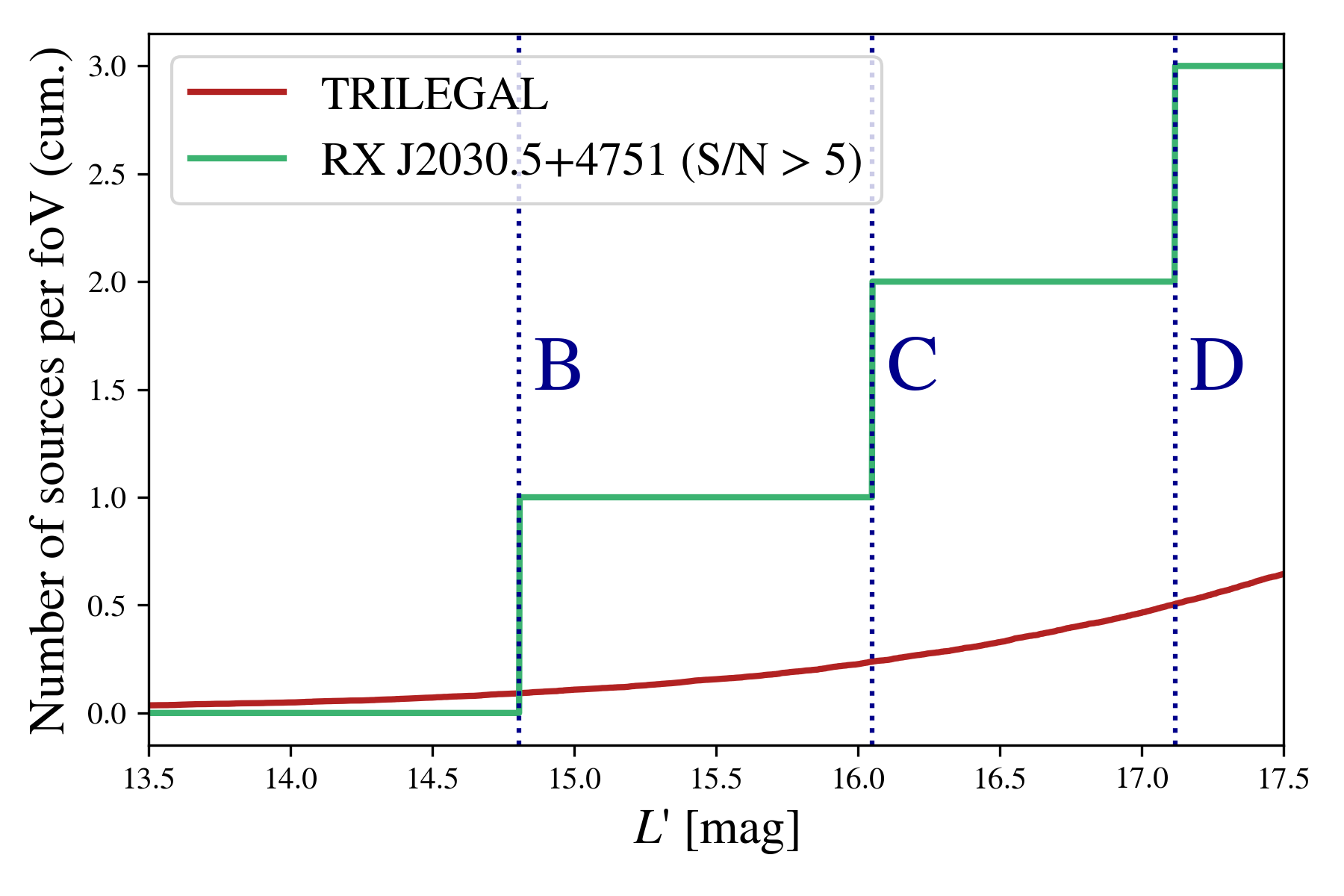}
    \includegraphics[width=0.45\textwidth]{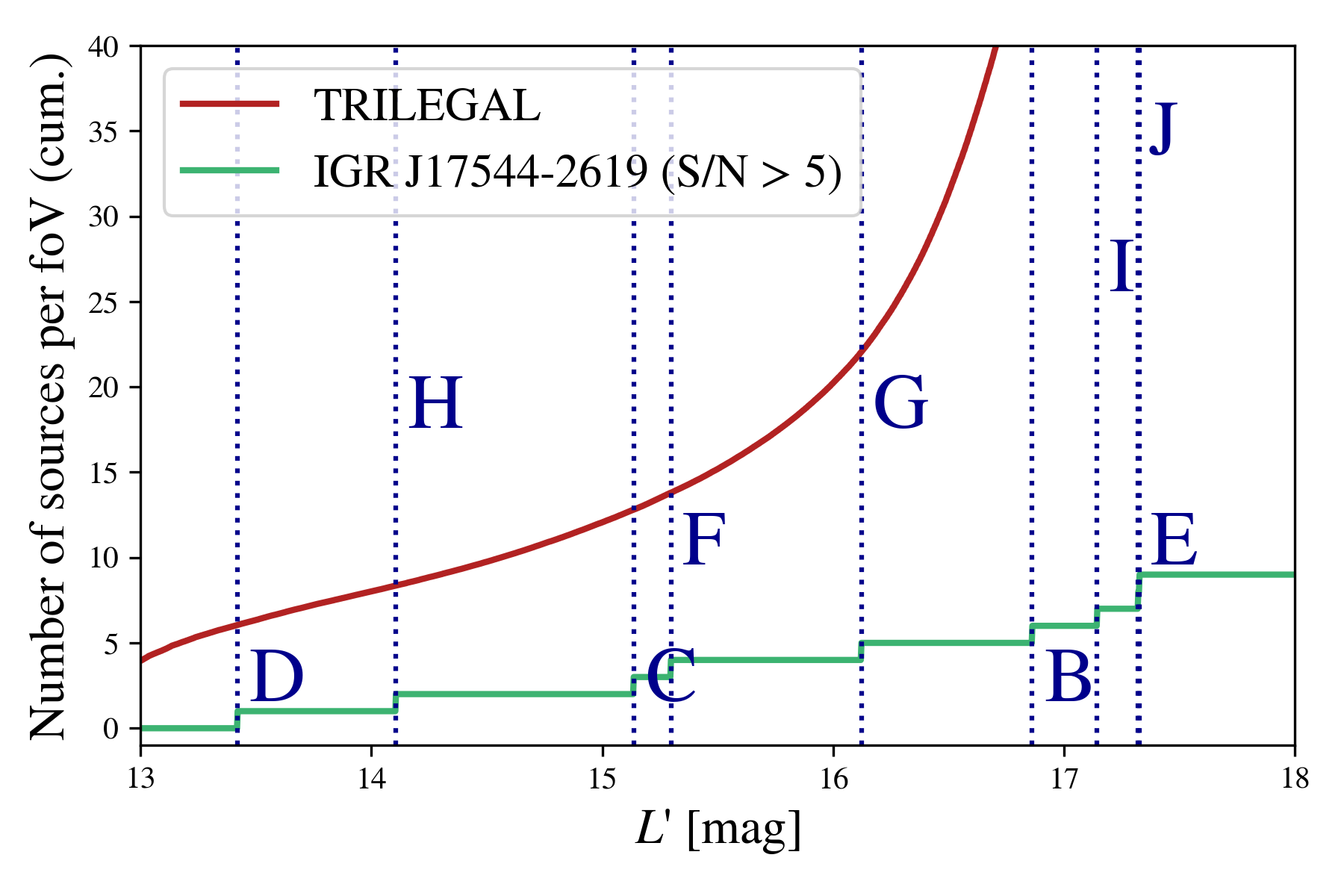}
    \includegraphics[width=0.45\textwidth]{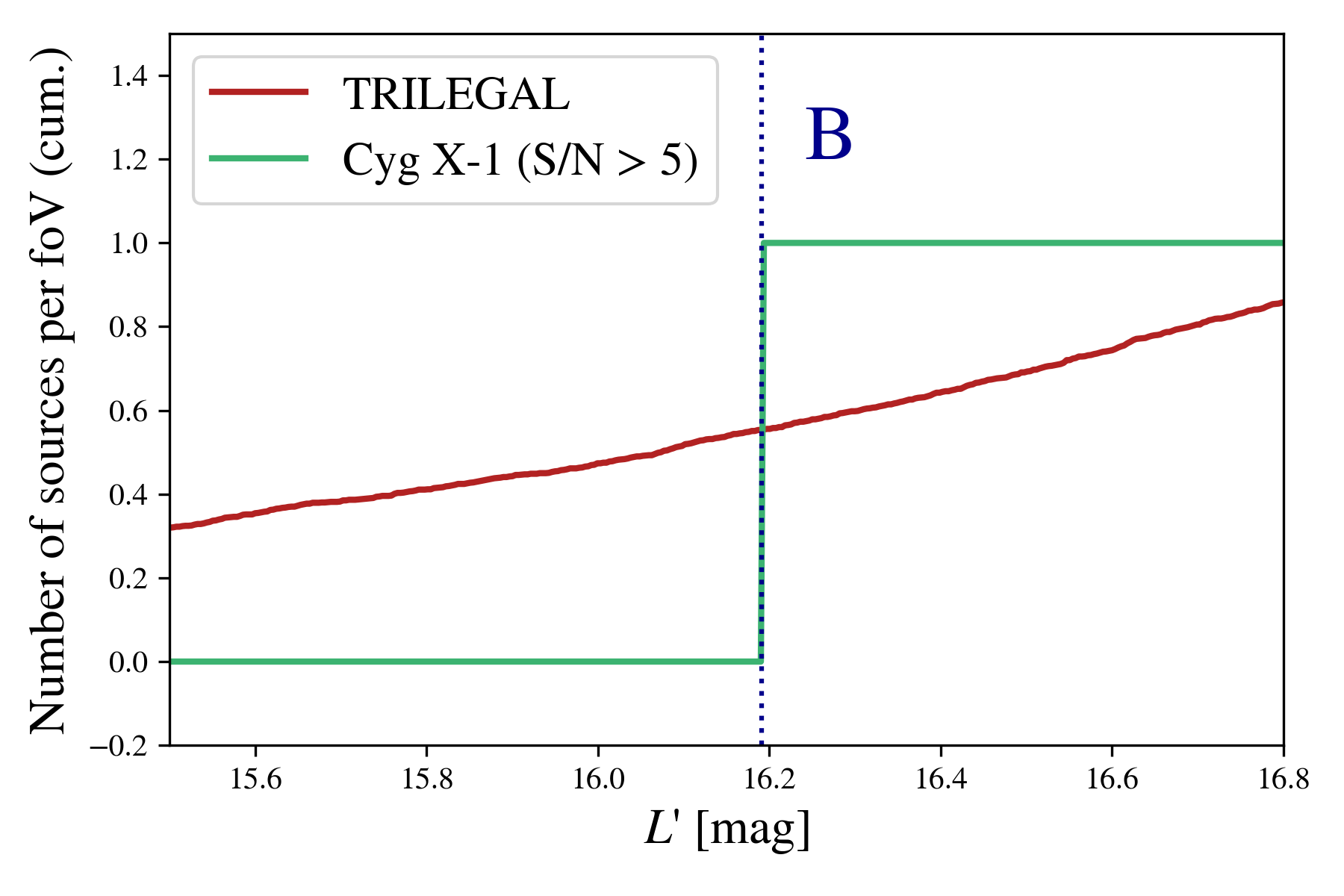}
    \includegraphics[width=0.45\textwidth]{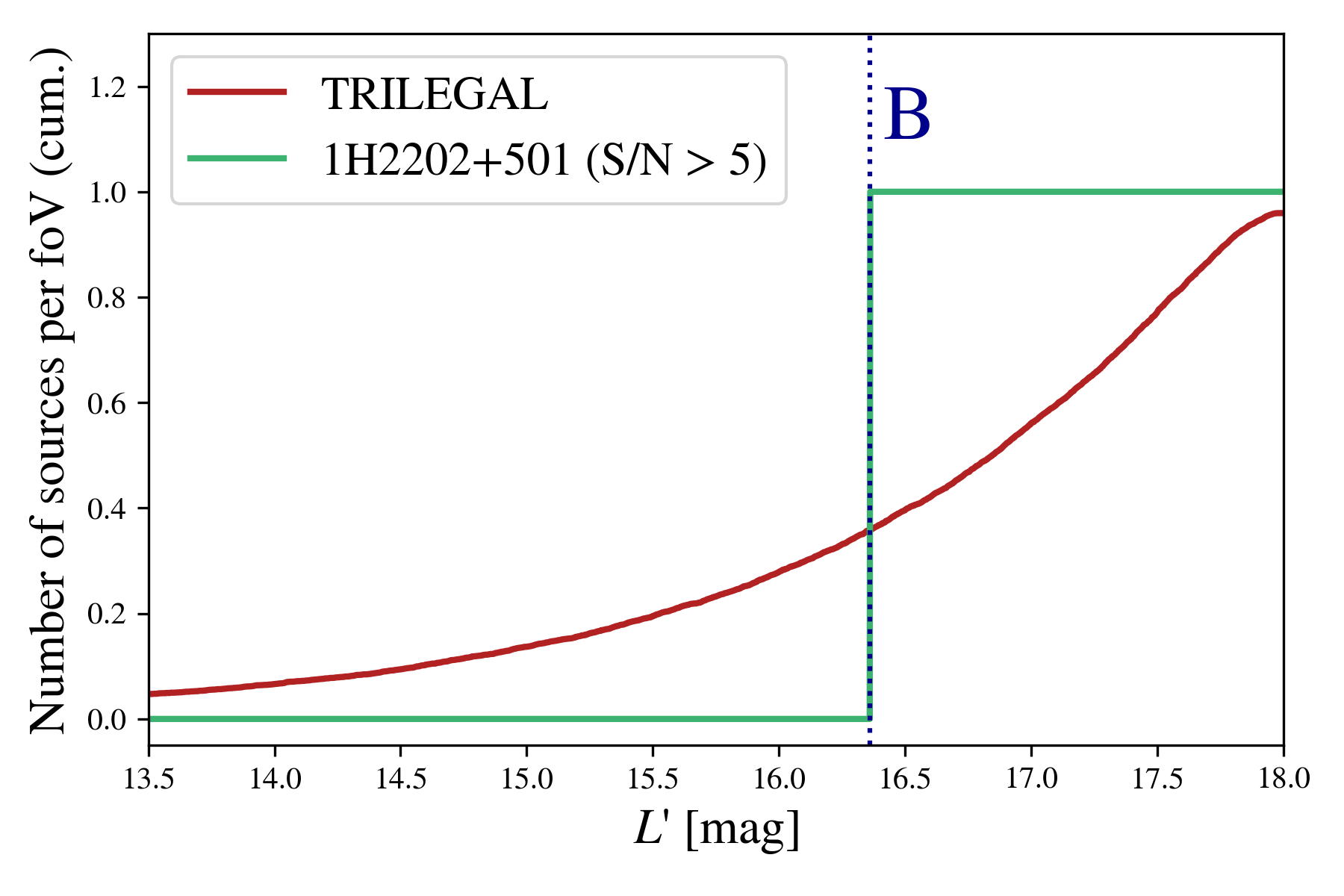}
    \includegraphics[width=0.45\textwidth]{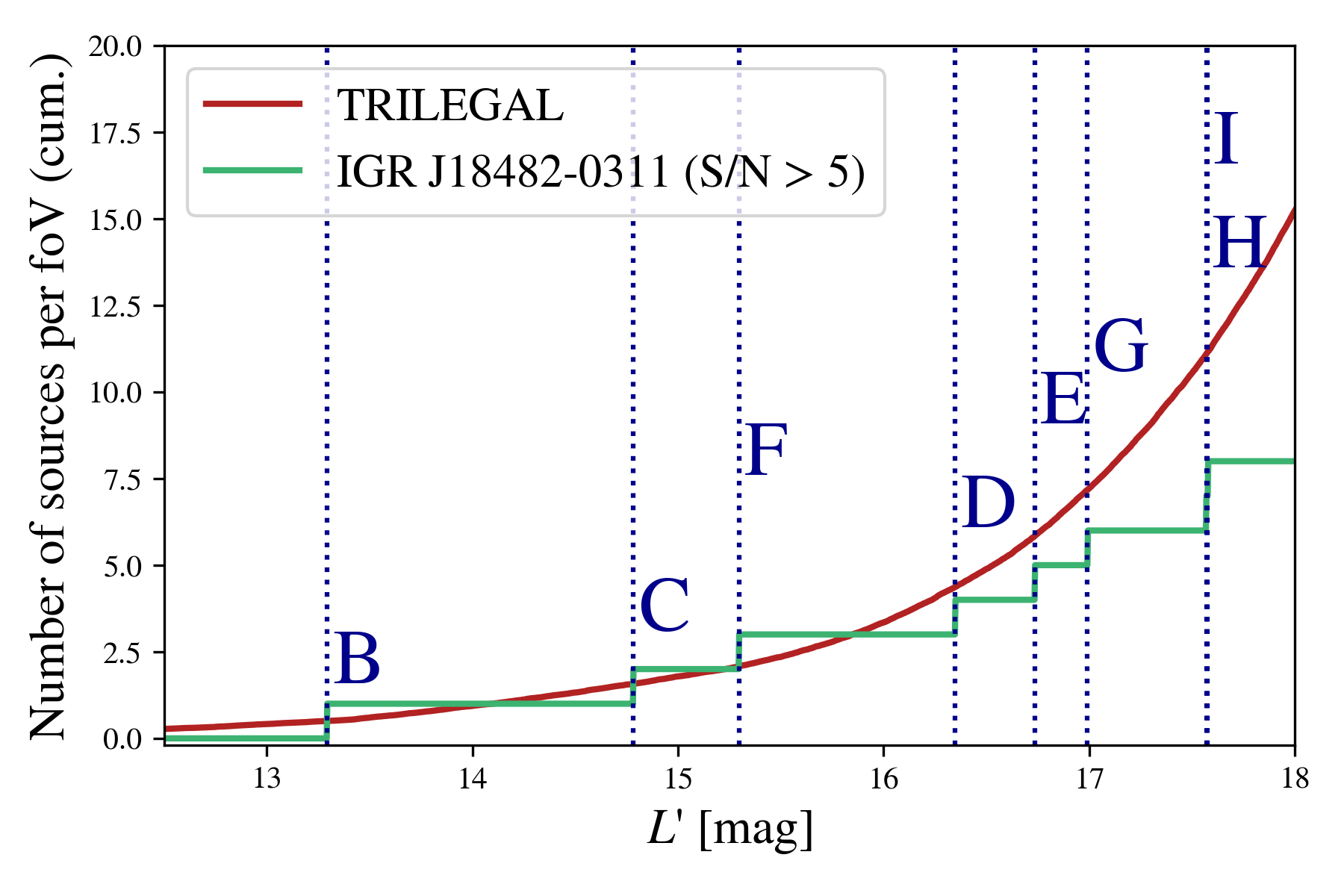}
    \includegraphics[width=0.45\textwidth]{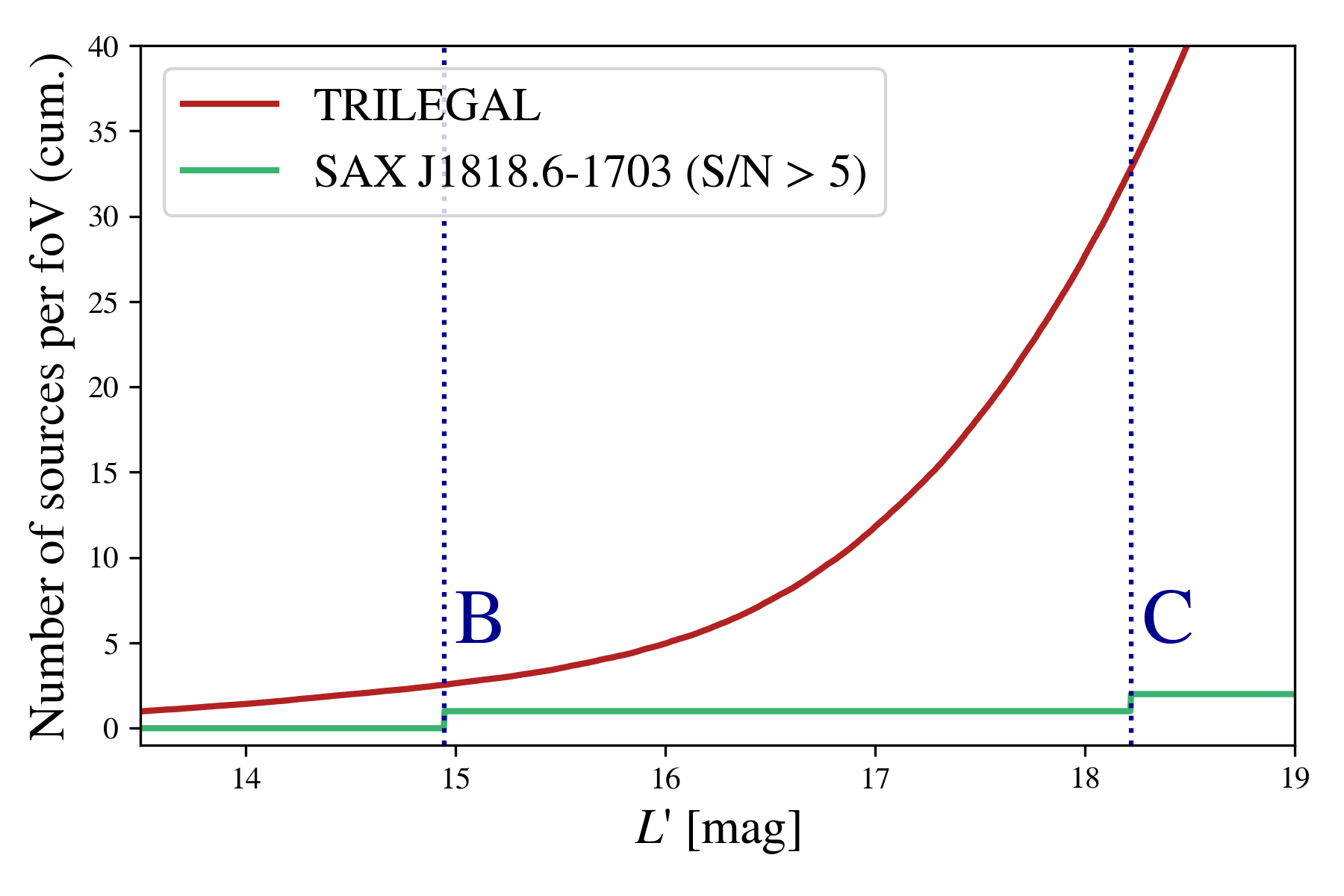}
    \includegraphics[width=0.45\textwidth]{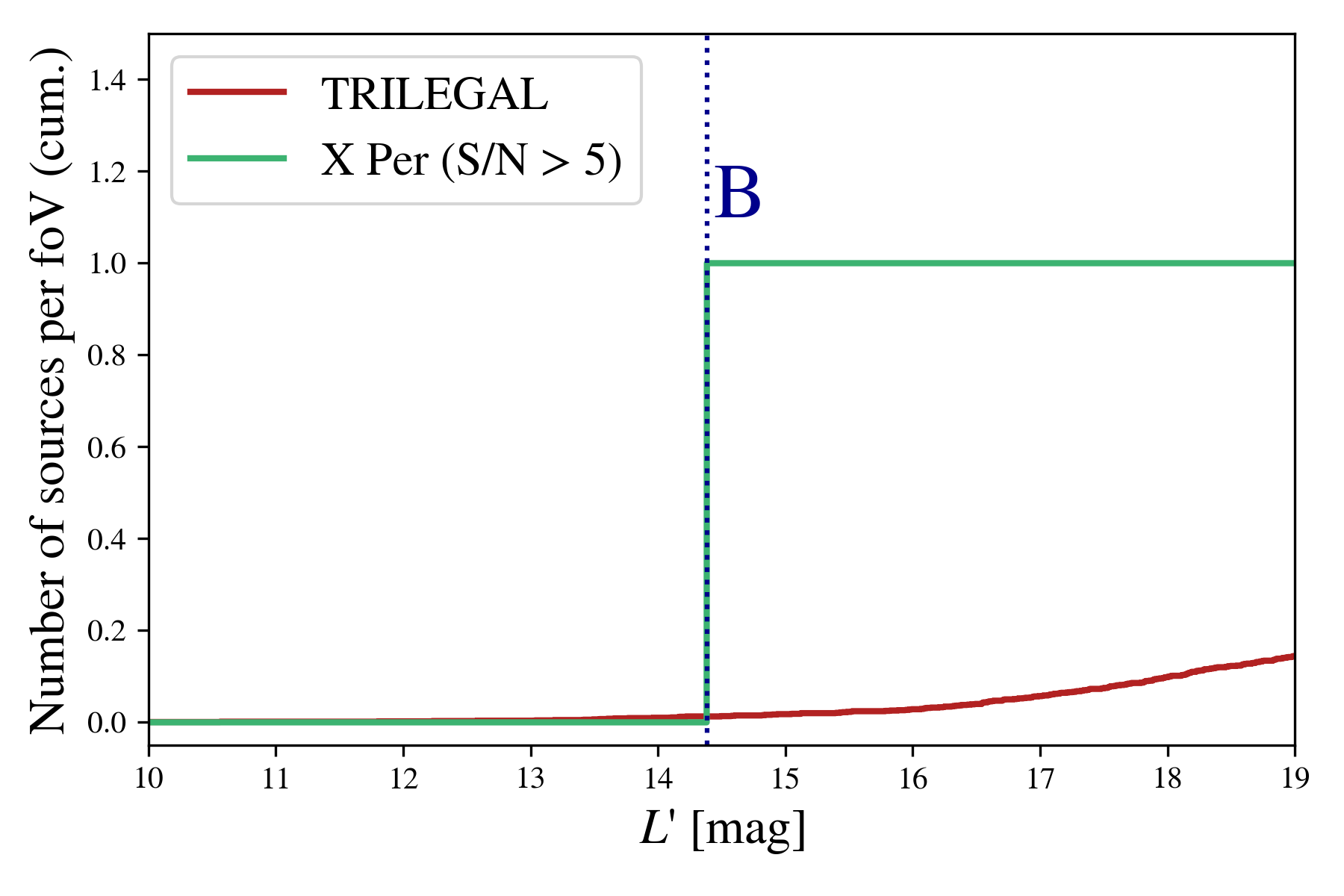}
    \caption{Cumulative distribution of the number of sources in the foV expected from TRILEGAL simulations (red) and detected with a SNR $>5$ in the high-contrast images (green) as a function of the apparent magnitude $m_{L'}$. It includes all X-ray binaries in which sources were detected except $\gamma$ Cas and RX J1744.7$-$2713 (see \citealt{2022AJ....164....7P}): 4U 1700$-$377, RX J2030.5+4751, IGR J17544-2619, Cygnus X-1, 1H2202+501, IGR J18483$-$0311, X Per, and SAX J1818.6$-$1703.}
    \label{fig:TRILEGAL}
\end{figure*}

Fig. \ref{fig:TRILEGAL} presents $n_\mathrm{foV} (\mathcal{L'} \le m_{L'})$ as a function of $m_{L'}$ for both TRILEGAL and the detected sources in the high-contrast images. It includes all X-ray binaries with detected sources, except $\gamma$ Cas (for which the confirmation of the nature of the CBCs relies primarily on proper motion analysis) and RX J1744.7$-$2713 (presented in \citealt{2022AJ....164....7P}). In Table \ref{tab:prop_sources_XRB}, we listed $n_\mathrm{foV} (\mathcal{L'} \le m_{L' \, \mathrm{source}})$ for each detected source. This represents the expected number of sources -- based on TRILEGAL simulations -- with apparent magnitudes ($m_{L'}$) below the magnitude of the corresponding source ($m_{L' \, \mathrm{source}}$) while accounting for errors. If the calculated value, including the upper limit, was lower than the total number of sources detected below $m_{L' \, \mathrm{source}}$, we reject the hypothesis of background/foreground contamination for that particular source.

Many sources were not predicted by TRILEGAL. Specifically, all detected sources in 4U 1700$-$377, RX J2030.5+4751, Cygnus X-1, X Per and 1H2202+501 were not expected from the model given their magnitudes and would thus suggest that they are bound to the X-ray binary. The foVs of IGR J17544$-$2619 and SAX J1818.6$-$1703 were expected to be more populated than what we detected, suggesting that the detected sources are likely background or foreground contaminants. The disparity between the predicted and observed number of sources might result from the elimination of stationary sources during the ADI process or from an insufficient S/N for detection. In the case of IGR J18483$-$0311, sources B, C, and F might be CBC candidates, but their status remains uncertain in the current stage of our study.

We extended our analysis by calculating the probability of chance alignment for each detected source. This probability represents the likelihood that these sources are not associated with the X-ray binary system based on the angular separation and the density on the sky of unrelated objects. This method assumes that the distribution of unrelated sources across the area follows a Poisson distribution. Note that this method usually relies on sky surveys such as the 2MASS Point Source Catalog (e.g., \citealt{2006A&A...459..909C, 2008ApJ...683..844L, 2014ApJ...785...47L, 2022AJ....164....7P}) to establish the distribution of unrelated objects. However, in this work, we used TRILEGAL as an alternative due to the absence of available $K_s$-band observations.

To calculate the aforementioned probability, we first divided the cumulative distribution of the number of sources $n_\mathrm{TRILEGAL} (\mathcal{L'} \le m_{L'})$ by the area from which the sources were retrieved (between 6 $\times$ 6 arcmin$^2$ and 15 $\times$ 15 arcmin$^2$ depending on the location of the X-ray binary). This division enabled us to derive a surface density denoted as $\Sigma$. Using the angular separation $\Theta$ in arcsec, the probability of a source being drawn from the TRILEGAL distribution -- thus indicating its lack of association with the central X-ray binary -- is given by:

\begin{equation}
    P_\mathrm{unrelated}(\Sigma, \Theta) = 1 - \exp(-\pi\Sigma\Theta^2)
    \label{eq:chance}
\end{equation}

In Table \ref{tab:prop_sources_XRB}, we listed $1 - P_\mathrm{unrelated}(\Sigma, \Theta)$ as percentages. Many sources have high probabilities ($>85\%$) of being associated with the central X-ray binary. However, some sources have lower probabilities (between 65\% and 75\%) which are not as statistically significant, but we nevertheless identified them as candidate CBCs given the early stage of the study. Sources with probabilities below 60\% (not statistically significant) were rejected as candidate CBCs.

\subsection{Proper Motion Analysis: $\gamma$ Cas}\label{sec:astrometry}
\begin{figure*}
    \centering
    \includegraphics[width=0.32\textwidth]{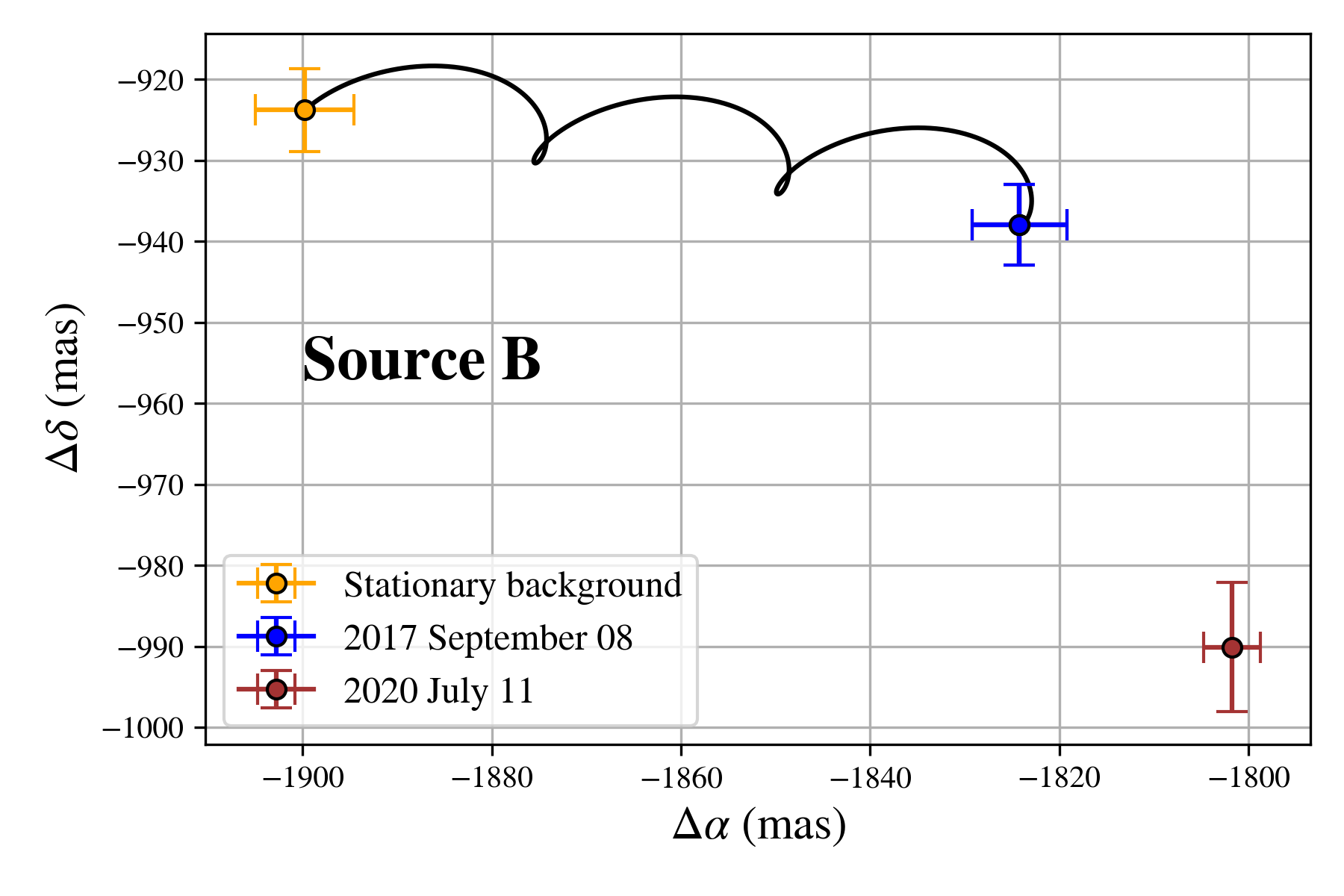}
    \includegraphics[width=0.32\textwidth]{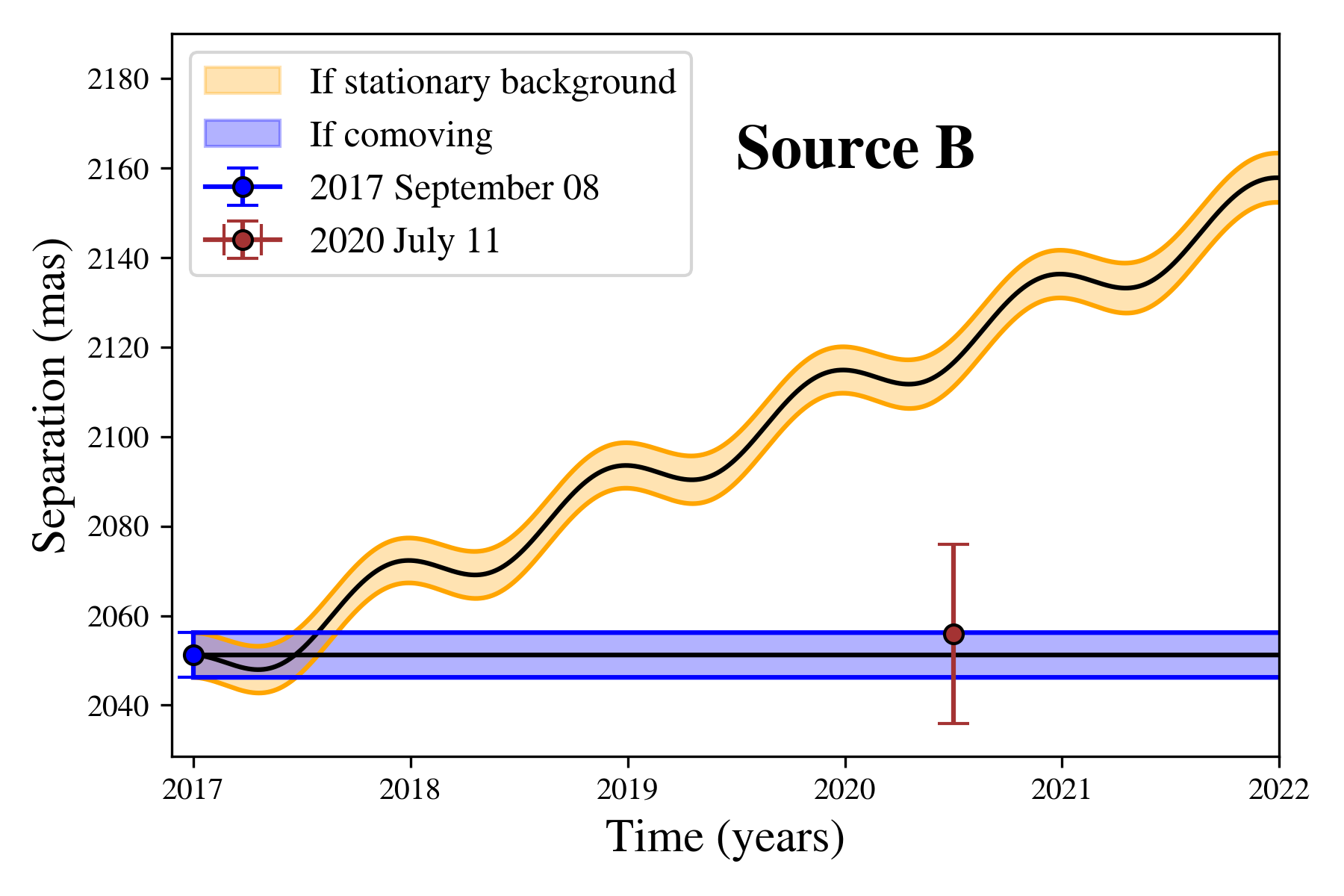}
    \includegraphics[width=0.32\textwidth]{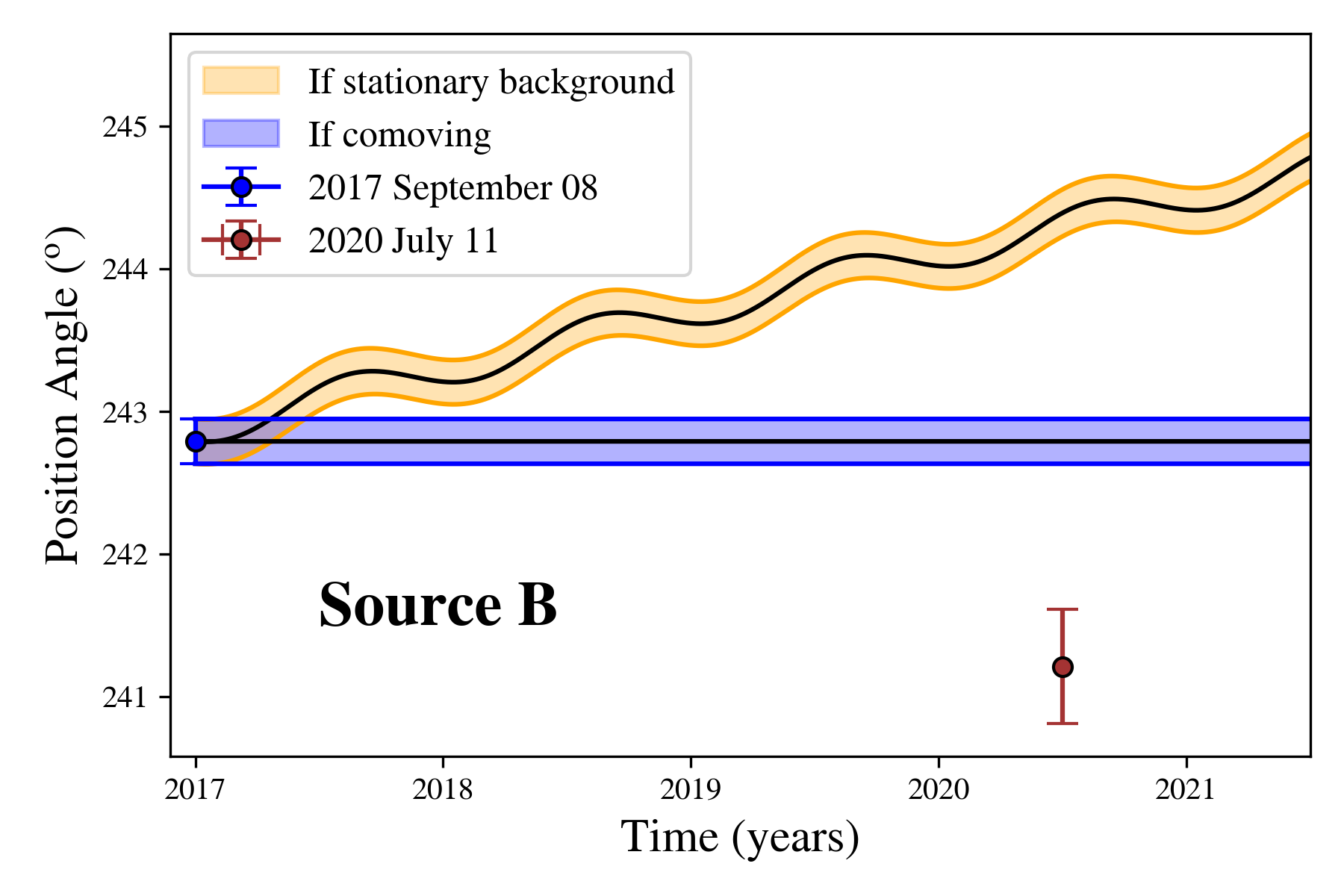}
    \includegraphics[width=0.32\textwidth]{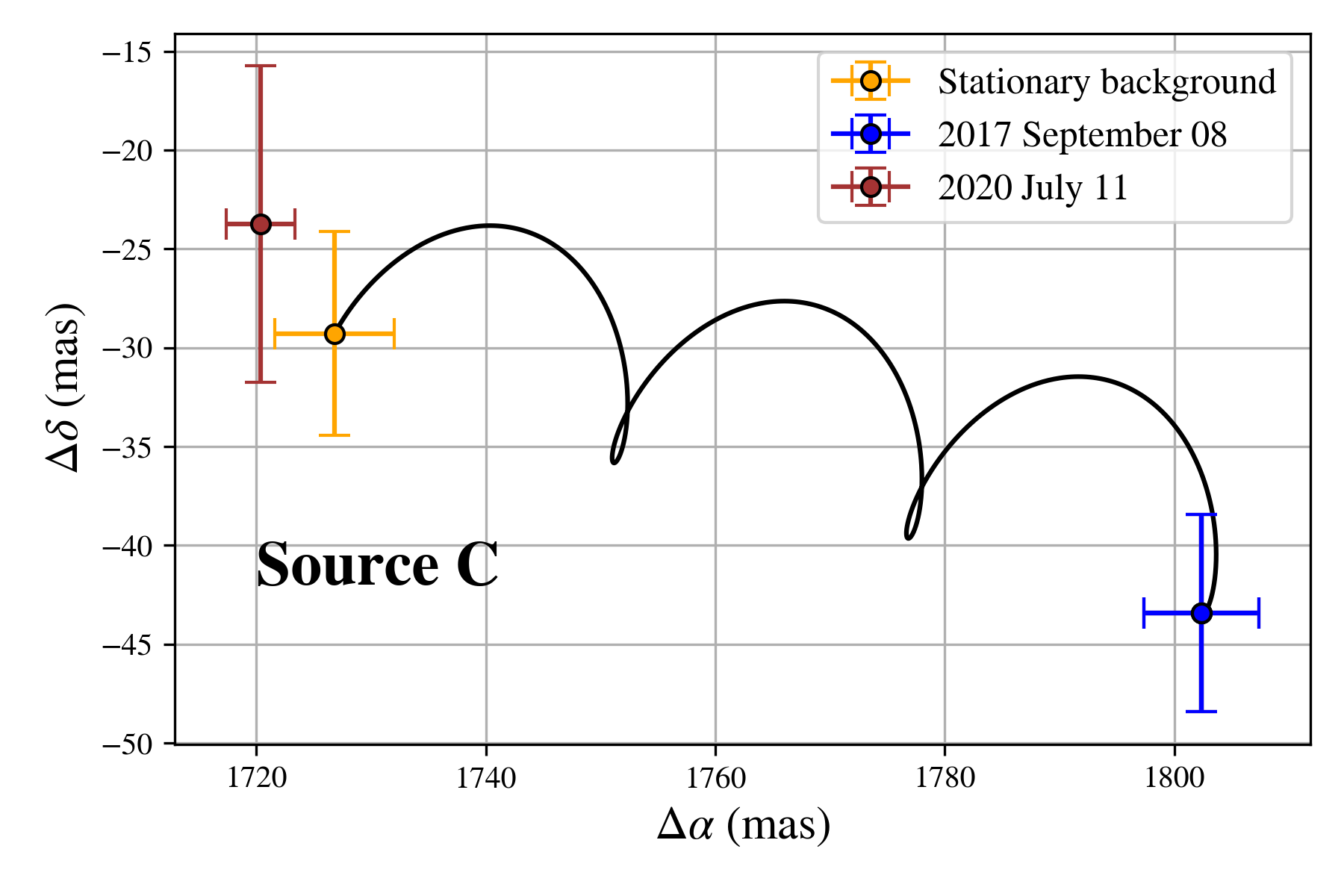}
    \includegraphics[width=0.32\textwidth]{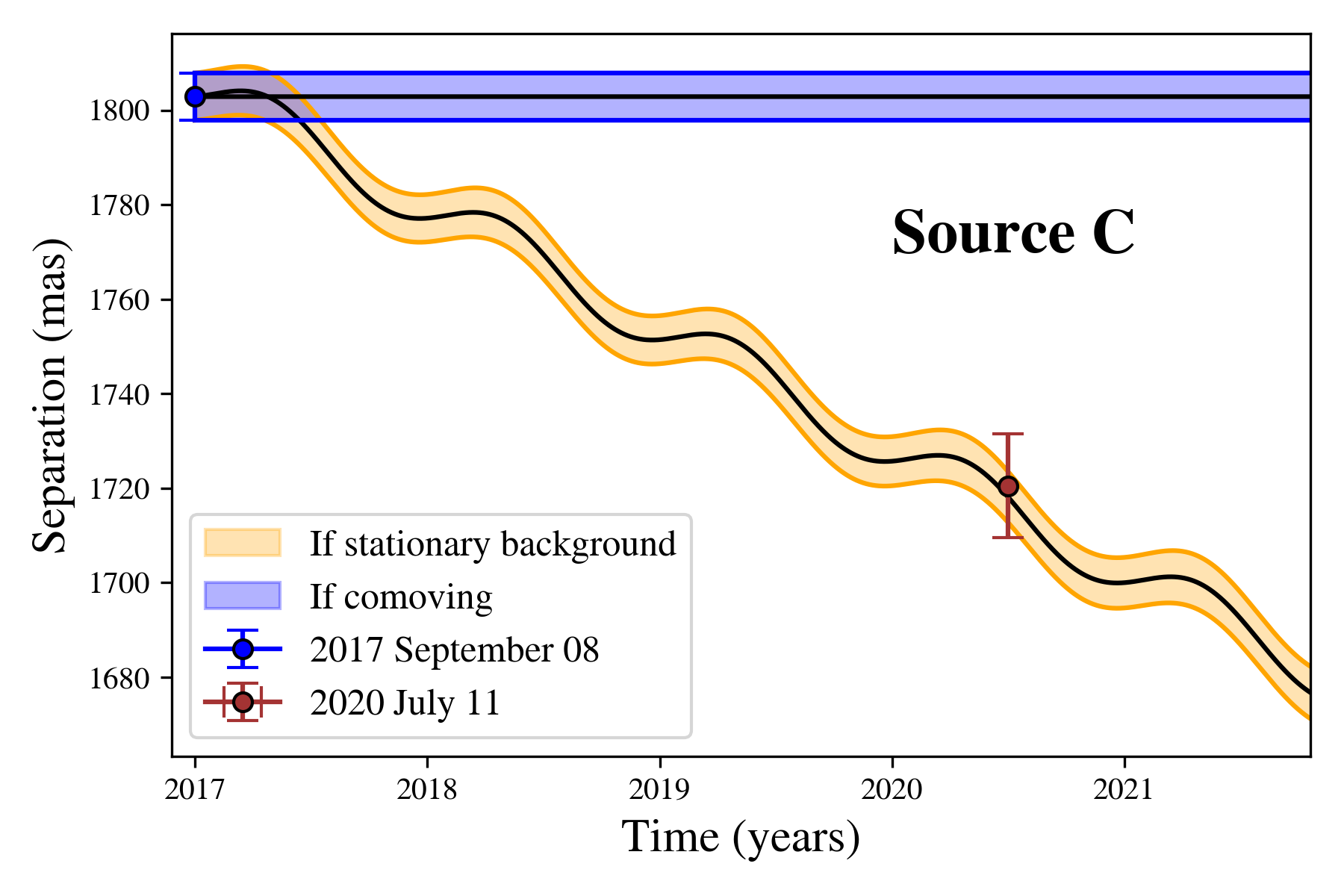}
    \includegraphics[width=0.32\textwidth]{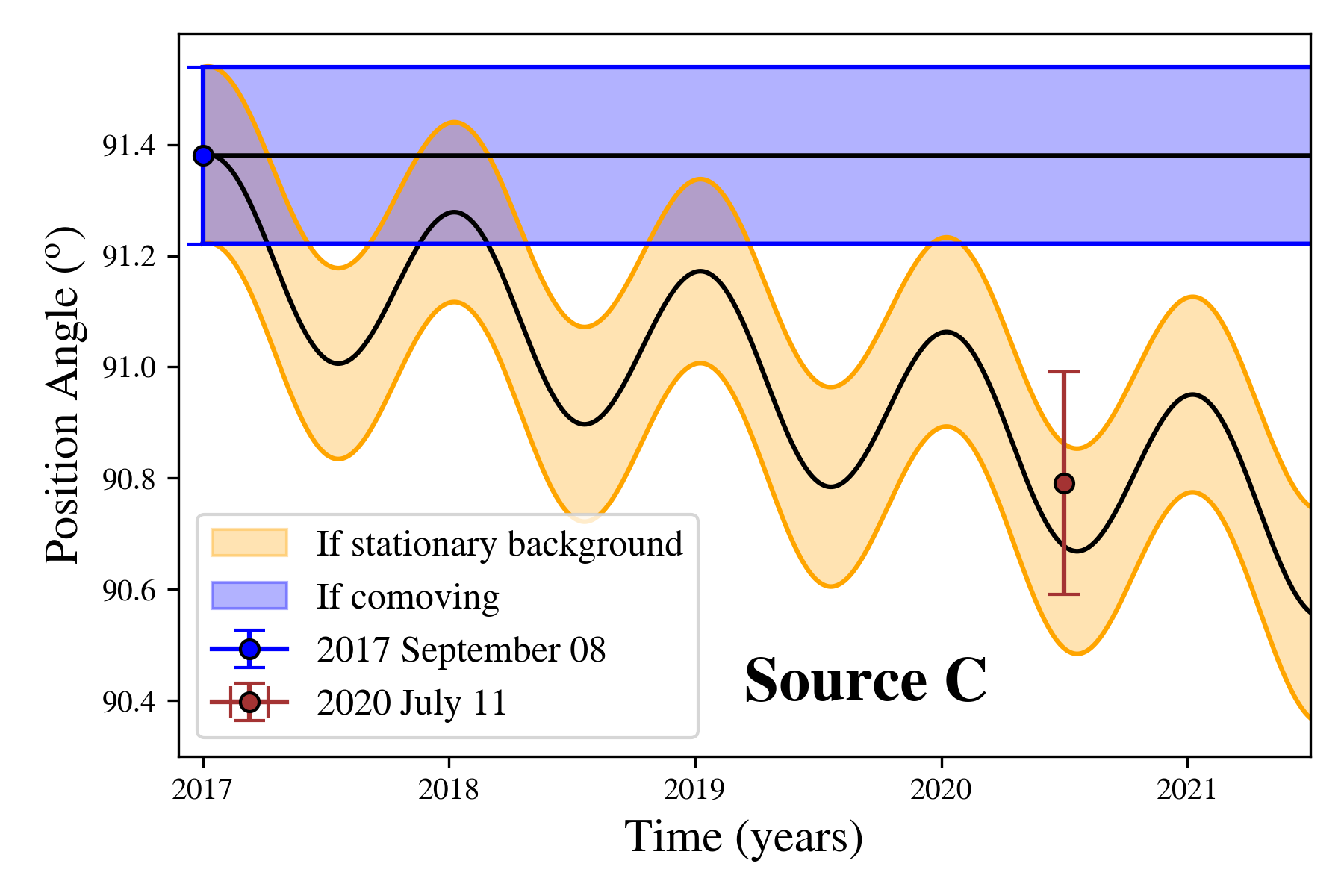}
    \caption{\textit{Left}: Relative separations between source B (top row), C (bottom row), and $\gamma$ Cas in right ascension ($\alpha$) and declination ($\delta$). The first epoch astrometric point is plotted in blue (September 08, 2017) and the second epoch astrometric point is plotted in red (July 11, 2020). The expected position for a stationary background object is plotted in yellow, along with its proper motion track. \textit{Middle}: Separation from $\gamma$ Cas in mas as a function of time. A background object with zero proper motion would follow the orange track, while a bounded and comoving CBC would lie within the blue zone. \textit{Right}: Same as middle, but for the position angle in degrees.}
    \label{fig:propermotion}
\end{figure*}

Conducting a proper motion analysis is among the most robust methods for confirming the gravitational association of a source with the system (e.g., \citealt{2021A&A...648A..73B}). Given the proximity of $\gamma$ Cas and its proper motions (see Section \ref{sec:gammaCas}), it was possible to conduct a statistically significant proper motion analysis between the two observed epochs (September 08, 2017, and July 11, 2020). Figure \ref{fig:propermotion} presents the relative separations between sources B, C, and $\gamma$ Cas in RA and Dec, alongside the expected position of a stationary background star. The figure also displays the angular separation and position angle over time, along with the expected tracks for both comoving and stationary background objects. 

Source B's trajectory implies an underlying motion that necessitates additional epochs of observation for validation. In 2020, the angular separation data point aligns with the comoving track, but the position angle data point deviates by $\sim 3\sigma$ from the same track. This trajectory suggests that source B is more likely to be bound to $\gamma$ Cas rather than an unrelated background or foreground object. Nonetheless, its motion suggests potential scenarios such as orbital motion, ejection from the system, or the presence of a non-stationary background or foreground object. Using the mass of $\gamma$ Cas ($\sim$ 13 $M_\odot$; \citealt{2012A&A...537A..59N}) and the radial separation of source B, we calculated the as $v_\mathrm{esc} = \sqrt{2GM/\rho} \approx 7680$ m/s. Additionally, the projected velocity between 2017 and 2020 was determined, resulting in the vector $v_\mathrm{proj} \approx (1670 \, \hat{\rho} -17964 \, \hat{\theta})$ m/s. Since the norm of the velocity vector is greater than the escape velocity, source B appears to be physically associated with $\gamma$ Cas, but is not bound (as also suggested in \citealt{2021ApJS..257...69H}). A comprehensive characterization of this motion using high-contrast imaging necessitates further epochs of observation.

Source C consistently follows the motion track of an unrelated object across all three plots. Thus, we excluded it from our list of candidate CBCs with a $\gg 3\sigma$ confidence level.

\subsection{Frequency of Multiple Systems and Companion Frequency}\label{sec:frequency}
Two key concepts are commonly defined in the stellar multiplicity literature (e.g., \citealt{2013ARA&A..51..269D}): the frequency of multiple systems (MF) and the companion frequency (CF; i.e., the average number of companions per target). While a proper motion analysis is required to confirm most of the sources, we calculated a first estimation of MF and CF for the observed X-ray binaries in our sample. Among the total of 14 observed X-ray binaries, we have identified candidate CBCs in eight systems: 4U 1700$-$377, RX J2030.5+4751, Cyg X-1, X Per, 1H2202+501, $\gamma$ Cas, RX J1744.7$-$2713, and IGR J18483$-$0311. Thus, based on these numbers and at this stage of the study, MF for triple or higher-tier systems would be $8/14 \approx 0.6$ ($\sim 60\%$). For CF, it is important to note that X-ray binaries inherently possess one companion, the donor star, which is included in our calculation of CF. Based on the status of the detected sources listed in Table \ref{tab:prop_sources_XRB}, the calculated average number of companions per compact object is $2.1 \pm 1.1$ ($210 \pm 110\%$). This means that, on average, the compact object has two companions (the donor star and one additional companion). However, this value reduces to $1.8 \pm 0.9$ ($180 \pm 90\%$) when considering only the sources that are most likely to be gravitationally bound ($1 - P_\mathrm{unrelated}(\Sigma, \Theta) > 85\%$). These values are subject to change as new observations become available and further analyses are conducted.

\section{Discussion} \label{sec:discussion}
\subsection{Stellar Multiplicity}
The discovery of candidate CBCs would imply that X-ray binaries can still be produced by multiple-star systems rather than exclusively binary systems. The total mass (i.e., compact object and donor star) of the HMXBs in our sample all exceeds $\sim$ 10 $M_\odot$, with some reaching up to around 60 $M_\odot$, placing them on the higher end of the mass spectrum. Stellar multiplicity is believed to be common in high-mass star systems (e.g., \citealt{2012MNRAS.424.1925C, 2013ARA&A..51..269D}), and high-order multiplicity is thought to increase with primary mass (e.g., \citealt{2012A&A...538A..74P}). However, surveys for high-mass stars remain incomplete.

For high-mass stars ($\gtrsim$ 16 $M_\odot$), MF and CF are estimated to be $\ge 80\%$ and $130 \pm 20\%$, respectively \citep{2012MNRAS.424.1925C, 2013ARA&A..51..269D}. The first estimation of MF for our sample ($\sim 60\%$; calculated in Section \ref{sec:frequency}) falls below this percentage. However, in this study, MF is constrained by the range of projected separations (up to $\sim 12,000$ au). This implies that increasing this limit could potentially lead to the discovery of more companions and hence increasing the estimation of MF. As for CF ($210 \pm 110\%$; calculated in Section \ref{sec:frequency}), it currently exceeds the estimate obtained from the literature ($130 \pm 20\%$; \citealt{2012MNRAS.424.1925C}). Further observations will likely lead to the rejection of candidate CBCs we have detected, which would lower the sample's CF (along with the associated uncertainty range). In both cases, our preliminary estimations (MF and CF) seem to be broadly in line with current estimates in the literature.

For solar-type stars, the frequency $N(n)$ of multiplicity $n$ follows a geometric distribution $N(n) \sim \beta^{-n}$ (up to $n=7$ with $\beta = 2.3$ or $3.4$; \citealt{2008MNRAS.389..869E}). Such a relation has not been derived for massive stars. Thus, the results from this study could be used to derive one, allowing us to predict the frequency of multiplicity in X-ray binaries and high-mass systems. The current sample size may be insufficient to infer a statistically significant relationship, but these results can still contribute valuable data to future surveys of high-mass systems.

This pilot study, in addition to showing evidence for potential additional components in X-ray binaries, could contribute to stellar multiplicity surveys for massive stars. It could also enable us to probe stellar multiplicity at low mass ratios (below $\sim 0.1$).

\subsection{Stability in Wide Orbit}
The projected separations within the scope of this study (from $\sim 350$ to $\sim$ 12,000 au) would suggest that CBCs orbit at very large distances from the central X-ray binary. Note that CBCs located closer to the X-ray binary within the 2-3 kpc distance range cannot currently be detected through direct imaging. To assess whether potential CBCs could be gravitationally bound to these systems, we calculated the binding energy $E_\mathrm{bind}$ for every source likely to be a CBC. Assuming circular orbits, $E_\mathrm{bind}$ is estimated using the following equation (e.g., \citealt{naud_discovery_2014, 2022AJ....164....7P}):

\begin{equation}
    E_\mathrm{bind} \sim -\frac{GM_\mathrm{XRB}M_\mathrm{comp}}{1.27r}
\end{equation}

where $G$ is the gravitational constant, $M_\mathrm{XRB}$ is the total mass of the central X-ray binary, $M_\mathrm{comp}$ is the mass of the CBC, $r$ is the projected separation between the CBC and the X-ray binary and 1.27 is the average projection factor between $r$ and the semi-major axis assuming a random viewing angle (e.g., \citealt{2006ApJ...652.1572B}). 

The binding energies for each candidate CBC range from $\sim -2 \times 10^{42}$ erg to $\sim -3 \times 10^{44}$ erg. A binding energy of $\sim -10^{41}$ erg was obtained for the comoving exoplanet GU Psc b around a M3 spectral type star ($\sim$ 0.46 $M_\odot$), with a mass of $\sim 9-13$ $M_\mathrm{Jup}$ and located at a distance of $\sim$ 2000 au \citep{naud_discovery_2014}. All candidate CBCs have binding energies largely exceeding this currently known lower limit. This suggests that these sources, if confirmed as CBCs, would fall within the gravitational binding range of the X-ray binary. 

We must also consider dynamic stability for systems containing more than one candidate CBC. $N$-body simulations have shown that a configuration of two CBCs at the same projected separation from the central system can lead to dynamic instability (e.g., \citealt{1994MNRAS.267..161K}). Thus, in the case of 4U 1700$-$377, where B and C have similar projected separations, configurations B+D or C+D are more likely than B+C+D. As for RX J2030.5+4751 and IGR J18483$-$0311, the candidate CBCs have distant projected separations, suggesting that configurations B+C and B+C+D, respectively, are plausible.

\subsection{Companion Formation and Capture Scenarios}
If our findings are confirmed, we hypothesize that CBCs orbiting X-ray binaries could originate through two main mechanisms: (1) formation within the same environment as the central X-ray binary or (2) capture by the system. On the one hand, as detailed in \cite{2022AJ....164....7P}, there are three scenarios in which CBCs could potentially form in the direct environment of X-ray binaries, and these scenarios may unfold at different times. Firstly, CBCs could have formed simultaneously with the initial stars that subsequently evolved to form the present X-ray binary, resulting from the direct fragmentation of a collapsing prestellar core (e.g., \citealt{2012MNRAS.419.3115B}). Note that this scenario is unlikely because we would expect companions with a mass similar to that of the main X-ray binary. Secondly, their formation might have occurred within the circumbinary disk surrounding the initial binary system, prior to the supernova explosion of the now compact object (e.g., \citealt{2010ApJ...708.1585K}). Thirdly, CBCs could have formed within the supernova fallback disk arising from the explosion (e.g., \citealt{1992Natur.355..145W}). Note that fallback disks do not contain enough mass to enable star formation, but they do possess enough mass for the formation of substellar objects. Additional observations are needed to detect and characterize such disks (see Section \ref{sec:followup}). 

On the other hand, stellar and substellar CBCs could be gravitationally captured by the X-ray binary. This is analogous to the case of the PSR B1620-26 system, where a giant exoplanet is thought to have been captured by a binary system containing a neutron star and a white dwarf (e.g., \citealt{2003Sci...301..193S}). Given that the HMXBs in our sample are massive ($>10$ $M_\odot$) and located in the galactic plane (see Fig. \ref{fig:skymap}), such events would not be unlikely.

\subsection{Follow-Up and Complementary Observations}\label{sec:followup}
The findings of our study primarily consist of intermediate results, and additional observations are necessary to conduct further analyses and confirm that the CBCs are bound to the system. In this section, we provide a list of recommendations for follow-up and complementary observations.

First, we recommend re-observing the systems in the same band ($L'$) using the same instrument (NIRC2) at one or multiple additional epochs. This will allow us to conduct proper motion analysis (see Section \ref{sec:astrometry}) for all the candidate CBCs identified in this study. Multi-epoch observations will also enable us to characterize the orbital motion of these CBCs. Since the HMXBs presented in this study are located within 2-3 kpc, the time interval between epochs ranges from a few months to a few years. Table \ref{tab:followup} provides an estimate of the recommended year of re-observation for each system, ensuring a statistically significant ($>3\sigma$) proper motion analysis.

Second, we recommend observing the remaining 5 sources, such as Scorpius X-1, 1A0620-00, and V404 Cyg, to complete the sample of all X-ray binaries within 2-3 kpc accessible with Keck/NIRC2. This would also enable us to incorporate LMXBs into the analysis and discussion.

Third, observations in other bands (e.g., $K_s$) would allow us to construct color-magnitude diagrams and employ evolutionary models to estimate the physical properties of the CBCs with greater constraints (e.g., \citealt{2022AJ....164....7P}). Similarly, obtaining the near-infrared spectrum of the candidate CBCs would enable us to characterize the nature of the source and extract additional physical properties. Finally, submillimeter observations of the continuum emission would allow us to detect and characterize potential AU-scale circumstellar or protoplanetary disks composed of dust and hot gas (e.g., \citealt{2013A&A...560A.108C,2017MNRAS.471..355I,2019A&A...623A..47W}).

Contrast curves for all systems with CBCs can be found in the Appendix to assess the limit that has been reached during these observations.

\begin{table}[ht!]
    \caption{Recommended year for follow-up observations for every X-ray binary with at least one candidate CBC -- calculated using the distance, proper motions, and astrometric errors on the 2020 observations -- enabling a proper motion analysis with at a $>3\sigma$ significance level.}
    \centering
    \begin{tabular}{cc}
    \hline\hline
    \textbf{Target} & \textbf{Recommended Year} \\ \hline
    X Per & 2026 \\
    Cyg X-1 & 2024 \\
    IGR J18483$-$0311 & 2024 \\
    1H2202+501 & 2027 \\
    4U 1700$-$377 & 2024 \\
    RX J2030.5+4751 & 2024 \\ \hline
    \end{tabular}
    \label{tab:followup}
\end{table}

\section{Summary and Conclusions} \label{sec:summary}
In this study, we presented the first $L'$-band high-contrast images of nearby, high-mass X-ray binaries using NIRC2 and the vortex coronagraph on the W. M. Keck Observatory. A total of 14 systems were observed from a sample of 19 X-ray binaries within $\sim$ 2-3 kpc and amenable for direct imaging. One or several sources with a SNR $>$ 5 were found in 8 of the observed X-ray binaries. To discern the nature of these sources -- whether unrelated objects or candidate CBCs -- we employed Galactic population models for all systems, and proper motion analysis for $\gamma$ Cas. We find that, if confirmed, these results would imply the presence of stellar and sub-stellar CBCs in the direct environment of X-ray binaries ($\sim 350-12,000$ au), which opens up the discussion on the binary nature of these systems. As a pilot study, this work presents a catalog of photometric and astrometric parameters for the first epochs of observations. Follow-up observations or additional characterization (e.g., infrared spectrum) will enable us to conduct proper motion analyses to discriminate more robustly background/foreground sources from comoving, gravitationally-bound CBCs. 

\section*{Acknowledgments}
The data presented herein were obtained at the W. M. Keck Observatory, which is operated as a scientific partnership among the California Institute of Technology, the University of California and the National Aeronautics and Space Administration. The Observatory was made possible by the generous financial support of the W. M. Keck Foundation. The authors wish to recognize and acknowledge the very significant cultural role and reverence that the summit of Mauna Kea has always had within the indigenous Hawaiian community.  We are most fortunate to have the opportunity to conduct observations from this mountain. We also want to thank V. Christiaens for his help with VIP. We thank Lauren M. Weiss, Marie-Eve Naud and Anjali Rao for their contribution to the Keck/NIRC2 proposal.

M.P.E. is supported by the Institute for Data Valorisation (IVADO) through the M. Sc. Excellence Scholarship, by the Department of Physics of the Université de Montréal and by the Institute for Research on Exoplanets (iREx). J.H.L. is supported by the Natural Sciences and Engineering Research Council of Canada (NSERC) through the Canada Research Chair programs and wishes to acknowledge the support of an NSERC Discovery Grant and the NSERC accelerator grant. 

\bibliography{bibliography}{}

\begin{thebibliography}{}
\expandafter\ifx\csname natexlab\endcsname\relax\def\natexlab#1{#1}\fi
\providecommand{\url}[1]{\href{#1}{#1}}
\providecommand{\dodoi}[1]{doi:~\href{http://doi.org/#1}{\nolinkurl{#1}}}
\providecommand{\doeprint}[1]{\href{http://ascl.net/#1}{\nolinkurl{http://ascl.net/#1}}}
\providecommand{\doarXiv}[1]{\href{https://arxiv.org/abs/#1}{\nolinkurl{https://arxiv.org/abs/#1}}}

\bibitem[{{Abate} {et~al.}(2013){Abate}, {Pols}, {Izzard}, {Mohamed}, \& {de Mink}}]{2013A&A...552A..26A}
{Abate}, C., {Pols}, O.~R., {Izzard}, R.~G., {Mohamed}, S.~S., \& {de Mink}, S.~E. 2013, \aap, 552, A26, \dodoi{10.1051/0004-6361/201220007}

\bibitem[{{Avakyan} {et~al.}(2023){Avakyan}, {Neumann}, {Zainab}, {Doroshenko}, {Wilms}, \& {Santangelo}}]{2023A&A...675A.199A}
{Avakyan}, A., {Neumann}, M., {Zainab}, A., {et~al.} 2023, \aap, 675, A199, \dodoi{10.1051/0004-6361/202346522}

\bibitem[{{Bailer-Jones} {et~al.}(2021){Bailer-Jones}, {Rybizki}, {Fouesneau}, {Demleitner}, \& {Andrae}}]{2021AJ....161..147B}
{Bailer-Jones}, C.~A.~L., {Rybizki}, J., {Fouesneau}, M., {Demleitner}, M., \& {Andrae}, R. 2021, \aj, 161, 147, \dodoi{10.3847/1538-3881/abd806}

\bibitem[{{Bakos} {et~al.}(2007){Bakos}, {Noyes}, {Kov{\'a}cs}, {Latham}, {Sasselov}, {Torres}, {Fischer}, {Stefanik}, {Sato}, {Johnson}, {P{\'a}l}, {Marcy}, {Butler}, {Esquerdo}, {Stanek}, {L{\'a}z{\'a}r}, {Papp}, {S{\'a}ri}, \& {Sip{\H{o}}cz}}]{2007ApJ...656..552B}
{Bakos}, G.~{\'A}., {Noyes}, R.~W., {Kov{\'a}cs}, G., {et~al.} 2007, \apj, 656, 552, \dodoi{10.1086/509874}

\bibitem[{{Bala} {et~al.}(2020){Bala}, {Roy}, \& {Bhattacharya}}]{2020MNRAS.493.3045B}
{Bala}, S., {Roy}, J., \& {Bhattacharya}, D. 2020, \mnras, 493, 3045, \dodoi{10.1093/mnras/staa437}

\bibitem[{{Bate}(2012)}]{2012MNRAS.419.3115B}
{Bate}, M.~R. 2012, \mnras, 419, 3115, \dodoi{10.1111/j.1365-2966.2011.19955.x}

\bibitem[{{Belczynski} \& {Ziolkowski}(2009)}]{2009ApJ...707..870B}
{Belczynski}, K., \& {Ziolkowski}, J. 2009, \apj, 707, 870, \dodoi{10.1088/0004-637X/707/2/870}

\bibitem[{{Bhalerao} {et~al.}(2015){Bhalerao}, {Romano}, {Tomsick}, {Natalucci}, {Smith}, {Bellm}, {Boggs}, {Chakrabarty}, {Christensen}, {Craig}, {Fuerst}, {Hailey}, {Harrison}, {Krivonos}, {Lu}, {Madsen}, {Stern}, {Younes}, \& {Zhang}}]{2015MNRAS.447.2274B}
{Bhalerao}, V., {Romano}, P., {Tomsick}, J., {et~al.} 2015, \mnras, 447, 2274, \dodoi{10.1093/mnras/stu2495}

\bibitem[{{Bikmaev} {et~al.}(2017){Bikmaev}, {Nikolaeva}, {Shimansky}, {Galeev}, {Zhuchkov}, {Irtuganov}, {Melnikov}, {Sakhibullin}, {Grebenev}, \& {Sharipova}}]{2017AstL...43..664B}
{Bikmaev}, I.~F., {Nikolaeva}, E.~A., {Shimansky}, V.~V., {et~al.} 2017, Astronomy Letters, 43, 664, \dodoi{10.1134/S1063773717100012}

\bibitem[{{Bird} {et~al.}(2009){Bird}, {Bazzano}, {Hill}, {McBride}, {Sguera}, {Shaw}, \& {Watkins}}]{2009MNRAS.393L..11B}
{Bird}, A.~J., {Bazzano}, A., {Hill}, A.~B., {et~al.} 2009, \mnras, 393, L11, \dodoi{10.1111/j.1745-3933.2008.00583.x}

\bibitem[{{Bird} {et~al.}(2004){Bird}, {Barlow}, {Bassani}, {Bazzano}, {Bodaghee}, {Capitanio}, {Cocchi}, {Del Santo}, {Dean}, {Hill}, {Lebrun}, {Malaguti}, {Malizia}, {Much}, {Shaw}, {Stephen}, {Terrier}, {Ubertini}, \& {Walter}}]{2004ApJ...607L..33B}
{Bird}, A.~J., {Barlow}, E.~J., {Bassani}, L., {et~al.} 2004, \apjl, 607, L33, \dodoi{10.1086/421772}

\bibitem[{{Bird} {et~al.}(2007){Bird}, {Malizia}, {Bazzano}, {Barlow}, {Bassani}, {Hill}, {B{\'e}langer}, {Capitanio}, {Clark}, {Dean}, {Fiocchi}, {G{\"o}tz}, {Lebrun}, {Molina}, {Produit}, {Renaud}, {Sguera}, {Stephen}, {Terrier}, {Ubertini}, {Walter}, {Winkler}, \& {Zurita}}]{2007ApJS..170..175B}
{Bird}, A.~J., {Malizia}, A., {Bazzano}, A., {et~al.} 2007, \apjs, 170, 175, \dodoi{10.1086/513148}

\bibitem[{{Bird} {et~al.}(2016){Bird}, {Bazzano}, {Malizia}, {Fiocchi}, {Sguera}, {Bassani}, {Hill}, {Ubertini}, \& {Winkler}}]{2016ApJS..223...15B}
{Bird}, A.~J., {Bazzano}, A., {Malizia}, A., {et~al.} 2016, \apjs, 223, 15, \dodoi{10.3847/0067-0049/223/1/15}

\bibitem[{{Blackman} {et~al.}(2021){Blackman}, {Beaulieu}, {Bennett}, {Danielski}, {Alard}, {Cole}, {Vandorou}, {Ranc}, {Terry}, {Bhattacharya}, {Bond}, {Bachelet}, {Veras}, {Koshimoto}, {Batista}, \& {Marquette}}]{2021Natur.598..272B}
{Blackman}, J.~W., {Beaulieu}, J.~P., {Bennett}, D.~P., {et~al.} 2021, \nat, 598, 272, \dodoi{10.1038/s41586-021-03869-6}

\bibitem[{{Blay} {et~al.}(2005){Blay}, {Rib{\'o}}, {Negueruela}, {Torrej{\'o}n}, {Reig}, {Camero}, {Mirabel}, \& {Reglero}}]{2005A&A...438..963B}
{Blay}, P., {Rib{\'o}}, M., {Negueruela}, I., {et~al.} 2005, \aap, 438, 963, \dodoi{10.1051/0004-6361:20042207}

\bibitem[{{Bohn} {et~al.}(2021){Bohn}, {Ginski}, {Kenworthy}, {Mamajek}, {Pecaut}, {Mugrauer}, {Vogt}, {Adam}, {Meshkat}, {Reggiani}, \& {Snik}}]{2021A&A...648A..73B}
{Bohn}, A.~J., {Ginski}, C., {Kenworthy}, M.~A., {et~al.} 2021, \aap, 648, A73, \dodoi{10.1051/0004-6361/202140508}

\bibitem[{{Bolton}(1972{\natexlab{a}})}]{1972NPhS..240..124B}
{Bolton}, C.~T. 1972{\natexlab{a}}, Nature Physical Science, 240, 124, \dodoi{10.1038/physci240124a0}

\bibitem[{{Bolton}(1972{\natexlab{b}})}]{1972Natur.235..271B}
---. 1972{\natexlab{b}}, \nat, 235, 271, \dodoi{10.1038/235271b0}

\bibitem[{Bonavita {et~al.}(2016)Bonavita, Desidera, Thalmann, Janson, Vigan, Chauvin, \& Lannier}]{bonavita_spots_2016}
Bonavita, M., Desidera, S., Thalmann, C., {et~al.} 2016, arXiv:1605.03962 [astro-ph], \dodoi{10.1051/0004-6361/201628231}

\bibitem[{{Bond} {et~al.}(2018){Bond}, {Wizinowich}, {Chun}, {Mawet}, {Lilley}, {Cetre}, {Jovanovic}, {Delorme}, {Wetherell}, {Jacobson}, {Lockhart}, {Warmbier}, {Wallace}, {Hall}, {Goebel}, {Guyon}, {Plantet}, {Agapito}, {Giordano}, {Esposito}, \& {Femenia-Castella}}]{2018SPIE10703E..1ZB}
{Bond}, C.~Z., {Wizinowich}, P., {Chun}, M., {et~al.} 2018, in Society of Photo-Optical Instrumentation Engineers (SPIE) Conference Series, Vol. 10703, Adaptive Optics Systems VI, ed. L.~M. {Close}, L.~{Schreiber}, \& D.~{Schmidt}, 107031Z, \dodoi{10.1117/12.2314121}

\bibitem[{{Boon} {et~al.}(2016){Boon}, {Bird}, {Hill}, {Sidoli}, {Sguera}, {Goossens}, {Fiocchi}, {McBride}, \& {Drave}}]{2016MNRAS.456.4111B}
{Boon}, C.~M., {Bird}, A.~J., {Hill}, A.~B., {et~al.} 2016, \mnras, 456, 4111, \dodoi{10.1093/mnras/stv2975}

\bibitem[{{Boroson} {et~al.}(2003){Boroson}, {Vrtilek}, {Kallman}, \& {Corcoran}}]{2003ApJ...592..516B}
{Boroson}, B., {Vrtilek}, S.~D., {Kallman}, T., \& {Corcoran}, M. 2003, \apj, 592, 516, \dodoi{10.1086/375636}

\bibitem[{{Bozzo} {et~al.}(2012){Bozzo}, {Pavan}, {Ferrigno}, {Falanga}, {Campana}, {Paltani}, {Stella}, \& {Walter}}]{2012A&A...544A.118B}
{Bozzo}, E., {Pavan}, L., {Ferrigno}, C., {et~al.} 2012, \aap, 544, A118, \dodoi{10.1051/0004-6361/201218900}

\bibitem[{{Bozzo} {et~al.}(2016){Bozzo}, {Bhalerao}, {Pradhan}, {Tomsick}, {Romano}, {Ferrigno}, {Chaty}, {Oskinova}, {Manousakis}, {Walter}, {Falanga}, {Campana}, {Stella}, {Ramolla}, \& {Chini}}]{2016A&A...596A..16B}
{Bozzo}, E., {Bhalerao}, V., {Pradhan}, P., {et~al.} 2016, \aap, 596, A16, \dodoi{10.1051/0004-6361/201629311}

\bibitem[{{Brandeker} {et~al.}(2006){Brandeker}, {Jayawardhana}, {Khavari}, {Haisch}, \& {Mardones}}]{2006ApJ...652.1572B}
{Brandeker}, A., {Jayawardhana}, R., {Khavari}, P., {Haisch}, Karl~E., J., \& {Mardones}, D. 2006, \apj, 652, 1572, \dodoi{10.1086/508483}

\bibitem[{{Burrows} {et~al.}(2001){Burrows}, {Hubbard}, {Lunine}, \& {Liebert}}]{2001RvMP...73..719B}
{Burrows}, A., {Hubbard}, W.~B., {Lunine}, J.~I., \& {Liebert}, J. 2001, Reviews of Modern Physics, 73, 719, \dodoi{10.1103/RevModPhys.73.719}

\bibitem[{{Caballero-Nieves} {et~al.}(2009){Caballero-Nieves}, {Gies}, {Bolton}, {Hadrava}, {Herrero}, {Hillwig}, {Howell}, {Huang}, {Kaper}, {Koubsk{\'y}}, \& {McSwain}}]{2009ApJ...701.1895C}
{Caballero-Nieves}, S.~M., {Gies}, D.~R., {Bolton}, C.~T., {et~al.} 2009, \apj, 701, 1895, \dodoi{10.1088/0004-637X/701/2/1895}

\bibitem[{{Charles} \& {Coe}(2006)}]{2006csxs.book..215C}
{Charles}, P.~A., \& {Coe}, M.~J. 2006, in Compact stellar X-ray sources, Vol.~39, 215--265

\bibitem[{{Charles} {et~al.}(1978){Charles}, {Mason}, {White}, {Culhane}, {Sanford}, \& {Moffat}}]{1978MNRAS.183..813C}
{Charles}, P.~A., {Mason}, K.~O., {White}, N.~E., {et~al.} 1978, \mnras, 183, 813, \dodoi{10.1093/mnras/183.4.813}

\bibitem[{{Chaty} {et~al.}(2008){Chaty}, {Rahoui}, {Foellmi}, {Tomsick}, {Rodriguez}, \& {Walter}}]{2008A&A...484..783C}
{Chaty}, S., {Rahoui}, F., {Foellmi}, C., {et~al.} 2008, \aap, 484, 783, \dodoi{10.1051/0004-6361:20078768}

\bibitem[{{Chen} \& {Podsiadlowski}(2016)}]{2016ApJ...830..131C}
{Chen}, W.-C., \& {Podsiadlowski}, P. 2016, \apj, 830, 131, \dodoi{10.3847/0004-637X/830/2/131}

\bibitem[{{Chernyakova} {et~al.}(2003){Chernyakova}, {Lutovinov}, {Capitanio}, {Lund}, \& {Gehrels}}]{2003ATel..157....1C}
{Chernyakova}, M., {Lutovinov}, A., {Capitanio}, F., {Lund}, N., \& {Gehrels}, N. 2003, The Astronomer's Telegram, 157, 1

\bibitem[{{Chevalier} \& {Ilovaisky}(1998)}]{1998A&A...330..201C}
{Chevalier}, C., \& {Ilovaisky}, S.~A. 1998, \aap, 330, 201.
\newblock \doarXiv{astro-ph/9710008}

\bibitem[{{Chini} {et~al.}(2012){Chini}, {Hoffmeister}, {Nasseri}, {Stahl}, \& {Zinnecker}}]{2012MNRAS.424.1925C}
{Chini}, R., {Hoffmeister}, V.~H., {Nasseri}, A., {Stahl}, O., \& {Zinnecker}, H. 2012, \mnras, 424, 1925, \dodoi{10.1111/j.1365-2966.2012.21317.x}

\bibitem[{{Choi} {et~al.}(2016){Choi}, {Dotter}, {Conroy}, {Cantiello}, {Paxton}, \& {Johnson}}]{2016ApJ...823..102C}
{Choi}, J., {Dotter}, A., {Conroy}, C., {et~al.} 2016, \apj, 823, 102, \dodoi{10.3847/0004-637X/823/2/102}

\bibitem[{{Chun} {et~al.}(2015){Chun}, {Jung}, {Kang}, {Kim}, \& {Sohn}}]{2015A&A...578A..51C}
{Chun}, S.-H., {Jung}, M., {Kang}, M., {Kim}, J.-W., \& {Sohn}, Y.-J. 2015, \aap, 578, A51, \dodoi{10.1051/0004-6361/201525849}

\bibitem[{{Coburn} {et~al.}(2001){Coburn}, {Heindl}, {Gruber}, {Rothschild}, {Staubert}, {Wilms}, \& {Kreykenbohm}}]{2001ApJ...552..738C}
{Coburn}, W., {Heindl}, W.~A., {Gruber}, D.~E., {et~al.} 2001, \apj, 552, 738, \dodoi{10.1086/320565}

\bibitem[{{Coburn} {et~al.}(2002){Coburn}, {Heindl}, {Rothschild}, {Gruber}, {Kreykenbohm}, {Wilms}, {Kretschmar}, \& {Staubert}}]{2002ApJ...580..394C}
{Coburn}, W., {Heindl}, W.~A., {Rothschild}, R.~E., {et~al.} 2002, \apj, 580, 394, \dodoi{10.1086/343033}

\bibitem[{Coleiro \& Chaty(2013)}]{coleiro_distribution_2013}
Coleiro, A., \& Chaty, S. 2013, ApJ, 764, 185, \dodoi{10.1088/0004-637X/764/2/185}

\bibitem[{{Coleiro} {et~al.}(2013){Coleiro}, {Chaty}, {Zurita Heras}, {Rahoui}, \& {Tomsick}}]{2013A&A...560A.108C}
{Coleiro}, A., {Chaty}, S., {Zurita Heras}, J.~A., {Rahoui}, F., \& {Tomsick}, J.~A. 2013, Astronomy and Astrophysics, 560, A108, \dodoi{10.1051/0004-6361/201322382}

\bibitem[{{Corbet} {et~al.}(2007){Corbet}, {Markwardt}, \& {Tueller}}]{2007ApJ...655..458C}
{Corbet}, R.~H.~D., {Markwardt}, C.~B., \& {Tueller}, J. 2007, \apj, 655, 458, \dodoi{10.1086/509319}

\bibitem[{{Corbet} \& {Peele}(2001)}]{2001ApJ...562..936C}
{Corbet}, R. H.~D., \& {Peele}, A.~G. 2001, \apj, 562, 936, \dodoi{10.1086/323849}

\bibitem[{{Correia} {et~al.}(2006){Correia}, {Zinnecker}, {Ratzka}, \& {Sterzik}}]{2006A&A...459..909C}
{Correia}, S., {Zinnecker}, H., {Ratzka}, T., \& {Sterzik}, M.~F. 2006, \aap, 459, 909, \dodoi{10.1051/0004-6361:20065545}

\bibitem[{{De Angeli} {et~al.}(2023){De Angeli}, {Weiler}, {Montegriffo}, {Evans}, {Riello}, {Andrae}, {Carrasco}, {Busso}, {Burgess}, {Cacciari}, {Davidson}, {Harrison}, {Hodgkin}, {Jordi}, {Osborne}, {Pancino}, {Altavilla}, {Barstow}, {Bailer-Jones}, {Bellazzini}, {Brown}, {Castellani}, {Cowell}, {Delchambre}, {De Luise}, {Diener}, {Fabricius}, {Fouesneau}, {Fr{\'e}mat}, {Gilmore}, {Giuffrida}, {Hambly}, {Hidalgo}, {Holland}, {Kostrzewa-Rutkowska}, {van Leeuwen}, {Lobel}, {Marinoni}, {Miller}, {Pagani}, {Palaversa}, {Piersimoni}, {Pulone}, {Ragaini}, {Rainer}, {Richards}, {Rixon}, {Ruz-Mieres}, {Sanna}, {Sarro}, {Rowell}, {Sordo}, {Walton}, \& {Yoldas}}]{2023A&A...674A...2D}
{De Angeli}, F., {Weiler}, M., {Montegriffo}, P., {et~al.} 2023, \aap, 674, A2, \dodoi{10.1051/0004-6361/202243680}

\bibitem[{{Delgado-Mart{\'\i}} {et~al.}(2001){Delgado-Mart{\'\i}}, {Levine}, {Pfahl}, \& {Rappaport}}]{2001ApJ...546..455D}
{Delgado-Mart{\'\i}}, H., {Levine}, A.~M., {Pfahl}, E., \& {Rappaport}, S.~A. 2001, \apj, 546, 455, \dodoi{10.1086/318236}

\bibitem[{{Desidera} \& {Barbieri}(2007)}]{2007A&A...462..345D}
{Desidera}, S., \& {Barbieri}, M. 2007, \aap, 462, 345, \dodoi{10.1051/0004-6361:20066319}

\bibitem[{{Di Salvo} {et~al.}(2001){Di Salvo}, {Done}, {{\.Z}ycki}, {Burderi}, \& {Robba}}]{2001ApJ...547.1024D}
{Di Salvo}, T., {Done}, C., {{\.Z}ycki}, P.~T., {Burderi}, L., \& {Robba}, N.~R. 2001, \apj, 547, 1024, \dodoi{10.1086/318396}

\bibitem[{{Dietrich} \& {Ginski}(2018)}]{2018A&A...620A.102D}
{Dietrich}, J., \& {Ginski}, C. 2018, \aap, 620, A102, \dodoi{10.1051/0004-6361/201731341}

\bibitem[{{Diez} {et~al.}(2022){Diez}, {Grinberg}, {F{\"u}rst}, {Sokolova-Lapa}, {Santangelo}, {Wilms}, {Pottschmidt}, {Mart{\'\i}nez-N{\'u}{\~n}ez}, {Malacaria}, \& {Kretschmar}}]{2022A&A...660A..19D}
{Diez}, C.~M., {Grinberg}, V., {F{\"u}rst}, F., {et~al.} 2022, \aap, 660, A19, \dodoi{10.1051/0004-6361/202141751}

\bibitem[{{Done} {et~al.}(2007){Done}, {Gierli{\'n}ski}, \& {Kubota}}]{2007A&ARv..15....1D}
{Done}, C., {Gierli{\'n}ski}, M., \& {Kubota}, A. 2007, \aapr, 15, 1, \dodoi{10.1007/s00159-007-0006-1}

\bibitem[{{Doroshenko} {et~al.}(2012){Doroshenko}, {Santangelo}, {Kreykenbohm}, \& {Doroshenko}}]{2012A&A...540L...1D}
{Doroshenko}, V., {Santangelo}, A., {Kreykenbohm}, I., \& {Doroshenko}, R. 2012, \aap, 540, L1, \dodoi{10.1051/0004-6361/201218878}

\bibitem[{{Dotter}(2016)}]{2016ApJS..222....8D}
{Dotter}, A. 2016, \apjs, 222, 8, \dodoi{10.3847/0067-0049/222/1/8}

\bibitem[{{Drave} {et~al.}(2014){Drave}, {Bird}, {Sidoli}, {Sguera}, {Bazzano}, {Hill}, \& {Goossens}}]{2014MNRAS.439.2175D}
{Drave}, S.~P., {Bird}, A.~J., {Sidoli}, L., {et~al.} 2014, \mnras, 439, 2175, \dodoi{10.1093/mnras/stu110}

\bibitem[{{Drave} {et~al.}(2012){Drave}, {Bird}, {Townsend}, {Hill}, {McBride}, {Sguera}, {Bazzano}, \& {Clark}}]{2012A&A...539A..21D}
{Drave}, S.~P., {Bird}, A.~J., {Townsend}, L.~J., {et~al.} 2012, \aap, 539, A21, \dodoi{10.1051/0004-6361/201117947}

\bibitem[{{Ducci} {et~al.}(2013){Ducci}, {Doroshenko}, {Sasaki}, {Santangelo}, {Esposito}, {Romano}, \& {Vercellone}}]{2013A&A...559A.135D}
{Ducci}, L., {Doroshenko}, V., {Sasaki}, M., {et~al.} 2013, \aap, 559, A135, \dodoi{10.1051/0004-6361/201322299}

\bibitem[{{Ducci} {et~al.}(2019){Ducci}, {Romano}, {Ji}, \& {Santangelo}}]{2019A&A...631A.135D}
{Ducci}, L., {Romano}, P., {Ji}, L., \& {Santangelo}, A. 2019, \aap, 631, A135, \dodoi{10.1051/0004-6361/201936544}

\bibitem[{{Duch{\^e}ne} \& {Kraus}(2013)}]{2013ARA&A..51..269D}
{Duch{\^e}ne}, G., \& {Kraus}, A. 2013, \araa, 51, 269, \dodoi{10.1146/annurev-astro-081710-102602}

\bibitem[{{Dupree} {et~al.}(1978){Dupree}, {Davis}, {Gursky}, {Hartmann}, {Raymond}, {Boggess}, {Holm}, {Kondo}, {Wu}, {Macchetto}, {Sandford}, {Willis}, {Wilson}, {Ciatti}, {Hutchings}, {Johnson}, {Jugaku}, {Morton}, {Treves}, \& {van den Heuvel}}]{1978Natur.275..400D}
{Dupree}, A.~K., {Davis}, R.~J., {Gursky}, H., {et~al.} 1978, \nat, 275, 400, \dodoi{10.1038/275400a0}

\bibitem[{{Eggleton} \& {Tokovinin}(2008)}]{2008MNRAS.389..869E}
{Eggleton}, P.~P., \& {Tokovinin}, A.~A. 2008, \mnras, 389, 869, \dodoi{10.1111/j.1365-2966.2008.13596.x}

\bibitem[{{El Mellah} {et~al.}(2019){El Mellah}, {Sundqvist}, \& {Keppens}}]{2019A&A...622L...3E}
{El Mellah}, I., {Sundqvist}, J.~O., \& {Keppens}, R. 2019, \aap, 622, L3, \dodoi{10.1051/0004-6361/201834543}

\bibitem[{{Eriksson} {et~al.}(2020){Eriksson}, {Asensio Torres}, {Janson}, {Aoyama}, {Marleau}, {Bonnefoy}, \& {Petrus}}]{2020A&A...638L...6E}
{Eriksson}, S.~C., {Asensio Torres}, R., {Janson}, M., {et~al.} 2020, \aap, 638, L6, \dodoi{10.1051/0004-6361/202038131}

\bibitem[{{Esin} {et~al.}(1998){Esin}, {Narayan}, {Cui}, {Grove}, \& {Zhang}}]{1998ApJ...505..854E}
{Esin}, A.~A., {Narayan}, R., {Cui}, W., {Grove}, J.~E., \& {Zhang}, S.-N. 1998, \apj, 505, 854, \dodoi{10.1086/306186}

\bibitem[{{Fabian} {et~al.}(1989){Fabian}, {Rees}, {Stella}, \& {White}}]{1989MNRAS.238..729F}
{Fabian}, A.~C., {Rees}, M.~J., {Stella}, L., \& {White}, N.~E. 1989, \mnras, 238, 729, \dodoi{10.1093/mnras/238.3.729}

\bibitem[{{Faherty} {et~al.}(2016){Faherty}, {Riedel}, {Cruz}, {Gagne}, {Filippazzo}, {Lambrides}, {Fica}, {Weinberger}, {Thorstensen}, {Tinney}, {Baldassare}, {Lemonier}, \& {Rice}}]{2016ApJS..225...10F}
{Faherty}, J.~K., {Riedel}, A.~R., {Cruz}, K.~L., {et~al.} 2016, \apjs, 225, 10, \dodoi{10.3847/0067-0049/225/1/10}

\bibitem[{{Fender} {et~al.}(2004){Fender}, {Belloni}, \& {Gallo}}]{2004MNRAS.355.1105F}
{Fender}, R.~P., {Belloni}, T.~M., \& {Gallo}, E. 2004, \mnras, 355, 1105, \dodoi{10.1111/j.1365-2966.2004.08384.x}

\bibitem[{{Finger} {et~al.}(2010){Finger}, {Ikhsanov}, {Wilson-Hodge}, \& {Patel}}]{2010ApJ...709.1249F}
{Finger}, M.~H., {Ikhsanov}, N.~R., {Wilson-Hodge}, C.~A., \& {Patel}, S.~K. 2010, \apj, 709, 1249, \dodoi{10.1088/0004-637X/709/2/1249}

\bibitem[{{Franson} {et~al.}(2022){Franson}, {Bowler}, {Brandt}, {Dupuy}, {Tran}, {Brandt}, {Li}, \& {Kraus}}]{2022AJ....163...50F}
{Franson}, K., {Bowler}, B.~P., {Brandt}, T.~D., {et~al.} 2022, \aj, 163, 50, \dodoi{10.3847/1538-3881/ac35e8}

\bibitem[{{F{\"u}rst} {et~al.}(2014){F{\"u}rst}, {Pottschmidt}, {Wilms}, {Tomsick}, {Bachetti}, {Boggs}, {Christensen}, {Craig}, {Grefenstette}, {Hailey}, {Harrison}, {Madsen}, {Miller}, {Stern}, {Walton}, \& {Zhang}}]{2014ApJ...780..133F}
{F{\"u}rst}, F., {Pottschmidt}, K., {Wilms}, J., {et~al.} 2014, \apj, 780, 133, \dodoi{10.1088/0004-637X/780/2/133}

\bibitem[{{Gandhi} {et~al.}(2022){Gandhi}, {Buckley}, {Charles}, {Hodgkin}, {Scaringi}, {Knigge}, {Rao}, {Paice}, \& {Zhao}}]{2022MNRAS.510.3885G}
{Gandhi}, P., {Buckley}, D.~A.~H., {Charles}, P.~A., {et~al.} 2022, \mnras, 510, 3885, \dodoi{10.1093/mnras/stab3771}

\bibitem[{{Giacconi} {et~al.}(1971){Giacconi}, {Kellogg}, {Gorenstein}, {Gursky}, \& {Tananbaum}}]{1971ApJ...165L..27G}
{Giacconi}, R., {Kellogg}, E., {Gorenstein}, P., {Gursky}, H., \& {Tananbaum}, H. 1971, \apjl, 165, L27, \dodoi{10.1086/180711}

\bibitem[{{Giacconi} {et~al.}(1972){Giacconi}, {Murray}, {Gursky}, {Kellogg}, {Schreier}, \& {Tananbaum}}]{1972ApJ...178..281G}
{Giacconi}, R., {Murray}, S., {Gursky}, H., {et~al.} 1972, \apj, 178, 281, \dodoi{10.1086/151790}

\bibitem[{{Gim{\'e}nez-Garc{\'\i}a} {et~al.}(2015){Gim{\'e}nez-Garc{\'\i}a}, {Torrej{\'o}n}, {Eikmann}, {Mart{\'\i}nez-N{\'u}{\~n}ez}, {Oskinova}, {Rodes-Roca}, \& {Bernab{\'e}u}}]{2015A&A...576A.108G}
{Gim{\'e}nez-Garc{\'\i}a}, A., {Torrej{\'o}n}, J.~M., {Eikmann}, W., {et~al.} 2015, \aap, 576, A108, \dodoi{10.1051/0004-6361/201425004}

\bibitem[{{Gim{\'e}nez-Garc{\'\i}a} {et~al.}(2016){Gim{\'e}nez-Garc{\'\i}a}, {Shenar}, {Torrej{\'o}n}, {Oskinova}, {Mart{\'\i}nez-N{\'u}{\~n}ez}, {Hamann}, {Rodes-Roca}, {Gonz{\'a}lez-Gal{\'a}n}, {Alonso-Santiago}, {Gonz{\'a}lez-Fern{\'a}ndez}, {Bernabeu}, \& {Sander}}]{2016A&A...591A..26G}
{Gim{\'e}nez-Garc{\'\i}a}, A., {Shenar}, T., {Torrej{\'o}n}, J.~M., {et~al.} 2016, \aap, 591, A26, \dodoi{10.1051/0004-6361/201527551}

\bibitem[{{Girardi} {et~al.}(2005){Girardi}, {Groenewegen}, {Hatziminaoglou}, \& {da Costa}}]{2005A&A...436..895G}
{Girardi}, L., {Groenewegen}, M.~A.~T., {Hatziminaoglou}, E., \& {da Costa}, L. 2005, \aap, 436, 895, \dodoi{10.1051/0004-6361:20042352}

\bibitem[{{Gomez Gonzalez} {et~al.}(2017){Gomez Gonzalez}, {Wertz}, {Absil}, {Christiaens}, {Defr{\`e}re}, {Mawet}, {Milli}, {Absil}, {Van Droogenbroeck}, {Cantalloube}, {Hinz}, {Skemer}, {Karlsson}, \& {Surdej}}]{2017AJ....154....7G}
{Gomez Gonzalez}, C.~A., {Wertz}, O., {Absil}, O., {et~al.} 2017, \aj, 154, 7, \dodoi{10.3847/1538-3881/aa73d7}

\bibitem[{{Gonz{\'a}lez-Riestra} {et~al.}(2004){Gonz{\'a}lez-Riestra}, {Oosterbroek}, {Kuulkers}, {Orr}, \& {Parmar}}]{2004A&A...420..589G}
{Gonz{\'a}lez-Riestra}, R., {Oosterbroek}, T., {Kuulkers}, E., {Orr}, A., \& {Parmar}, A.~N. 2004, \aap, 420, 589, \dodoi{10.1051/0004-6361:20035940}

\bibitem[{{Grebenev} {et~al.}(2003){Grebenev}, {Lutovinov}, \& {Sunyaev}}]{2003ATel..192....1G}
{Grebenev}, S.~A., {Lutovinov}, A.~A., \& {Sunyaev}, R.~A. 2003, The Astronomer's Telegram, 192, 1

\bibitem[{{Grebenev} {et~al.}(2004){Grebenev}, {Rodriguez}, {Westergaard}, {Sunyaev}, \& {Oosterbroek}}]{2004ATel..252....1G}
{Grebenev}, S.~A., {Rodriguez}, J., {Westergaard}, N.~J., {Sunyaev}, R.~A., \& {Oosterbroek}, T. 2004, The Astronomer's Telegram, 252, 1

\bibitem[{{Grebenev} \& {Sunyaev}(2005)}]{2005AstL...31..672G}
{Grebenev}, S.~A., \& {Sunyaev}, R.~A. 2005, Astronomy Letters, 31, 672, \dodoi{10.1134/1.2075310}

\bibitem[{{Haberl} {et~al.}(1989){Haberl}, {White}, \& {Kallman}}]{1989ApJ...343..409H}
{Haberl}, F., {White}, N.~E., \& {Kallman}, T.~R. 1989, \apj, 343, 409, \dodoi{10.1086/167714}

\bibitem[{{Hainich} {et~al.}(2020){Hainich}, {Oskinova}, {Torrej{\'o}n}, {Fuerst}, {Bodaghee}, {Shenar}, {Sander}, {Todt}, {Spetzer}, \& {Hamann}}]{2020A&A...634A..49H}
{Hainich}, R., {Oskinova}, L.~M., {Torrej{\'o}n}, J.~M., {et~al.} 2020, \aap, 634, A49, \dodoi{10.1051/0004-6361/201935498}

\bibitem[{{Hanson} {et~al.}(1996){Hanson}, {Conti}, \& {Rieke}}]{1996ApJS..107..281H}
{Hanson}, M.~M., {Conti}, P.~S., \& {Rieke}, M.~J. 1996, \apjs, 107, 281, \dodoi{10.1086/192366}

\bibitem[{{Hawkins} {et~al.}(1975){Hawkins}, {Mason}, \& {Sanford}}]{1975ApL....16...19H}
{Hawkins}, F.~J., {Mason}, K.~O., \& {Sanford}, P.~W. 1975, \aplett, 16, 19

\bibitem[{{Helfand} \& {Moran}(2001)}]{2001ApJ...554...27H}
{Helfand}, D.~J., \& {Moran}, E.~C. 2001, \apj, 554, 27, \dodoi{10.1086/321368}

\bibitem[{{Hiltner} {et~al.}(1972){Hiltner}, {Werner}, \& {Osmer}}]{1972ApJ...175L..19H}
{Hiltner}, W.~A., {Werner}, J., \& {Osmer}, P. 1972, \apjl, 175, L19, \dodoi{10.1086/180976}

\bibitem[{{Huby} {et~al.}(2017){Huby}, {Bottom}, {Femenia}, {Ngo}, {Mawet}, {Serabyn}, \& {Absil}}]{2017A&A...600A..46H}
{Huby}, E., {Bottom}, M., {Femenia}, B., {et~al.} 2017, \aap, 600, A46, \dodoi{10.1051/0004-6361/201630232}

\bibitem[{{Hutter} {et~al.}(2021){Hutter}, {Tycner}, {Zavala}, {Benson}, {Hummel}, \& {Zirm}}]{2021ApJS..257...69H}
{Hutter}, D.~J., {Tycner}, C., {Zavala}, R.~T., {et~al.} 2021, \apjs, 257, 69, \dodoi{10.3847/1538-4365/ac23cb}

\bibitem[{{Ikhsanov} \& {Beskrovnaya}(2010)}]{2010Ap.....53..237I}
{Ikhsanov}, N.~R., \& {Beskrovnaya}, N.~G. 2010, Astrophysics, 53, 237, \dodoi{10.1007/s10511-010-9115-z}

\bibitem[{Imara \& Di~Stefano(2018)}]{imara_searching_2018}
Imara, N., \& Di~Stefano, R. 2018, arXiv:1703.05762 [astro-ph], \dodoi{10.3847/1538-4357/aab903}

\bibitem[{{in 't Zand} {et~al.}(1998){in 't Zand}, {Heise}, {Smith}, {Muller}, {Ubertini}, \& {Bazzano}}]{1998IAUC.6840....2I}
{in 't Zand}, J., {Heise}, J., {Smith}, M., {et~al.} 1998, \iaucirc, 6840, 2

\bibitem[{{in't Zand} {et~al.}(2006){in't Zand}, {Jonker}, {Mendez}, \& {Markwardt}}]{2006ATel..915....1I}
{in't Zand}, J., {Jonker}, P., {Mendez}, M., \& {Markwardt}, C. 2006, The Astronomer's Telegram, 915, 1

\bibitem[{{in't Zand}(2005)}]{2005A&A...441L...1I}
{in't Zand}, J.~J.~M. 2005, \aap, 441, L1, \dodoi{10.1051/0004-6361:200500162}

\bibitem[{{Israel} {et~al.}(1997){Israel}, {Stella}, {Angelini}, {White}, {Kallman}, {Giommi}, \& {Treves}}]{1997ApJ...474L..53I}
{Israel}, G.~L., {Stella}, L., {Angelini}, L., {et~al.} 1997, \apjl, 474, L53, \dodoi{10.1086/310418}

\bibitem[{{Iyer} \& {Paul}(2017)}]{2017MNRAS.471..355I}
{Iyer}, N., \& {Paul}, B. 2017, Monthly Notices of the Royal Astronomical Society, 471, 355, \dodoi{10.1093/mnras/stx1575}

\bibitem[{{Jaschek} \& {Jaschek}(1963)}]{1963PASP...75..365J}
{Jaschek}, M., \& {Jaschek}, C. 1963, \pasp, 75, 365, \dodoi{10.1086/127968}

\bibitem[{{Jones} {et~al.}(1973){Jones}, {Forman}, {Tananbaum}, {Schreier}, {Gursky}, {Kellogg}, \& {Giacconi}}]{1973ApJ...181L..43J}
{Jones}, C., {Forman}, W., {Tananbaum}, H., {et~al.} 1973, \apjl, 181, L43, \dodoi{10.1086/181181}

\bibitem[{{Jones} {et~al.}(2021){Jones}, {Nally}, {Sharp}, {McDonald}, {Boyer}, {Meixner}, {Kemper}, {Ferguson}, {Goldman}, \& {Rich}}]{2021MNRAS.504..565J}
{Jones}, O.~C., {Nally}, C., {Sharp}, M.~J., {et~al.} 2021, \mnras, 504, 565, \dodoi{10.1093/mnras/stab923}

\bibitem[{{Justham} {et~al.}(2006){Justham}, {Rappaport}, \& {Podsiadlowski}}]{2006MNRAS.366.1415J}
{Justham}, S., {Rappaport}, S., \& {Podsiadlowski}, P. 2006, \mnras, 366, 1415, \dodoi{10.1111/j.1365-2966.2005.09907.x}

\bibitem[{Kara {et~al.}(2019)Kara, Steiner, Fabian, Cackett, Uttley, Remillard, Gendreau, Arzoumanian, Altamirano, Eikenberry, Enoto, Homan, Neilsen, \& Stevens}]{kara_corona_2019}
Kara, E., Steiner, J.~F., Fabian, A.~C., {et~al.} 2019, Nature, 565, 198, \dodoi{10.1038/s41586-018-0803-x}

\bibitem[{{Khargharia} {et~al.}(2010){Khargharia}, {Froning}, \& {Robinson}}]{2010ApJ...716.1105K}
{Khargharia}, J., {Froning}, C.~S., \& {Robinson}, E.~L. 2010, \apj, 716, 1105, \dodoi{10.1088/0004-637X/716/2/1105}

\bibitem[{{Kiseleva} {et~al.}(1994){Kiseleva}, {Eggleton}, \& {Anosova}}]{1994MNRAS.267..161K}
{Kiseleva}, G., {Eggleton}, P.~P., \& {Anosova}, J.~P. 1994, \mnras, 267, 161, \dodoi{10.1093/mnras/267.1.161}

\bibitem[{{Klare} \& {Neckel}(1977)}]{1977A&AS...27..215K}
{Klare}, G., \& {Neckel}, T. 1977, \aaps, 27, 215

\bibitem[{{Kratter} {et~al.}(2010){Kratter}, {Matzner}, {Krumholz}, \& {Klein}}]{2010ApJ...708.1585K}
{Kratter}, K.~M., {Matzner}, C.~D., {Krumholz}, M.~R., \& {Klein}, R.~I. 2010, \apj, 708, 1585, \dodoi{10.1088/0004-637X/708/2/1585}

\bibitem[{{Kravtsov} {et~al.}(2020){Kravtsov}, {Berdyugin}, {Piirola}, {Kosenkov}, {Tsygankov}, {Chernyakova}, {Malyshev}, {Sakanoi}, {Kagitani}, {Berdyugina}, \& {Poutanen}}]{2020A&A...643A.170K}
{Kravtsov}, V., {Berdyugin}, A.~V., {Piirola}, V., {et~al.} 2020, \aap, 643, A170, \dodoi{10.1051/0004-6361/202038745}

\bibitem[{{Kretschmar} {et~al.}(2021){Kretschmar}, {El Mellah}, {Mart{\'\i}nez-N{\'u}{\~n}ez}, {F{\"u}rst}, {Grinberg}, {Sander}, {van den Eijnden}, {Degenaar}, {Ma{\'\i}z Apell{\'a}niz}, {Jim{\'e}nez Esteban}, {Ramos-Lerate}, \& {Utrilla}}]{2021A&A...652A..95K}
{Kretschmar}, P., {El Mellah}, I., {Mart{\'\i}nez-N{\'u}{\~n}ez}, S., {et~al.} 2021, \aap, 652, A95, \dodoi{10.1051/0004-6361/202040272}

\bibitem[{{Kreykenbohm} {et~al.}(2002){Kreykenbohm}, {Coburn}, {Wilms}, {Kretschmar}, {Staubert}, {Heindl}, \& {Rothschild}}]{2002A&A...395..129K}
{Kreykenbohm}, I., {Coburn}, W., {Wilms}, J., {et~al.} 2002, \aap, 395, 129, \dodoi{10.1051/0004-6361:20021181}

\bibitem[{{Kreykenbohm} {et~al.}(2008){Kreykenbohm}, {Wilms}, {Kretschmar}, {Torrej{\'o}n}, {Pottschmidt}, {Hanke}, {Santangelo}, {Ferrigno}, \& {Staubert}}]{2008A&A...492..511K}
{Kreykenbohm}, I., {Wilms}, J., {Kretschmar}, P., {et~al.} 2008, \aap, 492, 511, \dodoi{10.1051/0004-6361:200809956}

\bibitem[{{Krimm} {et~al.}(2013){Krimm}, {Holland}, {Corbet}, {Pearlman}, {Romano}, {Kennea}, {Bloom}, {Barthelmy}, {Baumgartner}, {Cummings}, {Gehrels}, {Lien}, {Markwardt}, {Palmer}, {Sakamoto}, {Stamatikos}, \& {Ukwatta}}]{2013ApJS..209...14K}
{Krimm}, H.~A., {Holland}, S.~T., {Corbet}, R.~H.~D., {et~al.} 2013, \apjs, 209, 14, \dodoi{10.1088/0067-0049/209/1/14}

\bibitem[{{Kubo} {et~al.}(1998){Kubo}, {Murakami}, {Ishida}, \& {Corbet}}]{1998PASJ...50..417K}
{Kubo}, S., {Murakami}, T., {Ishida}, M., \& {Corbet}, R. H.~D. 1998, \pasj, 50, 417, \dodoi{10.1093/pasj/50.4.417}

\bibitem[{{Kuulkers} {et~al.}(2007){Kuulkers}, {Oneca}, {Brandt}, {Shaw}, {Beckmann}, {Chenevez}, {Courvoisier}, {Domingo}, {Ebisawa}, {Jonker}, {Kretschmar}, {Markwardt}, {Oosterbroek}, {Paizis}, {Sanchez-Fernandez}, \& {Wijnands}}]{2007ATel.1266....1K}
{Kuulkers}, E., {Oneca}, D.~R., {Brandt}, S., {et~al.} 2007, The Astronomer's Telegram, 1266, 1

\bibitem[{{La Barbera} {et~al.}(2003){La Barbera}, {Santangelo}, {Orlandini}, \& {Segreto}}]{2003A&A...400..993L}
{La Barbera}, A., {Santangelo}, A., {Orlandini}, M., \& {Segreto}, A. 2003, \aap, 400, 993, \dodoi{10.1051/0004-6361:20030010}

\bibitem[{{La Palombara} {et~al.}(2015){La Palombara}, {Esposito}, {Mereghetti}, {Novara}, \& {Tiengo}}]{2015A&A...580A..56L}
{La Palombara}, N., {Esposito}, P., {Mereghetti}, S., {Novara}, G., \& {Tiengo}, A. 2015, \aap, 580, A56, \dodoi{10.1051/0004-6361/201526141}

\bibitem[{{La Palombara} \& {Mereghetti}(2007)}]{2007A&A...474..137L}
{La Palombara}, N., \& {Mereghetti}, S. 2007, \aap, 474, 137, \dodoi{10.1051/0004-6361:20077970}

\bibitem[{{Lafreni{\`e}re} {et~al.}(2008){Lafreni{\`e}re}, {Jayawardhana}, {Brandeker}, {Ahmic}, \& {van Kerkwijk}}]{2008ApJ...683..844L}
{Lafreni{\`e}re}, D., {Jayawardhana}, R., {Brandeker}, A., {Ahmic}, M., \& {van Kerkwijk}, M.~H. 2008, \apj, 683, 844, \dodoi{10.1086/590239}

\bibitem[{{Lafreni{\`e}re} {et~al.}(2014){Lafreni{\`e}re}, {Jayawardhana}, {van Kerkwijk}, {Brandeker}, \& {Janson}}]{2014ApJ...785...47L}
{Lafreni{\`e}re}, D., {Jayawardhana}, R., {van Kerkwijk}, M.~H., {Brandeker}, A., \& {Janson}, M. 2014, \apj, 785, 47, \dodoi{10.1088/0004-637X/785/1/47}

\bibitem[{{Lagrange} {et~al.}(2010){Lagrange}, {Bonnefoy}, {Chauvin}, {Apai}, {Ehrenreich}, {Boccaletti}, {Gratadour}, {Rouan}, {Mouillet}, {Lacour}, \& {Kasper}}]{2010Sci...329...57L}
{Lagrange}, A.~M., {Bonnefoy}, M., {Chauvin}, G., {et~al.} 2010, Science, 329, 57, \dodoi{10.1126/science.1187187}

\bibitem[{{Laurent} {et~al.}(1992){Laurent}, {Goldwurm}, {Lebrun}, {Paul}, {Mereghetti}, {Barret}, {Bouchet}, {Roques}, {Churazov}, {Gilfanov}, {Kuznetzov}, {Sunyaev}, {Chulkov}, {Dyachkov}, {Khavenson}, \& {Novikov}}]{1992A&A...260..237L}
{Laurent}, P., {Goldwurm}, A., {Lebrun}, F., {et~al.} 1992, \aap, 260, 237

\bibitem[{{Liu} {et~al.}(2011){Liu}, {Chaty}, \& {Yan}}]{2011MNRAS.415.3349L}
{Liu}, Q.~Z., {Chaty}, S., \& {Yan}, J.~Z. 2011, \mnras, 415, 3349, \dodoi{10.1111/j.1365-2966.2011.18949.x}

\bibitem[{{Liu} {et~al.}(2000){Liu}, {van Paradijs}, \& {van den Heuvel}}]{2000A&AS..147...25L}
{Liu}, Q.~Z., {van Paradijs}, J., \& {van den Heuvel}, E.~P.~J. 2000, \aaps, 147, 25, \dodoi{10.1051/aas:2000288}

\bibitem[{{Liu} {et~al.}(2006){Liu}, {van Paradijs}, \& {van den Heuvel}}]{2006A&A...455.1165L}
---. 2006, \aap, 455, 1165, \dodoi{10.1051/0004-6361:20064987}

\bibitem[{{Liu} {et~al.}(2007){Liu}, {van Paradijs}, \& {van den Heuvel}}]{2007A&A...469..807L}
---. 2007, \aap, 469, 807, \dodoi{10.1051/0004-6361:20077303}

\bibitem[{{Lopes de Oliveira} {et~al.}(2006){Lopes de Oliveira}, {Motch}, {Haberl}, {Negueruela}, \& {Janot-Pacheco}}]{2006A&A...454..265L}
{Lopes de Oliveira}, R., {Motch}, C., {Haberl}, F., {Negueruela}, I., \& {Janot-Pacheco}, E. 2006, \aap, 454, 265, \dodoi{10.1051/0004-6361:20054589}

\bibitem[{{Lutovinov} {et~al.}(2005){Lutovinov}, {Revnivtsev}, {Gilfanov}, {Shtykovskiy}, {Molkov}, \& {Sunyaev}}]{2005A&A...444..821L}
{Lutovinov}, A., {Revnivtsev}, M., {Gilfanov}, M., {et~al.} 2005, \aap, 444, 821, \dodoi{10.1051/0004-6361:20042392}

\bibitem[{{Lutovinov} {et~al.}(2012){Lutovinov}, {Tsygankov}, \& {Chernyakova}}]{2012MNRAS.423.1978L}
{Lutovinov}, A., {Tsygankov}, S., \& {Chernyakova}, M. 2012, \mnras, 423, 1978, \dodoi{10.1111/j.1365-2966.2012.21036.x}

\bibitem[{{Lyubimkov} {et~al.}(1997){Lyubimkov}, {Rostopchin}, {Roche}, \& {Tarasov}}]{1997MNRAS.286..549L}
{Lyubimkov}, L.~S., {Rostopchin}, S.~I., {Roche}, P., \& {Tarasov}, A.~E. 1997, \mnras, 286, 549, \dodoi{10.1093/mnras/286.3.549}

\bibitem[{{Maitra} {et~al.}(2017){Maitra}, {Raichur}, {Pradhan}, \& {Paul}}]{2017MNRAS.470..713M}
{Maitra}, C., {Raichur}, H., {Pradhan}, P., \& {Paul}, B. 2017, \mnras, 470, 713, \dodoi{10.1093/mnras/stx1281}

\bibitem[{{Markoff} {et~al.}(2001){Markoff}, {Falcke}, \& {Fender}}]{2001A&A...372L..25M}
{Markoff}, S., {Falcke}, H., \& {Fender}, R. 2001, \aap, 372, L25, \dodoi{10.1051/0004-6361:20010420}

\bibitem[{{Marois} {et~al.}(2006){Marois}, {Lafreni{\`e}re}, {Doyon}, {Macintosh}, \& {Nadeau}}]{2006ApJ...641..556M}
{Marois}, C., {Lafreni{\`e}re}, D., {Doyon}, R., {Macintosh}, B., \& {Nadeau}, D. 2006, \apj, 641, 556, \dodoi{10.1086/500401}

\bibitem[{{Martin} {et~al.}(2011){Martin}, {Pringle}, {Tout}, \& {Lubow}}]{2011MNRAS.416.2827M}
{Martin}, R.~G., {Pringle}, J.~E., {Tout}, C.~A., \& {Lubow}, S.~H. 2011, \mnras, 416, 2827, \dodoi{10.1111/j.1365-2966.2011.19231.x}

\bibitem[{{Mart{\'\i}nez-Chicharro} {et~al.}(2021){Mart{\'\i}nez-Chicharro}, {Grinberg}, {Torrej{\'o}n}, {Schulz}, {Oskinova}, {Nowak}, {F{\"u}rst}, {Hell}, \& {Hainich}}]{2021MNRAS.501.5646M}
{Mart{\'\i}nez-Chicharro}, M., {Grinberg}, V., {Torrej{\'o}n}, J.~M., {et~al.} 2021, \mnras, 501, 5646, \dodoi{10.1093/mnras/staa3956}

\bibitem[{{Mart{\'\i}nez-N{\'u}{\~n}ez} {et~al.}(2014){Mart{\'\i}nez-N{\'u}{\~n}ez}, {Torrej{\'o}n}, {K{\"u}hnel}, {Kretschmar}, {Stuhlinger}, {Rodes-Roca}, {F{\"u}rst}, {Kreykenbohm}, {Martin-Carrillo}, {Pollock}, \& {Wilms}}]{2014A&A...563A..70M}
{Mart{\'\i}nez-N{\'u}{\~n}ez}, S., {Torrej{\'o}n}, J.~M., {K{\"u}hnel}, M., {et~al.} 2014, \aap, 563, A70, \dodoi{10.1051/0004-6361/201322404}

\bibitem[{{Masetti} {et~al.}(2004){Masetti}, {Dal Fiume}, {Amati}, {Del Sordo}, {Frontera}, {Orlandini}, \& {Palazzi}}]{2004A&A...423..311M}
{Masetti}, N., {Dal Fiume}, D., {Amati}, L., {et~al.} 2004, \aap, 423, 311, \dodoi{10.1051/0004-6361:20040273}

\bibitem[{{Mason} {et~al.}(1976){Mason}, {White}, {Sanford}, {Hawkins}, {Drake}, \& {York}}]{1976MNRAS.176..193M}
{Mason}, K.~O., {White}, N.~E., {Sanford}, P.~W., {et~al.} 1976, \mnras, 176, 193, \dodoi{10.1093/mnras/176.1.193}

\bibitem[{{Mastroserio} {et~al.}(2019){Mastroserio}, {Ingram}, \& {van der Klis}}]{2019MNRAS.488..348M}
{Mastroserio}, G., {Ingram}, A., \& {van der Klis}, M. 2019, \mnras, 488, 348, \dodoi{10.1093/mnras/stz1727}

\bibitem[{{Mawet} {et~al.}(2005){Mawet}, {Riaud}, {Absil}, \& {Surdej}}]{2005ApJ...633.1191M}
{Mawet}, D., {Riaud}, P., {Absil}, O., \& {Surdej}, J. 2005, \apj, 633, 1191, \dodoi{10.1086/462409}

\bibitem[{{Mereghetti} {et~al.}(2011){Mereghetti}, {La Palombara}, {Tiengo}, {Pizzolato}, {Esposito}, {Woudt}, {Israel}, \& {Stella}}]{2011ApJ...737...51M}
{Mereghetti}, S., {La Palombara}, N., {Tiengo}, A., {et~al.} 2011, \apj, 737, 51, \dodoi{10.1088/0004-637X/737/2/51}

\bibitem[{{Mereghetti} {et~al.}(2016){Mereghetti}, {Pintore}, {Esposito}, {La Palombara}, {Tiengo}, {Israel}, \& {Stella}}]{2016MNRAS.458.3523M}
{Mereghetti}, S., {Pintore}, F., {Esposito}, P., {et~al.} 2016, \mnras, 458, 3523, \dodoi{10.1093/mnras/stw536}

\bibitem[{{Mereghetti} {et~al.}(2009){Mereghetti}, {Tiengo}, {Esposito}, {La Palombara}, {Israel}, \& {Stella}}]{2009Sci...325.1222M}
{Mereghetti}, S., {Tiengo}, A., {Esposito}, P., {et~al.} 2009, Science, 325, 1222, \dodoi{10.1126/science.1176252}

\bibitem[{{Mereghetti} {et~al.}(2021){Mereghetti}, {Pintore}, {Rauch}, {La Palombara}, {Esposito}, {Geier}, {Pelisoli}, {Rigoselli}, {Schaffenroth}, \& {Tiengo}}]{2021MNRAS.504..920M}
{Mereghetti}, S., {Pintore}, F., {Rauch}, T., {et~al.} 2021, \mnras, 504, 920, \dodoi{10.1093/mnras/stab1004}

\bibitem[{{Miller-Jones} {et~al.}(2021){Miller-Jones}, {Bahramian}, {Orosz}, {Mandel}, {Gou}, {Maccarone}, {Neijssel}, {Zhao}, {Zi{\'o}{\l}kowski}, {Reid}, {Uttley}, {Zheng}, {Byun}, {Dodson}, {Grinberg}, {Jung}, {Kim}, {Marcote}, {Markoff}, {Rioja}, {Rushton}, {Russell}, {Sivakoff}, {Tetarenko}, {Tudose}, \& {Wilms}}]{2021Sci...371.1046M}
{Miller-Jones}, J. C.~A., {Bahramian}, A., {Orosz}, J.~A., {et~al.} 2021, Science, 371, 1046, \dodoi{10.1126/science.abb3363}

\bibitem[{{Mirabel} \& {Rodrigues}(2003)}]{2003Sci...300.1119M}
{Mirabel}, I.~F., \& {Rodrigues}, I. 2003, Science, 300, 1119, \dodoi{10.1126/science.1083451}

\bibitem[{{Moffat} {et~al.}(1973){Moffat}, {Haupt}, \& {Schmidt-Kaler}}]{1973A&A....23..433M}
{Moffat}, A.~F.~J., {Haupt}, W., \& {Schmidt-Kaler}, T. 1973, \aap, 23, 433

\bibitem[{{Mohamed} \& {Podsiadlowski}(2007)}]{2007ASPC..372..397M}
{Mohamed}, S., \& {Podsiadlowski}, P. 2007, in Astronomical Society of the Pacific Conference Series, Vol. 372, 15th European Workshop on White Dwarfs, ed. R.~{Napiwotzki} \& M.~R. {Burleigh}, 397

\bibitem[{{Montegriffo} {et~al.}(2023){Montegriffo}, {De Angeli}, {Andrae}, {Riello}, {Pancino}, {Sanna}, {Bellazzini}, {Evans}, {Carrasco}, {Sordo}, {Busso}, {Cacciari}, {Jordi}, {van Leeuwen}, {Vallenari}, {Altavilla}, {Barstow}, {Brown}, {Burgess}, {Castellani}, {Cowell}, {Davidson}, {De Luise}, {Delchambre}, {Diener}, {Fabricius}, {Fr{\'e}mat}, {Fouesneau}, {Gilmore}, {Giuffrida}, {Hambly}, {Harrison}, {Hidalgo}, {Hodgkin}, {Holland}, {Marinoni}, {Osborne}, {Pagani}, {Palaversa}, {Piersimoni}, {Pulone}, {Ragaini}, {Rainer}, {Richards}, {Rowell}, {Ruz-Mieres}, {Sarro}, {Walton}, \& {Yoldas}}]{2023A&A...674A...3M}
{Montegriffo}, P., {De Angeli}, F., {Andrae}, R., {et~al.} 2023, \aap, 674, A3, \dodoi{10.1051/0004-6361/202243880}

\bibitem[{{Mooley} {et~al.}(2018){Mooley}, {Nakar}, {Hotokezaka}, {Hallinan}, {Corsi}, {Frail}, {Horesh}, {Murphy}, {Lenc}, {Kaplan}, {de}, {Dobie}, {Chandra}, {Deller}, {Gottlieb}, {Kasliwal}, {Kulkarni}, {Myers}, {Nissanke}, {Piran}, {Lynch}, {Bhalerao}, {Bourke}, {Bannister}, \& {Singer}}]{2018Natur.554..207M}
{Mooley}, K.~P., {Nakar}, E., {Hotokezaka}, K., {et~al.} 2018, \nat, 554, 207, \dodoi{10.1038/nature25452}

\bibitem[{{Motch} {et~al.}(1997){Motch}, {Haberl}, {Dennerl}, {Pakull}, \& {Janot-Pacheco}}]{1997A&A...323..853M}
{Motch}, C., {Haberl}, F., {Dennerl}, K., {Pakull}, M., \& {Janot-Pacheco}, E. 1997, \aap, 323, 853.
\newblock \doarXiv{astro-ph/9611122}

\bibitem[{{Murdin} \& {Webster}(1971)}]{1971Natur.233..110M}
{Murdin}, P., \& {Webster}, B.~L. 1971, \nat, 233, 110, \dodoi{10.1038/233110a0}

\bibitem[{Naud {et~al.}(2014)Naud, Artigau, Malo, Albert, Doyon, Lafrenière, Gagné, Saumon, Morley, Allard, Homeier, Beichman, Gelino, \& Boucher}]{naud_discovery_2014}
Naud, M.-E., Artigau, {\'E}., Malo, L., {et~al.} 2014, ApJ, 787, 5, \dodoi{10.1088/0004-637X/787/1/5}

\bibitem[{Naze \& Motch(2018)}]{naze_hot_2018}
Naze, Y., \& Motch, C. 2018, A\&A, 619, A148, \dodoi{10.1051/0004-6361/201833842}

\bibitem[{{Negueruela} \& {Reig}(2001)}]{2001A&A...371.1056N}
{Negueruela}, I., \& {Reig}, P. 2001, \aap, 371, 1056, \dodoi{10.1051/0004-6361:20010476}

\bibitem[{{Negueruela} \& {Smith}(2006)}]{2006ATel..831....1N}
{Negueruela}, I., \& {Smith}, D.~M. 2006, The Astronomer's Telegram, 831, 1

\bibitem[{{Negueruela} {et~al.}(2006){Negueruela}, {Smith}, {Reig}, {Chaty}, \& {Torrej{\'o}n}}]{2006ESASP.604..165N}
{Negueruela}, I., {Smith}, D.~M., {Reig}, P., {Chaty}, S., \& {Torrej{\'o}n}, J.~M. 2006, in ESA Special Publication, Vol. 604, The X-ray Universe 2005, ed. A.~{Wilson}, 165.
\newblock \doarXiv{astro-ph/0511088}

\bibitem[{{Nemravov{\'a}} {et~al.}(2012){Nemravov{\'a}}, {Harmanec}, {Koubsk{\'y}}, {Miroshnichenko}, {Yang}, {{\v{S}}lechta}, {Buil}, {Kor{\v{c}}{\'a}kov{\'a}}, \& {Votruba}}]{2012A&A...537A..59N}
{Nemravov{\'a}}, J., {Harmanec}, P., {Koubsk{\'y}}, P., {et~al.} 2012, \aap, 537, A59, \dodoi{10.1051/0004-6361/201117922}

\bibitem[{{Neumann} {et~al.}(2023){Neumann}, {Avakyan}, {Doroshenko}, \& {Santangelo}}]{2023A&A...677A.134N}
{Neumann}, M., {Avakyan}, A., {Doroshenko}, V., \& {Santangelo}, A. 2023, \aap, 677, A134, \dodoi{10.1051/0004-6361/202245728}

\bibitem[{{Nikolov} {et~al.}(2017){Nikolov}, {Zamanov}, {Stoyanov}, \& {Mart{\'\i}}}]{2017BlgAJ..27...10N}
{Nikolov}, Y.~M., {Zamanov}, R.~K., {Stoyanov}, K.~A., \& {Mart{\'\i}}, J. 2017, Bulgarian Astronomical Journal, 27, 10

\bibitem[{{Okazaki} {et~al.}(2002){Okazaki}, {Bate}, {Ogilvie}, \& {Pringle}}]{2002MNRAS.337..967O}
{Okazaki}, A.~T., {Bate}, M.~R., {Ogilvie}, G.~I., \& {Pringle}, J.~E. 2002, \mnras, 337, 967, \dodoi{10.1046/j.1365-8711.2002.05960.x}

\bibitem[{{Orosz} {et~al.}(2011){Orosz}, {McClintock}, {Aufdenberg}, {Remillard}, {Reid}, {Narayan}, \& {Gou}}]{2011ApJ...742...84O}
{Orosz}, J.~A., {McClintock}, J.~E., {Aufdenberg}, J.~P., {et~al.} 2011, \apj, 742, 84, \dodoi{10.1088/0004-637X/742/2/84}

\bibitem[{{Paczy{\'n}ski}(1967)}]{1967AcA....17..287P}
{Paczy{\'n}ski}, B. 1967, \actaa, 17, 287

\bibitem[{{Parker} {et~al.}(2015){Parker}, {Tomsick}, {Miller}, {Yamaoka}, {Lohfink}, {Nowak}, {Fabian}, {Alston}, {Boggs}, {Christensen}, {Craig}, {F{\"u}rst}, {Gandhi}, {Grefenstette}, {Grinberg}, {Hailey}, {Harrison}, {Kara}, {King}, {Stern}, {Walton}, {Wilms}, \& {Zhang}}]{2015ApJ...808....9P}
{Parker}, M.~L., {Tomsick}, J.~A., {Miller}, J.~M., {et~al.} 2015, \apj, 808, 9, \dodoi{10.1088/0004-637X/808/1/9}

\bibitem[{{Paxton} {et~al.}(2011){Paxton}, {Bildsten}, {Dotter}, {Herwig}, {Lesaffre}, \& {Timmes}}]{2011ApJS..192....3P}
{Paxton}, B., {Bildsten}, L., {Dotter}, A., {et~al.} 2011, \apjs, 192, 3, \dodoi{10.1088/0067-0049/192/1/3}

\bibitem[{{Paxton} {et~al.}(2013){Paxton}, {Cantiello}, {Arras}, {Bildsten}, {Brown}, {Dotter}, {Mankovich}, {Montgomery}, {Stello}, {Timmes}, \& {Townsend}}]{2013ApJS..208....4P}
{Paxton}, B., {Cantiello}, M., {Arras}, P., {et~al.} 2013, \apjs, 208, 4, \dodoi{10.1088/0067-0049/208/1/4}

\bibitem[{{Paxton} {et~al.}(2015){Paxton}, {Marchant}, {Schwab}, {Bauer}, {Bildsten}, {Cantiello}, {Dessart}, {Farmer}, {Hu}, {Langer}, {Townsend}, {Townsley}, \& {Timmes}}]{2015ApJS..220...15P}
{Paxton}, B., {Marchant}, P., {Schwab}, J., {et~al.} 2015, \apjs, 220, 15, \dodoi{10.1088/0067-0049/220/1/15}

\bibitem[{{Paxton} {et~al.}(2018){Paxton}, {Schwab}, {Bauer}, {Bildsten}, {Blinnikov}, {Duffell}, {Farmer}, {Goldberg}, {Marchant}, {Sorokina}, {Thoul}, {Townsend}, \& {Timmes}}]{2018ApJS..234...34P}
{Paxton}, B., {Schwab}, J., {Bauer}, E.~B., {et~al.} 2018, \apjs, 234, 34, \dodoi{10.3847/1538-4365/aaa5a8}

\bibitem[{{Pellizza} {et~al.}(2006){Pellizza}, {Chaty}, \& {Negueruela}}]{2006A&A...455..653P}
{Pellizza}, L.~J., {Chaty}, S., \& {Negueruela}, I. 2006, \aap, 455, 653, \dodoi{10.1051/0004-6361:20054436}

\bibitem[{{Perryman} {et~al.}(1997){Perryman}, {Lindegren}, {Kovalevsky}, {Hog}, {Bastian}, {Bernacca}, {Creze}, {Donati}, {Grenon}, {Grewing}, {van Leeuwen}, {van der Marel}, {Mignard}, {Murray}, {Le Poole}, {Schrijver}, {Turon}, {Arenou}, {Froeschle}, \& {Petersen}}]{1997A&A...323L..49P}
{Perryman}, M.~A.~C., {Lindegren}, L., {Kovalevsky}, J., {et~al.} 1997, \aap, 500, 501

\bibitem[{{Peter} {et~al.}(2012){Peter}, {Feldt}, {Henning}, \& {Hormuth}}]{2012A&A...538A..74P}
{Peter}, D., {Feldt}, M., {Henning}, T., \& {Hormuth}, F. 2012, \aap, 538, A74, \dodoi{10.1051/0004-6361/201015027}

\bibitem[{{Popov} {et~al.}(2018){Popov}, {Mereghetti}, {Blinnikov}, {Kuranov}, \& {Yungelson}}]{2018MNRAS.474.2750P}
{Popov}, S.~B., {Mereghetti}, S., {Blinnikov}, S.~I., {Kuranov}, A.~G., \& {Yungelson}, L.~R. 2018, \mnras, 474, 2750, \dodoi{10.1093/mnras/stx2910}

\bibitem[{{Popper}(1950)}]{1950ApJ...111..495P}
{Popper}, D.~M. 1950, \apj, 111, 495, \dodoi{10.1086/145292}

\bibitem[{{Postnov} {et~al.}(2017){Postnov}, {Oskinova}, \& {Torrej{\'o}n}}]{2017MNRAS.465L.119P}
{Postnov}, K., {Oskinova}, L., \& {Torrej{\'o}n}, J.~M. 2017, \mnras, 465, L119, \dodoi{10.1093/mnrasl/slw223}

\bibitem[{{Prasow-{\'E}mond} {et~al.}(2022){Prasow-{\'E}mond}, {Hlavacek-Larrondo}, {Fogarty}, {Rameau}, {Guit{\'e}}, {Mawet}, {Gandhi}, {Rao}, {Steiner}, {Artigau}, {Lafreni{\`e}re}, {Fabian}, {Walton}, {Weiss}, {Doyon}, {Ren}, {Rhea}, {B{\'e}gin}, {Vigneron}, \& {Naud}}]{2022AJ....164....7P}
{Prasow-{\'E}mond}, M., {Hlavacek-Larrondo}, J., {Fogarty}, K., {et~al.} 2022, \aj, 164, 7, \dodoi{10.3847/1538-3881/ac6d57}

\bibitem[{{Raguzova}(2007)}]{2007BeSN...38...24R}
{Raguzova}, N.~V. 2007, Be Star Newsletter, 38, 24

\bibitem[{{Raguzova} \& {Popov}(2005)}]{2005A&AT...24..151R}
{Raguzova}, N.~V., \& {Popov}, S.~B. 2005, Astronomical and Astrophysical Transactions, 24, 151, \dodoi{10.1080/10556790500497311}

\bibitem[{{Rahoui} \& {Chaty}(2008)}]{2008A&A...492..163R}
{Rahoui}, F., \& {Chaty}, S. 2008, \aap, 492, 163, \dodoi{10.1051/0004-6361:200810695}

\bibitem[{{Rampy} {et~al.}(2009){Rampy}, {Smith}, \& {Negueruela}}]{2009ApJ...707..243R}
{Rampy}, R.~A., {Smith}, D.~M., \& {Negueruela}, I. 2009, \apj, 707, 243, \dodoi{10.1088/0004-637X/707/1/243}

\bibitem[{{Reig} {et~al.}(2009){Reig}, {Torrej{\'o}n}, {Negueruela}, {Blay}, {Rib{\'o}}, \& {Wilms}}]{2009A&A...494.1073R}
{Reig}, P., {Torrej{\'o}n}, J.~M., {Negueruela}, I., {et~al.} 2009, \aap, 494, 1073, \dodoi{10.1051/0004-6361:200810950}

\bibitem[{{Reynolds} {et~al.}(1999){Reynolds}, {Owens}, {Kaper}, {Parmar}, \& {Segreto}}]{1999A&A...349..873R}
{Reynolds}, A.~P., {Owens}, A., {Kaper}, L., {Parmar}, A.~N., \& {Segreto}, A. 1999, \aap, 349, 873.
\newblock \doarXiv{astro-ph/9904349}

\bibitem[{{Rib{\'o}} {et~al.}(2006){Rib{\'o}}, {Negueruela}, {Blay}, {Torrej{\'o}n}, \& {Reig}}]{2006A&A...449..687R}
{Rib{\'o}}, M., {Negueruela}, I., {Blay}, P., {Torrej{\'o}n}, J.~M., \& {Reig}, P. 2006, \aap, 449, 687, \dodoi{10.1051/0004-6361:20054206}

\bibitem[{{R{\'\i}mulo} {et~al.}(2018){R{\'\i}mulo}, {Carciofi}, {Vieira}, {Rivinius}, {Faes}, {Figueiredo}, {Bjorkman}, {Georgy}, {Ghoreyshi}, \& {Soszy{\'n}ski}}]{2018MNRAS.476.3555R}
{R{\'\i}mulo}, L.~R., {Carciofi}, A.~C., {Vieira}, R.~G., {et~al.} 2018, \mnras, 476, 3555, \dodoi{10.1093/mnras/sty431}

\bibitem[{{Robinson} \& {Smith}(2000)}]{2000ApJ...540..474R}
{Robinson}, R.~D., \& {Smith}, M.~A. 2000, \apj, 540, 474, \dodoi{10.1086/309310}

\bibitem[{{Romano} {et~al.}(2010){Romano}, {Sidoli}, {Ducci}, {Cusumano}, {La Parola}, {Pagani}, {Page}, {Kennea}, {Burrows}, {Gehrels}, {Sguera}, \& {Bazzano}}]{2010MNRAS.401.1564R}
{Romano}, P., {Sidoli}, L., {Ducci}, L., {et~al.} 2010, \mnras, 401, 1564, \dodoi{10.1111/j.1365-2966.2009.15789.x}

\bibitem[{{Romano} {et~al.}(2015){Romano}, {Bozzo}, {Mangano}, {Esposito}, {Israel}, {Tiengo}, {Campana}, {Ducci}, {Ferrigno}, \& {Kennea}}]{2015A&A...576L...4R}
{Romano}, P., {Bozzo}, E., {Mangano}, V., {et~al.} 2015, \aap, 576, L4, \dodoi{10.1051/0004-6361/201525749}

\bibitem[{{Rubin} {et~al.}(1996){Rubin}, {Finger}, {Harmon}, {Paciesas}, {Fishman}, {Wilson}, {Wilson}, {Brock}, {Briggs}, {Pendleton}, {Cominsky}, \& {Roberts}}]{1996ApJ...459..259R}
{Rubin}, B.~C., {Finger}, M.~H., {Harmon}, B.~A., {et~al.} 1996, \apj, 459, 259, \dodoi{10.1086/176889}

\bibitem[{{Saraswat} \& {Apparao}(1992)}]{1992ApJ...401..678S}
{Saraswat}, P., \& {Apparao}, K. M.~V. 1992, \apj, 401, 678, \dodoi{10.1086/172095}

\bibitem[{Sarty {et~al.}(2011)Sarty, Pilecki, Reichart, Ivarsen, Haislip, Nysewander, LaCluyze, Johnston, Shobbrook, Kiss, \& Wu}]{sarty_photometric_2011}
Sarty, G.~E., Pilecki, B., Reichart, D.~E., {et~al.} 2011, Res. Astron. Astrophys., 11, 947, \dodoi{10.1088/1674-4527/11/8/007}

\bibitem[{{Savonije}(1978)}]{1978A&A....62..317S}
{Savonije}, G.~J. 1978, \aap, 62, 317

\bibitem[{{Schlafly} \& {Finkbeiner}(2011)}]{2011ApJ...737..103S}
{Schlafly}, E.~F., \& {Finkbeiner}, D.~P. 2011, \apj, 737, 103, \dodoi{10.1088/0004-637X/737/2/103}

\bibitem[{{Sell} {et~al.}(2015){Sell}, {Heinz}, {Richards}, {Maccarone}, {Russell}, {Gallo}, {Fender}, {Markoff}, \& {Nowak}}]{2015MNRAS.446.3579S}
{Sell}, P.~H., {Heinz}, S., {Richards}, E., {et~al.} 2015, \mnras, 446, 3579, \dodoi{10.1093/mnras/stu2320}

\bibitem[{{Serabyn} {et~al.}(2016){Serabyn}, {Liewer}, \& {Mawet}}]{2016OptCo.379...64S}
{Serabyn}, E., {Liewer}, K., \& {Mawet}, D. 2016, Optics Communications, 379, 64, \dodoi{10.1016/j.optcom.2016.05.042}

\bibitem[{{Service} {et~al.}(2016){Service}, {Lu}, {Campbell}, {Sitarski}, {Ghez}, \& {Anderson}}]{2016PASP..128i5004S}
{Service}, M., {Lu}, J.~R., {Campbell}, R., {et~al.} 2016, \pasp, 128, 095004, \dodoi{10.1088/1538-3873/128/967/095004}

\bibitem[{{Servillat} {et~al.}(2012){Servillat}, {Tang}, {Grindlay}, \& {Los}}]{2012int..workE..23S}
{Servillat}, M., {Tang}, S., {Grindlay}, J.~E., \& {Los}, E. 2012, in Proceedings of ``An INTEGRAL view of the high-energy sky (the first 10 years)'' - 9th INTEGRAL Workshop and celebration of the 10th anniversary of the launch (INTEGRAL 2012). 15-19 October 2012. Bibliotheque Nationale de France, 23.
\newblock \doarXiv{1303.1179}

\bibitem[{{Seto}(2018)}]{2018MNRAS.475.1392S}
{Seto}, N. 2018, \mnras, 475, 1392, \dodoi{10.1093/mnras/stx3301}

\bibitem[{{Sguera} {et~al.}(2010){Sguera}, {Ducci}, {Sidoli}, {Bazzano}, \& {Bassani}}]{2010MNRAS.402L..49S}
{Sguera}, V., {Ducci}, L., {Sidoli}, L., {Bazzano}, A., \& {Bassani}, L. 2010, \mnras, 402, L49, \dodoi{10.1111/j.1745-3933.2009.00798.x}

\bibitem[{{Sguera} {et~al.}(2015){Sguera}, {Sidoli}, {Bird}, \& {Bazzano}}]{2015MNRAS.449.1228S}
{Sguera}, V., {Sidoli}, L., {Bird}, A.~J., \& {Bazzano}, A. 2015, \mnras, 449, 1228, \dodoi{10.1093/mnras/stv341}

\bibitem[{{Sguera} {et~al.}(2005){Sguera}, {Barlow}, {Bird}, {Clark}, {Dean}, {Hill}, {Moran}, {Shaw}, {Willis}, {Bazzano}, {Ubertini}, \& {Malizia}}]{2005A&A...444..221S}
{Sguera}, V., {Barlow}, E.~J., {Bird}, A.~J., {et~al.} 2005, \aap, 444, 221, \dodoi{10.1051/0004-6361:20053103}

\bibitem[{{Sguera} {et~al.}(2007){Sguera}, {Hill}, {Bird}, {Dean}, {Bazzano}, {Ubertini}, {Masetti}, {Landi}, {Malizia}, {Clark}, \& {Molina}}]{2007A&A...467..249S}
{Sguera}, V., {Hill}, A.~B., {Bird}, A.~J., {et~al.} 2007, \aap, 467, 249, \dodoi{10.1051/0004-6361:20066762}

\bibitem[{{Shrader} {et~al.}(2015){Shrader}, {Hamaguchi}, {Sturner}, {Oskinova}, {Almeyda}, \& {Petre}}]{2015ApJ...799...84S}
{Shrader}, C.~R., {Hamaguchi}, K., {Sturner}, S.~J., {et~al.} 2015, \apj, 799, 84, \dodoi{10.1088/0004-637X/799/1/84}

\bibitem[{{Sidoli}(2013)}]{2013arXiv1301.7574S}
{Sidoli}, L. 2013, arXiv e-prints, arXiv:1301.7574.
\newblock \doarXiv{1301.7574}

\bibitem[{{Sidoli} {et~al.}(2016){Sidoli}, {Paizis}, \& {Postnov}}]{2016MNRAS.457.3693S}
{Sidoli}, L., {Paizis}, A., \& {Postnov}, K. 2016, \mnras, 457, 3693, \dodoi{10.1093/mnras/stw237}

\bibitem[{{Sidoli} {et~al.}(2009){Sidoli}, {Romano}, {Esposito}, {Parola}, {Kennea}, {Krimm}, {Chester}, {Bazzano}, {Burrows}, \& {Gehrels}}]{2009MNRAS.400..258S}
{Sidoli}, L., {Romano}, P., {Esposito}, P., {et~al.} 2009, \mnras, 400, 258, \dodoi{10.1111/j.1365-2966.2009.15445.x}

\bibitem[{{Sigurdsson} {et~al.}(2003){Sigurdsson}, {Richer}, {Hansen}, {Stairs}, \& {Thorsett}}]{2003Sci...301..193S}
{Sigurdsson}, S., {Richer}, H.~B., {Hansen}, B.~M., {Stairs}, I.~H., \& {Thorsett}, S.~E. 2003, Science, 301, 193, \dodoi{10.1126/science.1086326}

\bibitem[{{Simon} {et~al.}(2019){Simon}, {Metlova}, {Godunova}, \& {Vasylenko}}]{2019KPCB...35...38S}
{Simon}, A.~O., {Metlova}, N.~V., {Godunova}, V.~G., \& {Vasylenko}, V.~V. 2019, Kinematics and Physics of Celestial Bodies, 35, 38, \dodoi{10.3103/S0884591319010069}

\bibitem[{{Smith} {et~al.}(2012){Smith}, {Markwardt}, {Swank}, \& {Negueruela}}]{2012MNRAS.422.2661S}
{Smith}, D.~M., {Markwardt}, C.~B., {Swank}, J.~H., \& {Negueruela}, I. 2012, \mnras, 422, 2661, \dodoi{10.1111/j.1365-2966.2012.20836.x}

\bibitem[{{Sota} {et~al.}(2014){Sota}, {Ma{\'\i}z Apell{\'a}niz}, {Morrell}, {Barb{\'a}}, {Walborn}, {Gamen}, {Arias}, \& {Alfaro}}]{2014ApJS..211...10S}
{Sota}, A., {Ma{\'\i}z Apell{\'a}niz}, J., {Morrell}, N.~I., {et~al.} 2014, \apjs, 211, 10, \dodoi{10.1088/0067-0049/211/1/10}

\bibitem[{{Sota} {et~al.}(2011){Sota}, {Ma{\'\i}z Apell{\'a}niz}, {Walborn}, {Alfaro}, {Barb{\'a}}, {Morrell}, {Gamen}, \& {Arias}}]{2011ApJS..193...24S}
{Sota}, A., {Ma{\'\i}z Apell{\'a}niz}, J., {Walborn}, N.~R., {et~al.} 2011, \apjs, 193, 24, \dodoi{10.1088/0067-0049/193/2/24}

\bibitem[{{Spiewak} {et~al.}(2018){Spiewak}, {Bailes}, {Barr}, {Bhat}, {Burgay}, {Cameron}, {Champion}, {Flynn}, {Jameson}, {Johnston}, {Keith}, {Kramer}, {Kulkarni}, {Levin}, {Lyne}, {Morello}, {Ng}, {Possenti}, {Ravi}, {Stappers}, {van Straten}, \& {Tiburzi}}]{2018MNRAS.475..469S}
{Spiewak}, R., {Bailes}, M., {Barr}, E.~D., {et~al.} 2018, \mnras, 475, 469, \dodoi{10.1093/mnras/stx3157}

\bibitem[{{Steele}(2016{\natexlab{a}})}]{2016ATel.8927....1S}
{Steele}, I.~A. 2016{\natexlab{a}}, The Astronomer's Telegram, 8927, 1

\bibitem[{{Steele}(2016{\natexlab{b}})}]{2016ATel.9265....1S}
---. 2016{\natexlab{b}}, The Astronomer's Telegram, 9265, 1

\bibitem[{{Steele}(2016{\natexlab{c}})}]{2016ATel.9487....1S}
---. 2016{\natexlab{c}}, The Astronomer's Telegram, 9487, 1

\bibitem[{{Steele} {et~al.}(1999){Steele}, {Negueruela}, \& {Clark}}]{1999A&AS..137..147S}
{Steele}, I.~A., {Negueruela}, I., \& {Clark}, J.~S. 1999, \aaps, 137, 147, \dodoi{10.1051/aas:1999478}

\bibitem[{{Steiner} {et~al.}(1984){Steiner}, {Ferrara}, {Garcia}, {Patterson}, {Schwartz}, {Warwick}, {Watson}, \& {McClintock}}]{1984ApJ...280..688S}
{Steiner}, J.~E., {Ferrara}, A., {Garcia}, M., {et~al.} 1984, \apj, 280, 688, \dodoi{10.1086/162042}

\bibitem[{{Strohmayer}(2002)}]{2002ApJ...581..577S}
{Strohmayer}, T.~E. 2002, \apj, 581, 577, \dodoi{10.1086/344101}

\bibitem[{{Sunyaev} {et~al.}(2003){Sunyaev}, {Grebenev}, {Lutovinov}, {Rodriguez}, {Mereghetti}, {Gotz}, \& {Courvoisier}}]{2003ATel..190....1S}
{Sunyaev}, R.~A., {Grebenev}, S.~A., {Lutovinov}, A.~A., {et~al.} 2003, The Astronomer's Telegram, 190, 1

\bibitem[{{Tauris} \& {van den Heuvel}(2006)}]{2006csxs.book..623T}
{Tauris}, T.~M., \& {van den Heuvel}, E.~P.~J. 2006, {Formation and evolution of compact stellar X-ray sources}, Vol.~39, 623--665

\bibitem[{{Tetarenko} {et~al.}(2021){Tetarenko}, {Shaw}, {Manrow}, {Charles}, {Miller}, {Russell}, \& {Tetarenko}}]{2021MNRAS.501.3406T}
{Tetarenko}, B.~E., {Shaw}, A.~W., {Manrow}, E.~R., {et~al.} 2021, \mnras, 501, 3406, \dodoi{10.1093/mnras/staa3861}

\bibitem[{{Tomsick} {et~al.}(2014){Tomsick}, {Nowak}, {Parker}, {Miller}, {Fabian}, {Harrison}, {Bachetti}, {Barret}, {Boggs}, {Christensen}, {Craig}, {Forster}, {F{\"u}rst}, {Grefenstette}, {Hailey}, {King}, {Madsen}, {Natalucci}, {Pottschmidt}, {Ross}, {Stern}, {Walton}, {Wilms}, \& {Zhang}}]{2014ApJ...780...78T}
{Tomsick}, J.~A., {Nowak}, M.~A., {Parker}, M., {et~al.} 2014, \apj, 780, 78, \dodoi{10.1088/0004-637X/780/1/78}

\bibitem[{{Torrej{\'o}n} {et~al.}(2004){Torrej{\'o}n}, {Kreykenbohm}, {Orr}, {Titarchuk}, \& {Negueruela}}]{2004A&A...423..301T}
{Torrej{\'o}n}, J.~M., {Kreykenbohm}, I., {Orr}, A., {Titarchuk}, L., \& {Negueruela}, I. 2004, \aap, 423, 301, \dodoi{10.1051/0004-6361:20035743}

\bibitem[{{Torrej{\'o}n} {et~al.}(2010){Torrej{\'o}n}, {Negueruela}, {Smith}, \& {Harrison}}]{2010A&A...510A..61T}
{Torrej{\'o}n}, J.~M., {Negueruela}, I., {Smith}, D.~M., \& {Harrison}, T.~E. 2010, \aap, 510, A61, \dodoi{10.1051/0004-6361/200912619}

\bibitem[{{Torrej{\'o}n} \& {Orr}(2001)}]{2001A&A...377..148T}
{Torrej{\'o}n}, J.~M., \& {Orr}, A. 2001, \aap, 377, 148, \dodoi{10.1051/0004-6361:20011070}

\bibitem[{{Torrej{\'o}n} {et~al.}(2018){Torrej{\'o}n}, {Reig}, {F{\"u}rst}, {Martinez-Chicharro}, {Postnov}, \& {Oskinova}}]{2018MNRAS.479.3366T}
{Torrej{\'o}n}, J.~M., {Reig}, P., {F{\"u}rst}, F., {et~al.} 2018, \mnras, 479, 3366, \dodoi{10.1093/mnras/sty1628}

\bibitem[{{Valencic} \& {Smith}(2013)}]{2013ApJ...770...22V}
{Valencic}, L.~A., \& {Smith}, R.~K. 2013, \apj, 770, 22, \dodoi{10.1088/0004-637X/770/1/22}

\bibitem[{{Van} {et~al.}(2019){Van}, {Ivanova}, \& {Heinke}}]{2019MNRAS.483.5595V}
{Van}, K.~X., {Ivanova}, N., \& {Heinke}, C.~O. 2019, \mnras, 483, 5595, \dodoi{10.1093/mnras/sty3489}

\bibitem[{{van der Klis} \& {Bonnet-Bidaud}(1984)}]{1984A&A...135..155V}
{van der Klis}, M., \& {Bonnet-Bidaud}, J.~M. 1984, \aap, 135, 155

\bibitem[{{van der Meer} {et~al.}(2005){van der Meer}, {Kaper}, {di Salvo}, {M{\'e}ndez}, {van der Klis}, {Barr}, \& {Trams}}]{2005A&A...432..999V}
{van der Meer}, A., {Kaper}, L., {di Salvo}, T., {et~al.} 2005, \aap, 432, 999, \dodoi{10.1051/0004-6361:20041288}

\bibitem[{{van der Meij} {et~al.}(2021){van der Meij}, {Guo}, {Kaper}, \& {Renzo}}]{2021A&A...655A..31V}
{van der Meij}, V., {Guo}, D., {Kaper}, L., \& {Renzo}, M. 2021, \aap, 655, A31, \dodoi{10.1051/0004-6361/202040114}

\bibitem[{{van Kerkwijk} {et~al.}(1995){van Kerkwijk}, {van Paradijs}, {Zuiderwijk}, {Hammerschlag-Hensberge}, {Kaper}, \& {Sterken}}]{1995A&A...303..483V}
{van Kerkwijk}, M.~H., {van Paradijs}, J., {Zuiderwijk}, E.~J., {et~al.} 1995, \aap, 303, 483.
\newblock \doarXiv{astro-ph/9505070}

\bibitem[{{Vanderburg} {et~al.}(2020){Vanderburg}, {Rappaport}, {Xu}, {Crossfield}, {Becker}, {Gary}, {Murgas}, {Blouin}, {Kaye}, {Palle}, {Melis}, {Morris}, {Kreidberg}, {Gorjian}, {Morley}, {Mann}, {Parviainen}, {Pearce}, {Newton}, {Carrillo}, {Zuckerman}, {Nelson}, {Zeimann}, {Brown}, {Tronsgaard}, {Klein}, {Ricker}, {Vanderspek}, {Latham}, {Seager}, {Winn}, {Jenkins}, {Adams}, {Benneke}, {Berardo}, {Buchhave}, {Caldwell}, {Christiansen}, {Collins}, {Col{\'o}n}, {Daylan}, {Doty}, {Doyle}, {Dragomir}, {Dressing}, {Dufour}, {Fukui}, {Glidden}, {Guerrero}, {Guo}, {Heng}, {Henriksen}, {Huang}, {Kaltenegger}, {Kane}, {Lewis}, {Lissauer}, {Morales}, {Narita}, {Pepper}, {Rose}, {Smith}, {Stassun}, \& {Yu}}]{2020Natur.585..363V}
{Vanderburg}, A., {Rappaport}, S.~A., {Xu}, S., {et~al.} 2020, \nat, 585, 363, \dodoi{10.1038/s41586-020-2713-y}

\bibitem[{{Verbunt} \& {Zwaan}(1981)}]{1981A&A...100L...7V}
{Verbunt}, F., \& {Zwaan}, C. 1981, \aap, 100, L7

\bibitem[{{Waisberg} {et~al.}(2019){Waisberg}, {Dexter}, {Petrucci}, {Dubus}, \& {Perraut}}]{2019A&A...623A..47W}
{Waisberg}, I., {Dexter}, J., {Petrucci}, P.-O., {Dubus}, G., \& {Perraut}, K. 2019, Astronomy and Astrophysics, 623, A47, \dodoi{10.1051/0004-6361/201834746}

\bibitem[{{Walter} \& {Zurita Heras}(2007)}]{2007A&A...476..335W}
{Walter}, R., \& {Zurita Heras}, J. 2007, \aap, 476, 335, \dodoi{10.1051/0004-6361:20078353}

\bibitem[{{Walton} {et~al.}(2016){Walton}, {Tomsick}, {Madsen}, {Grinberg}, {Barret}, {Boggs}, {Christensen}, {Clavel}, {Craig}, {Fabian}, {Fuerst}, {Hailey}, {Harrison}, {Miller}, {Parker}, {Rahoui}, {Stern}, {Tao}, {Wilms}, \& {Zhang}}]{2016ApJ...826...87W}
{Walton}, D.~J., {Tomsick}, J.~A., {Madsen}, K.~K., {et~al.} 2016, \apj, 826, 87, \dodoi{10.3847/0004-637X/826/1/87}

\bibitem[{{Weisskopf} {et~al.}(1984){Weisskopf}, {Elsner}, {Darbro}, {Naranan}, {Weisskopf}, {Williams}, {White}, {Grindlay}, \& {Sutherland}}]{1984ApJ...278..711W}
{Weisskopf}, M.~C., {Elsner}, R.~C., {Darbro}, W., {et~al.} 1984, \apj, 278, 711, \dodoi{10.1086/161840}

\bibitem[{{White} {et~al.}(1977){White}, {Mason}, \& {Sanford}}]{1977Natur.267..229W}
{White}, N.~E., {Mason}, K.~O., \& {Sanford}, P.~W. 1977, \nat, 267, 229, \dodoi{10.1038/267229a0}

\bibitem[{{Williams} {et~al.}(2021){Williams}, {Durbin}, {Dalcanton}, {Lang}, {Girardi}, {Smercina}, {Dolphin}, {Weisz}, {Choi}, {Bell}, {Rosolowsky}, {Skillman}, {Koch}, {Lindberg}, {Hagen}, {Gordon}, {Seth}, {Gilbert}, {Guhathakurta}, {Lauer}, \& {Bianchi}}]{2021ApJS..253...53W}
{Williams}, B.~F., {Durbin}, M.~J., {Dalcanton}, J.~J., {et~al.} 2021, \apjs, 253, 53, \dodoi{10.3847/1538-4365/abdf4e}

\bibitem[{{Wizinowich} {et~al.}(2000){Wizinowich}, {Acton}, {Shelton}, {Stomski}, {Gathright}, {Ho}, {Lupton}, {Tsubota}, {Lai}, {Max}, {Brase}, {An}, {Avicola}, {Olivier}, {Gavel}, {Macintosh}, {Ghez}, \& {Larkin}}]{2000PASP..112..315W}
{Wizinowich}, P., {Acton}, D.~S., {Shelton}, C., {et~al.} 2000, \pasp, 112, 315, \dodoi{10.1086/316543}

\bibitem[{{Wolszczan} \& {Frail}(1992)}]{1992Natur.355..145W}
{Wolszczan}, A., \& {Frail}, D.~A. 1992, \nat, 355, 145, \dodoi{10.1038/355145a0}

\bibitem[{{Wood} {et~al.}(1984){Wood}, {Meekins}, {Yentis}, {Smathers}, {McNutt}, {Bleach}, {Byram}, {Chupp}, {Friedman}, \& {Meidav}}]{1984ApJS...56..507W}
{Wood}, K.~S., {Meekins}, J.~F., {Yentis}, D.~J., {et~al.} 1984, \apjs, 56, 507, \dodoi{10.1086/190992}

\bibitem[{{Wright} {et~al.}(2010){Wright}, {Eisenhardt}, {Mainzer}, {Ressler}, {Cutri}, {Jarrett}, {Kirkpatrick}, {Padgett}, {McMillan}, {Skrutskie}, {Stanford}, {Cohen}, {Walker}, {Mather}, {Leisawitz}, {Gautier}, {McLean}, {Benford}, {Lonsdale}, {Blain}, {Mendez}, {Irace}, {Duval}, {Liu}, {Royer}, {Heinrichsen}, {Howard}, {Shannon}, {Kendall}, {Walsh}, {Larsen}, {Cardon}, {Schick}, {Schwalm}, {Abid}, {Fabinsky}, {Naes}, \& {Tsai}}]{2010AJ....140.1868W}
{Wright}, E.~L., {Eisenhardt}, P. R.~M., {Mainzer}, A.~K., {et~al.} 2010, \aj, 140, 1868, \dodoi{10.1088/0004-6256/140/6/1868}

\bibitem[{{Yatabe} {et~al.}(2018){Yatabe}, {Makishima}, {Mihara}, {Nakajima}, {Sugizaki}, {Kitamoto}, {Yoshida}, \& {Takagi}}]{2018PASJ...70...89Y}
{Yatabe}, F., {Makishima}, K., {Mihara}, T., {et~al.} 2018, \pasj, 70, 89, \dodoi{10.1093/pasj/psy088}

\bibitem[{{Zhao} {et~al.}(2023){Zhao}, {Gandhi}, {Dashwood Brown}, {Knigge}, {Charles}, {Maccarone}, \& {Nuchvanichakul}}]{2023MNRAS.525.1498Z}
{Zhao}, Y., {Gandhi}, P., {Dashwood Brown}, C., {et~al.} 2023, \mnras, 525, 1498, \dodoi{10.1093/mnras/stad2226}

\bibitem[{{Zorec} {et~al.}(2005){Zorec}, {Fr{\'e}mat}, \& {Cidale}}]{2005A&A...441..235Z}
{Zorec}, J., {Fr{\'e}mat}, Y., \& {Cidale}, L. 2005, \aap, 441, 235, \dodoi{10.1051/0004-6361:20053051}

\bibitem[{{Zurita Heras} \& {Chaty}(2009)}]{2009A&A...493L...1Z}
{Zurita Heras}, J.~A., \& {Chaty}, S. 2009, \aap, 493, L1, \dodoi{10.1051/0004-6361:200811179}

\end{thebibliography}
\bibliographystyle{aasjournal}

\appendix

\begin{figure*}
    \centering
    \includegraphics[width=0.45\textwidth]{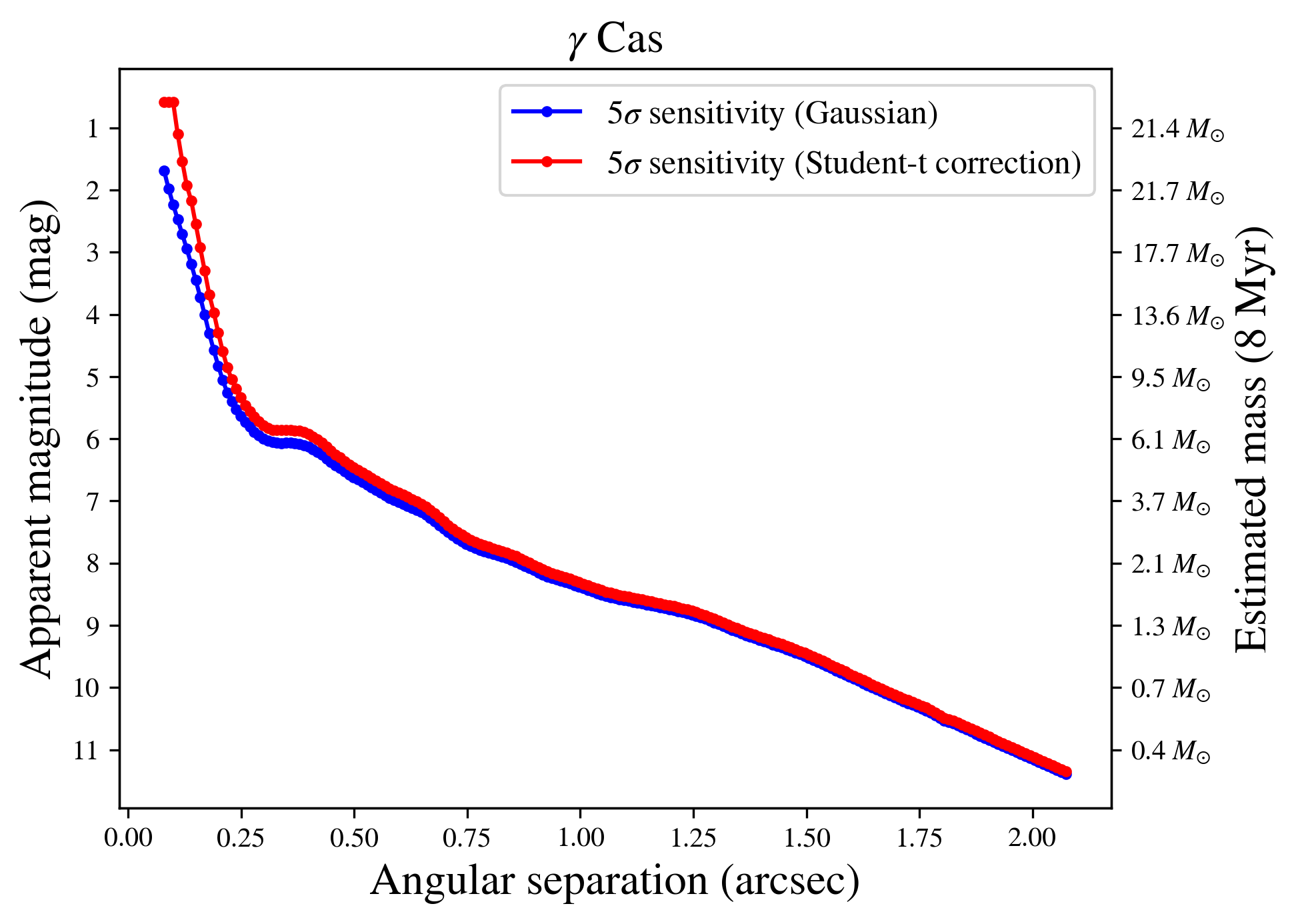}
    \includegraphics[width=0.45\textwidth]{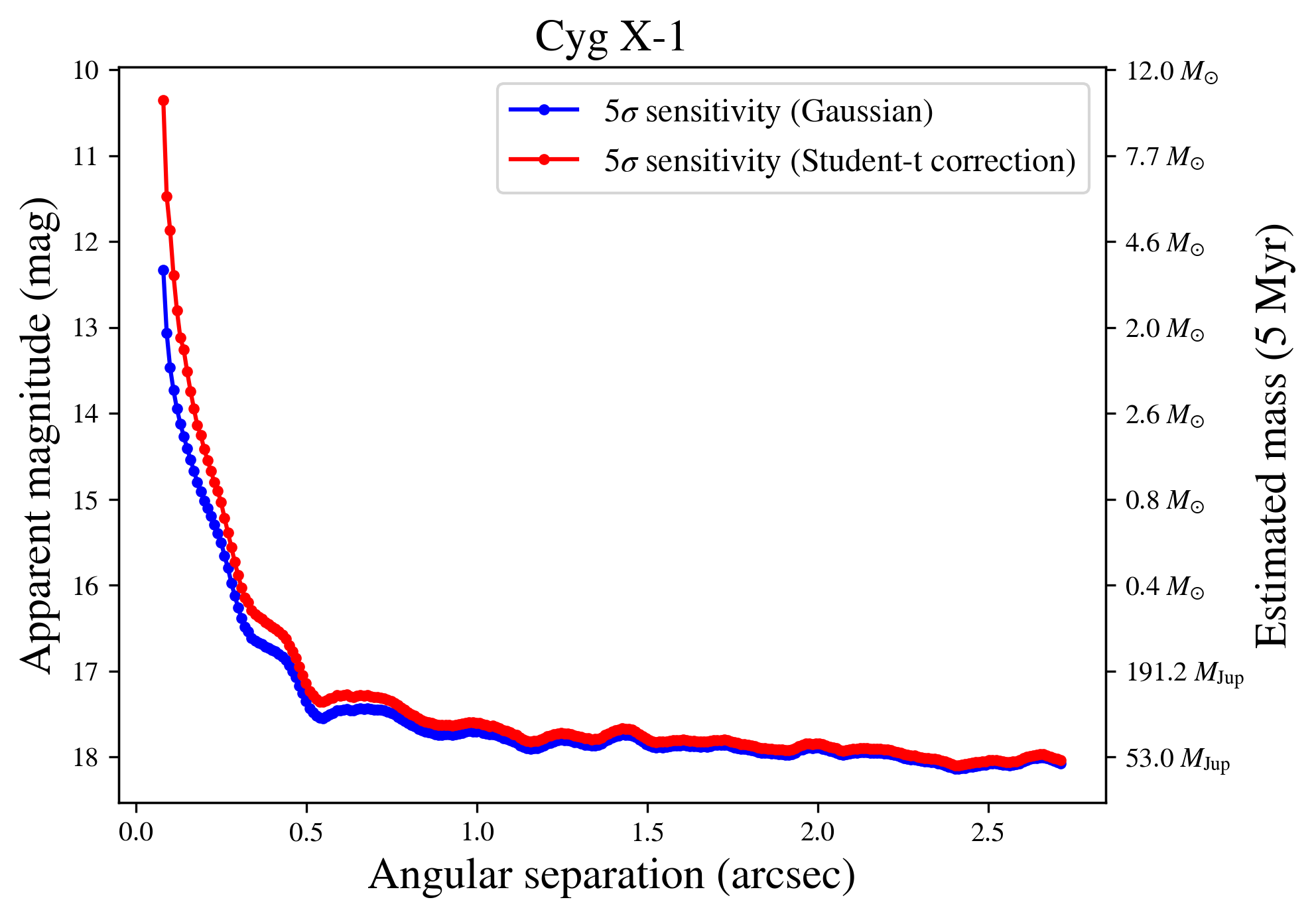}
    \includegraphics[width=0.45\textwidth]{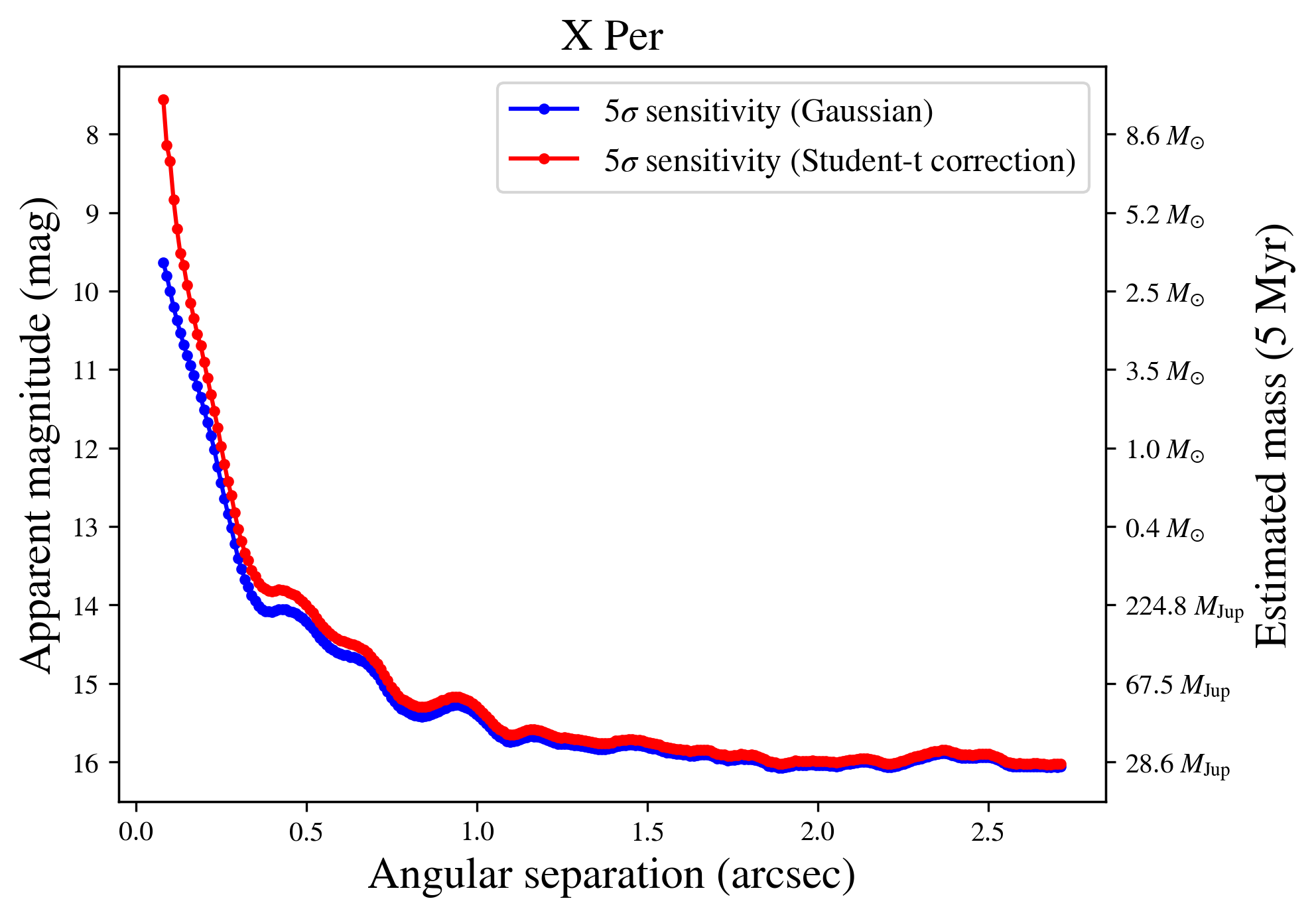}
    \includegraphics[width=0.45\textwidth]{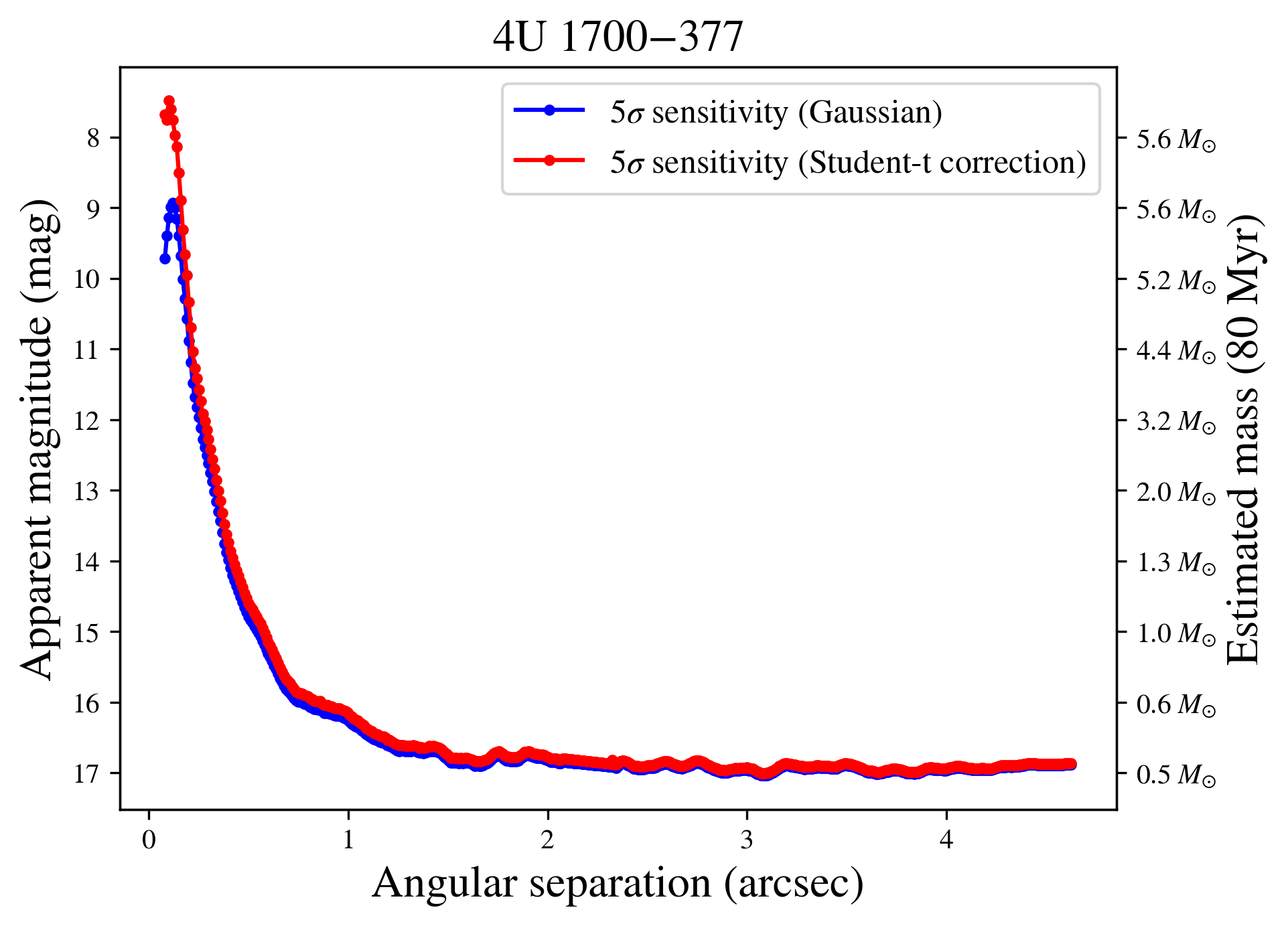}    
    \includegraphics[width=0.45\textwidth]{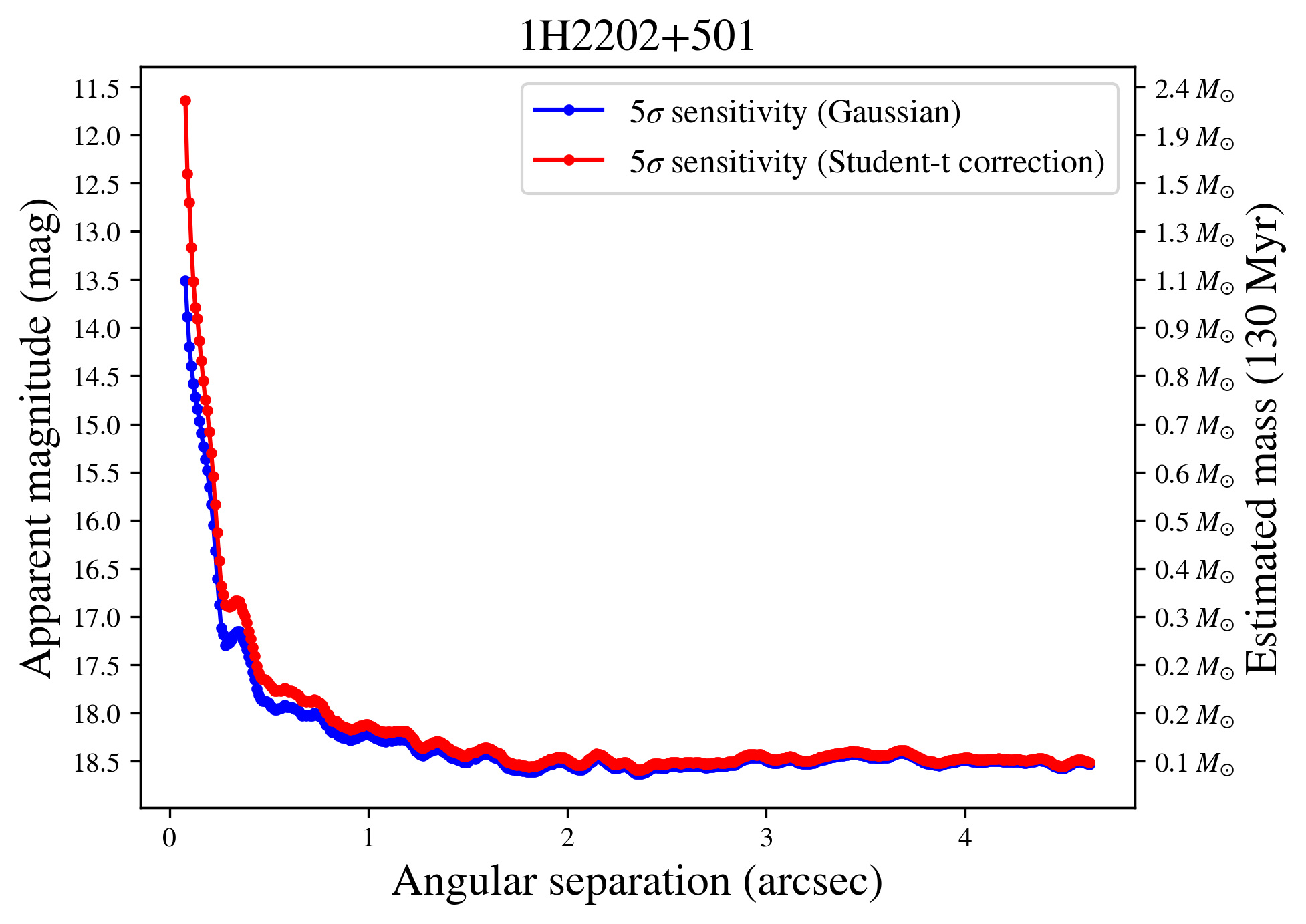}        
    \includegraphics[width=0.45\textwidth]{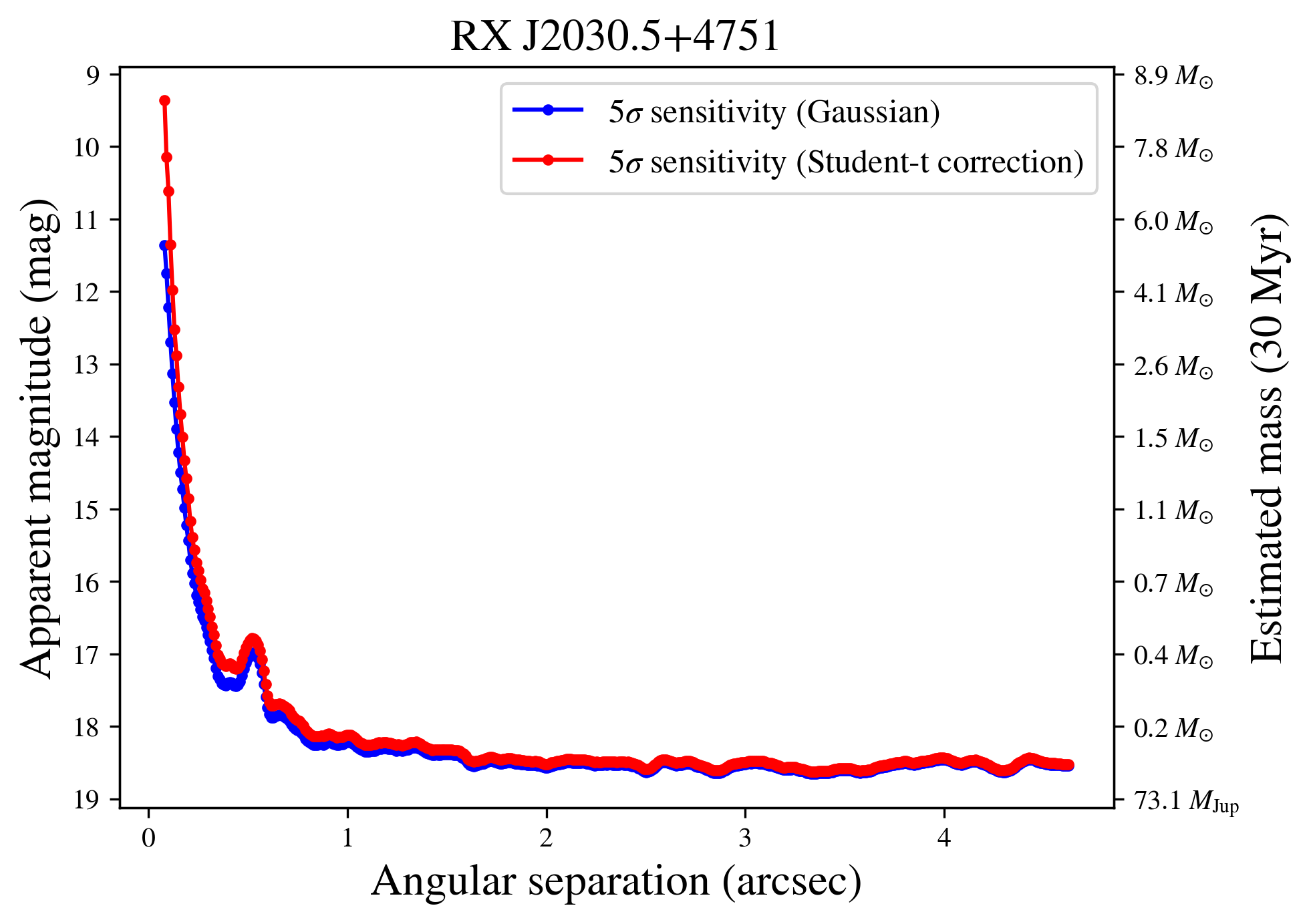} 
    \includegraphics[width=0.45\textwidth]{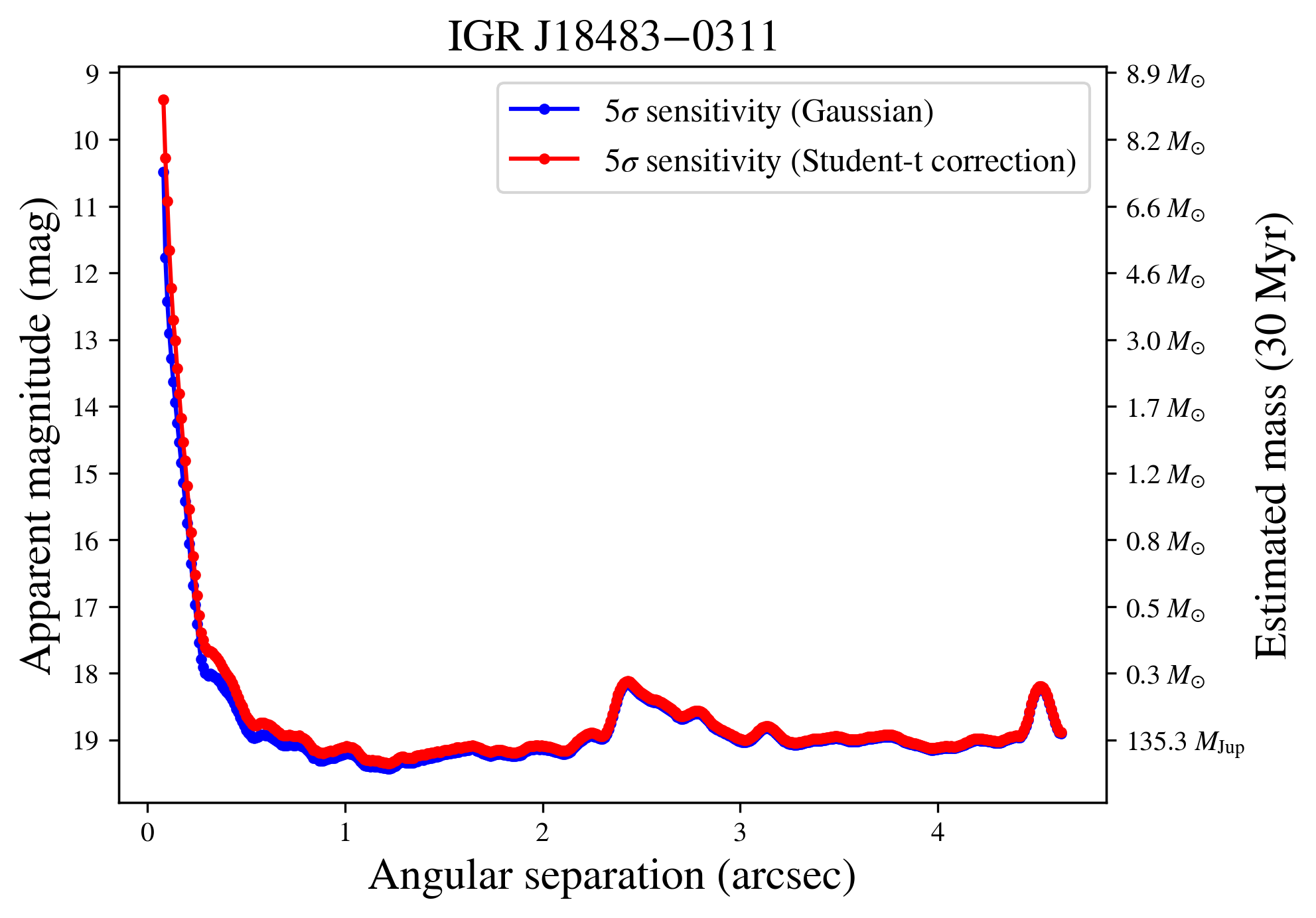} 
    \caption{5$\sigma$ contrast curves in apparent magnitude (left axis) and estimated mass from the age of the XRB (right axis) for $\gamma$ Cas, Cyg X-1, X Per, 4U 1700$-$377, 1H2202+501, RX J2030.5+4751 and IGR J18483$-$0311. The curves were generated using \texttt{VIP} for a Gaussian distribution (blue) and with a Student-t correction (red).}
    \label{fig:contrast_curves}
\end{figure*}

This Appendix presents additional data related to the main text. Figure \ref{fig:contrast_curves} presents the 5$\sigma$ contrast curves for all X-ray binaries with CBCs. Table \ref{tab:XRB_add_properties} presents additional relevant physical properties for the 14 observed X-ray binaries. Table \ref{tab:prop_sources_XRB} presents the properties of the detected sources in the high-contrast image, including the status (background/foreground source or candidate CBC), optimization parameters, physical properties (astrometric and photometric parameters), and estimated mass.

\begin{longrotatetable}
\begin{table*}

\caption{Additional relevant physical properties for the 14 observed X-ray binaries. The columns are: 1. Name of the target; 2. Distance in kpc (see Section \ref{sec:distance}); 3. Proper motion in right ascension in mas yr$^{-1}$; 4. Proper motion in declination in mas yr$^{-1}$; 5. Apparent magnitude in the $L'$-band (see Section \ref{sec:appmag}); 6. Estimated age of the X-ray binary in Myr (see Section \ref{sec:age}); 7. Value of the extinction and 8. the reference; 9. Observed values of the X-ray luminosity in erg s$^{-1}$ and 10. the reference; 11. The variability indicator, as found in \cite{2016ApJS..223...15B}. \textbf{Reference code}: (1) \cite{2011ApJ...737..103S}, (2) \cite{naze_hot_2018}, (3) \cite{2010A&A...510A..61T}, (4) \cite{2010MNRAS.401.1564R}, (5) \cite{2015MNRAS.449.1228S}, (6) \cite{1973A&A....23..433M}, (7) \cite{2005A&AT...24..151R}, (8) \cite{2012A&A...544A.118B}, (9) \cite{2019KPCB...35...38S}, (10) \cite{1998A&A...330..201C}, (11) \cite{2017BlgAJ..27...10N}, (12) \cite{2001A&A...371.1056N}, (13) \cite{2006A&A...449..687R}, (14) \cite{2020A&A...634A..49H}, (15) \cite{1992A&A...260..237L}, (16) \cite{2015A&A...576L...4R}, (17) \cite{2016A&A...596A..16B}, (18) \cite{2006A&A...455.1165L}, (19) \cite{2007BeSN...38...24R}, (20) \cite{2009ApJ...701.1895C}, (21) \cite{2001ApJ...547.1024D}, (22) \cite{1997MNRAS.286..549L}, (23) \cite{2001ApJ...552..738C}, (24) \cite{2001ApJ...554...27H}, (25) \cite{2021MNRAS.504..920M}, (26) \cite{1977A&AS...27..215K}, (27) \cite{2008A&A...492..511K}.
}
\centering
\footnotesize
\renewcommand{\arraystretch}{1.7}
\begin{tabular}{ccccccccccc}
\hline \hline
Target & Distance & PM RA & PM Dec & $m_\mathrm{L}$ & Age & $E(B-V)$ & Ref. & $L_\mathrm{X}$ & Ref. & Var. ind. \\
 & (kpc) & (mas yr$^{-1}$) & (mas yr$^{-1}$) & ($\pm$ 0.5 mag) & (Myr) &  &   &
(erg s$^{-1}$) &  & \\ \hline
\textbf{RX J1744.7$-$2713} & 1.22 $\pm$ 0.04 & -0.87 $\pm$ 0.02 & -2.30 $\pm$ 0.02 & 5.809 & Up to $\sim$60 & $\sim$3.4 & (1) & (3.08 $\pm$ 0.49) $\times$ 10$^{32}$ & (2) & N/A \\ \hline
\textbf{IGR J18483$-$0311} & 2.6 $\pm$ 0.7 & -1.7 $\pm$ 0.2 & -3.7 $\pm$ 0.1 & 7.907 & Up to $\sim$30-50 & 5.22 $\pm$ 0.02 & (3) & From $\sim$10$^{33}$ to 10$^{36}$ & (3, 4, 5) & N \\ \hline
\textbf{$\gamma$ Cas} & 0.19 $\pm$ 0.02 & 25.7 $\pm$ 0.5 & -3.8 $\pm$ 0.4 & -0.912 & 8.0 $\pm$ 0.4 & -0.15 & (6) & $\sim$10$^{32}$--10$^{33}$ & (7) & N \\ \hline
\textbf{SAX J1818.6$-$1703} & 2.3 $\pm$ 0.8 & -1.6 $\pm$ 0.2 & -4.6 $\pm$ 0.1 & 8.964 & Up to $\sim$30--50 & 5.08 $\pm$ 0.05 & (3) & From $\sim$10$^{32}$ to 10$^{35}$ & (3, 8) & Y \\\hline
\textbf{1H2202+501} & 1.10 $\pm$ 0.01 & 2.36 $\pm$ 0.01 & -0.29 $\pm$ 0.01 & 8.285 & Up to $\sim$130 & 0.36 $\pm$ 0.03 & (9) & $\sim$9 $\times$ 10$^{32}$ & (10) & N/A \\ \hline
\textbf{4U 2206+543} & 3.1 $\pm$ 0.1 & -4.17 $\pm$ 0.02 & -3.32 $\pm$ 0.01 & 8.700 & Up to $\sim$8 & 0.547 $\pm$ 0.066 & (11) & $\sim$10$^{35}$--10$^{36}$ & (12, 13) & N \\ \hline
\textbf{4U 1700$-$377} & 1.5 $\pm$ 0.1 & 2.41 $\pm$ 0.03 & 5.02 $\pm$ 0.02 & 5.36 & $\sim$80 & 0.50 $\pm$ 0.01 & (14) & Up to $\sim$7 $\times$ 10$^{36}$ & (15) & N \\ \hline
\textbf{IGR J17544$-$2619} & 2.4 $\pm$ 0.2 & -0.51 $\pm$ 0.03 & -0.67 $\pm$ 0.02 & 7.67 & Up to $\sim$8 & N/A &  & From $\sim 10^{32}$ to $10^{38}$ & (16, 17) & Y \\ \hline
\textbf{RX J2030.5+4751} & 2.30 $\pm$ 0.07 & -2.71 $\pm$ 0.02 & -4.54 $\pm$ 0.02 & 7.088 & Up to $\sim$30-50 & N/A &  & Up to $\sim$10$^{33}$ & (18, 19) & N/A \\ \hline
\textbf{Cyg X-1} & 2.1 $\pm$ 0.1 & -3.81 $\pm$ 0.01 & -6.31 $\pm$ 0.02 & 6.406 & 5 $\pm$ 1.5 & 1.11 $\pm$ 0.03 & (20) & $\sim$3 $\times$ 10$^{37}$ & (21) & N \\ \hline
\textbf{X Per} & 0.6 $\pm$ 0.1 & -1.28 $\pm$ 0.05 & -1.87 $\pm$ 0.03 & 4.596 & $\sim$5 & $\sim$0.4 & (22) & $\sim$4 $\times$ 10$^{34}$ & (23) & N \\ \hline
\textbf{1H0556+286} & 1.5 $\pm$ 0.1 & 0.63 $\pm$ 0.03 & -2.19 $\pm$ 0.02 & 7.618 & Up to $\sim$150 & N/A &  & Up to $\sim$4 $\times$ 10$^{35}$ & (24) & N/A \\ \hline
\textbf{RX J0648.1$-$4419} & 0.52 $\pm$ 0.01 & -4.16 $\pm$ 0.07 & 5.93 $\pm$ 0.06 & 9.150 & N/A & N/A &  & $\sim$10$^{32}$ & (25) & N/A \\ \hline
\textbf{Vela X-1} & 2.0 $\pm$ 0.1 & -4.82 $\pm$ 0.02 & 9.28 $\pm$ 0.02 & 5.458 & Up to $\sim$30-50 & $\sim$0.8 & (26) & Up to $\sim$4 $\times$ 10$^{36}$ & (27) & N \\ \hline
\end{tabular}
\label{tab:XRB_add_properties}
\end{table*}
\end{longrotatetable}

\begin{longrotatetable}
\begin{table*}
\caption{Physical properties of the detected sources in the high-contrast images. The columns are: 1. The target; 2. The label of the detected source (S/N $>$ 5); 3. The current status of the source, either background/foreground object (bkg) or candidate CBC (cc); 4. The optimal number of principal components used for the fitting ($n_\mathrm{comp}$; 5. The expected number of sources in the foV below the apparent magnitude ($n_\mathrm{foV} (\mathcal{L'} \le m_{L'})$; see Section \ref{sec:contamination}); 6. One minus the probability of being unrelated with the central X-ray binary ($1 - P_\mathrm{unrelated}(\Sigma, \Theta)$; see Section \ref{sec:contamination}); 7. The radial separation from the X-ray binary in mas; 8. The position angle in the image in degrees; 9. The apparent magnitude in the $L'$-band (no extinction correction was applied); 10. The mass of the source, in $M_\odot$ or $M_\mathrm{Jup}$, estimated from evolutionary models (MIST or COND/DUSTY); 11. The projected separation from the X-ray binary in au.}
\centering
\renewcommand{\arraystretch}{1.1}
\footnotesize
\begin{tabular}{ccccccccccc}
\hline \hline
Target & Source & Status & $n_\mathrm{comp}$ & $n_\mathrm{foV} (\mathcal{L'} \le m_{L'})$ & $1 - P_\mathrm{unrelated}(\Sigma, \Theta)$ & $\rho$ & $\theta$ & $m_{L'}$ & Est. Mass & Proj. sep. \\
 & (S/N $>$ 5) & & & & (\%) & (mas) & (deg) & (mag) &  & (au) \\ \hline
\textbf{X Per} & B & cc & 20 & $\sim$0.1 & $>$ 99 & 587 $\pm$  18 & 112.3 $\pm$  2.0 & 14.4 $\pm$  0.6 & $\sim$45$-$110 $M_\mathrm{Jup}$ & 350 \\ \hline
\textbf{Cyg X-1} & B & cc & 26 & $\sim$0.6 & 93 $\pm$ 1 & 1853 $\pm$ 12 & 170.1 $\pm$ 0.6 & 16.2 $\pm$ 0.6 & $\sim$0.2$-$0.3   $M_\mathrm{\odot}$ & 4000 \\ \hline
\multirow{8}{*}{\textbf{IGR J18483$-$0311}} & B & cc & 2 & 0.3$-$0.8 & 90 $\pm$ 4 & 2527 $\pm$ 12 & 291.5 $\pm$ 0.2 & 13.3 $\pm$ 0.8 & $\sim$2.5 $M_\mathrm{\odot}$ & 6440 \\
 & C & cc & 3 & 0.9$-$1.8 & 89 $\pm$ 4 & 1516 $\pm$ 13 & 21.6 $\pm$ 0.4 & 14.8 $\pm$ 0.8 & $\sim$1.1$-$1.3 $M_\mathrm{\odot}$ & 3865 \\
 & D & bkg & 7 & 3.4$-$7.0 & 32 $\pm$ 15 & 2787 $\pm$ 14 & 16.9 $\pm$ 0.2 & 16.3 $\pm$ 0.8 & -- & 7100 \\
 & E & bkg & 13 & 4.1$-$8.3 & 14 $\pm$ 10 & 3127 $\pm$ 15 & 13.8 $\pm$ 0.2 & 16.7 $\pm$ 0.8 & -- & 7975 \\
 & F & bkg & 17 & 1.3$-$3.1 & 23 $\pm$ 10 & 4522 $\pm$ 14 & 15.0 $\pm$ 0.1 & 15.3 $\pm$ 0.8 & -- & 11530 \\
 & G & bkg & 25 & $\sim$7 & $<$ 5 & 3761 $\pm$ 12 & 69.7 $\pm$ 0.2 & 17.0 $\pm$ 0.8 & -- & 9600 \\
 & H & bkg & 23 & $\sim$11 & $<$ 5 & 3358 $\pm$ 13 & 83.3 $\pm$ 0.2 & 17.6 $\pm$ 0.8 & -- & 8560 \\
 & I & bkg & 19 & $\sim$11 & $<$ 5 & 3168 $\pm$ 15 & 106.8 $\pm$ 0.2 & 17.6 $\pm$ 0.8 & -- & 8080 \\ \hline
 \multirow{2}{*}{\textbf{SAX J1818.6$-$1703}} & B & bkg & 3 & 1.5$-$3.5 & 25 $\pm$ 10 & 2820 $\pm$ 11 & 96.9 $\pm$  0.2 & 18.2 $\pm$ 0.8 & -- & 6600 \\
 & C & bkg & 20 & $>$ 30 & $<$ 5 & 4069 $\pm$  12 & 353.9 $\pm$  0.2 & 14.9 $\pm$  0.8 & -- & 9525 \\ \hline
 \textbf{1H2202+501} & B & cc & 14 & 0.2$-$0.5 & 98 $\pm$ 1 & 1222 $\pm$ 15 & 162.4 $\pm$ 0.5 & 16.4 $\pm$ 0.9 & $\sim$70 $M_\mathrm{Jup}$ to $\sim$0.4 $M_\mathrm{\odot}$ & 1370 \\ \hline
 \multirow{3}{*}{\textbf{4U 1700$-$377}} & B & cc & 8 & 0.5$-$0.7 & 91 $\pm$ 3 & 2238 $\pm$ 12 & 353.0 $\pm$ 0.3 & 14.0 $\pm$ 0.6 & $\sim$1.2$-$1.3   $M_\mathrm{\odot}$ & 4075 \\
 & C & cc & 7 & 1.3$-$1.9 & 68 $\pm$ 4 & 2616 $\pm$ 12 & 274.4 $\pm$ 0.2 & 13.4 $\pm$ 0.6 & $\sim$1.5$-$1.6 $M_\mathrm{\odot}$ & 4760 \\
 & D & cc & 3 & 0.7$-$0.9 & 65 $\pm$ 6 & 4157 $\pm$ 15 & 283.4 $\pm$ 0.1 & 15.4 $\pm$ 0.7 & $\sim$0.4$-$0.8 $M_\mathrm{\odot}$ & 7570 \\ \hline
 \multirow{9}{*}{\textbf{IGR J17544$-$2619}} & B & bkg & 5 & $\sim$48 & $<$ 5 & 827 $\pm$ 12 & 19.4 $\pm$ 0.7 & 16.9 $\pm$ 0.8 & -- & 2350 \\
 & C & bkg & 12 & $\sim$13 & $<$ 5 & 2993 $\pm$ 13 & 10.8 $\pm$ 0.2 & 15.1 $\pm$ 0.8 & -- & 8500 \\
 & D & bkg & 3 & $\sim$6 & $<$ 5 & 4099 $\pm$ 11 & 72.5 $\pm$ 0.1 & 13.4 $\pm$ 0.8 & -- & 11640 \\
 & E & bkg & 15 & $\sim$86 & $<$ 5 & 3170 $\pm$ 14 & 70.7 $\pm$ 0.2 & 17.3 $\pm$ 0.8 & -- & 9000 \\
 & F & bkg & 11 & $\sim$14 & $\sim$7 & 2392 $\pm$ 14 & 103.5 $\pm$ 0.2 & 15.3 $\pm$ 0.8 & -- & 6790 \\
 & G & bkg & 3 & $\sim$22 & $<$ 5 & 2947 $\pm$ 12 & 118.7 $\pm$ 0.2 & 16.1 $\pm$ 0.8 & -- & 8370 \\
 & H & bkg & 8 & $\sim$8 & $<$ 5 & 3935 $\pm$ 13 & 147.7 $\pm$ 0.1 & 14.1 $\pm$ 0.8 & -- & 11180 \\
 & I & bkg & 10 & $\sim$68 & $<$ 5 & 2093 $\pm$ 11 & 235.1 $\pm$ 0.3 & 17.1 $\pm$ 0.8 & -- & 5950 \\
 & J & bkg & 15 & $\sim$85 & $<$ 5 & 3537 $\pm$ 15 & 267.1 $\pm$ 0.2 & 17.3 $\pm$ 0.8 & -- & 10050 \\ \hline
\multirow{3}{*}{\textbf{RX J2030.5+4751}} & B & cc & 7 & $\sim$0.1 & $>$ 99 & 513 $\pm$ 11 & 302.9 $\pm$ 0.9 & 14.8 $\pm$ 0.8 & $\sim$0.3$-$1.1 $M_\mathrm{\odot}$ & 1130 \\
 & C & cc & 17 & 0.2$-$0.3 & 92 $\pm$ 3 & 3251 $\pm$ 15 & 132.1 $\pm$ 0.3 & 16.1 $\pm$ 0.8 & $\sim$0.1$-$0.6 $M_\mathrm{\odot}$ & 7150 \\
 & D & cc & 28 & 0.3$-$0.7 & 74 $\pm$ 8 & 4208 $\pm$ 13 & 343.1 $\pm$ 0.3 & 17.1 $\pm$ 0.8 & $\sim$60$-$400 $M_\mathrm{Jup}$ & 9250 \\ \hline
 \multirow{2}{*}{\textbf{$\gamma$ Cas (2017)}} & B & -- & 1 & -- & -- & 2051 $\pm$ 20 & 242.8 $\pm$ 0.3 & -- & -- & 390 \\
 & C & -- & 24 & -- & -- & 1803 $\pm$ 20 & 91.4 $\pm$ 0.3 & -- & -- & 343 \\ \hline
\multirow{2}{*}{\textbf{$\gamma$ Cas (2020)}} & B & cc & 1 & -- & -- & 2056 $\pm$ 20 & 241.2 $\pm$ 0.3 & 3.1 $\pm$ 0.9 & $\sim$ 13 $M_\mathrm{\odot}$ & 391 \\
 & C & bkg & 24 & -- & -- & 1721 $\pm$ 20 & 90.8 $\pm$ 0.3 & 10.4 $\pm$ 0.9 & -- & 327 \\ \hline
\end{tabular}
\label{tab:prop_sources_XRB}
\end{table*}
\end{longrotatetable}

\end{document}